\documentclass[aps,preprint,floatfix,nofootinbib,showpacs]{revtex4-1}
\pdfoutput=1
\usepackage{dcolumn}
\usepackage{bm}
\usepackage{graphicx}
\usepackage{amssymb,amsmath}
\usepackage{multirow}
\usepackage{color,url}
\usepackage{tabu}
\usepackage{array}
\usepackage[colorlinks=true,urlcolor=blue,anchorcolor=blue
,citecolor=blue,filecolor=blue,linkcolor=blue,menucolor=blue
,linktocpage=true,pdfproducer=medialab,pdfa=true]{hyperref}
\usepackage{colordvi}
\def\ga{\mathrel{\raise.3ex\hbox{$>$\kern-.75em\lower1ex\hbox{$\sim$}}}}
\def\la{\mathrel{\raise.3ex\hbox{$<$\kern-.75em\lower1ex\hbox{$\sim$}}}}
\def\beqa{\begin{eqnarray}}
\def\eeqa{\end{eqnarray}}

\begin{document}

\title{
  Same-sign Charged Higgs Pair Production in bosonic decay channels
  at the HL-LHC and HE-LHC }
\def\slash#1{#1\!\!/}

\renewcommand{\thefootnote}{\arabic{footnote}}
\renewcommand{\thefootnote}{\arabic{footnote}}

\author{
  Abdesslam  Arhrib$^{1,2}$, Kingman Cheung$^{2,3,4}$, Chih-Ting Lu$^5$ }
\affiliation{
  $^1$ D\'{e}partement de Math\'ematique, Facult\'e des Sciences et Techniques,
  Universit\'{e}  Abdelmalek Essaadi, B. 416, Tangier, Morocco \\
  $^2$ Physics Division, National Center for Theoretical Sciences, Hsinchu, Taiwan 300 \\
 $^3$ Department of Physics, National Tsing Hua University,
Hsinchu 300, Taiwan \\
$^4$ Division of Quantum Phases and Devices, School of Physics, 
Konkuk University, Seoul 143-701, Republic of Korea \\
$^5$ School of Physics, KIAS, Seoul 130-722, Republic of Korea
}
\date{\today}

\begin{abstract}
  Same-sign charged Higgs pair production via vector-boson scattering
  is a useful probe of the mass spectrum among the heavier scalar,
  pseudoscalar, and charged Higgs bosons 
  in two-Higgs-doublet models. It has been shown
  that the production cross section scales as the square of the mass
  difference $\Delta m = (m_{H^0} - m_{A^0})$ in the alignment limit
  $(\cos(\beta - \alpha)=0)$.  We study the potential measurement of
  this same-sign charged Higgs pair production at the high-luminosity
  LHC (HL-LHC) and the proposed 27 TeV $pp$ collider, with emphasis in
  the decay channel $H^\pm H^\pm \to (W^\pm A^0 ) (W^\pm A^0)$, which
  is in general the dominant mode when the charged Higgs mass is above
  the $W^\pm A^0$ threshold.
  We also examine the current allowed parameter space taking into
  account the theoretical constraints on the model, the electroweak
  precision test (EWPT) measurements, $B$ decays, and direct searches in the
  $H^\pm \to \tau^\pm \nu_\tau$ and $H^\pm \to W ^\pm A^0 \to (\ell^\pm
  \nu_{\ell} ) (\mu \mu)$.
\end{abstract}

\maketitle

\section{Introduction}

Since the discovery of a Higgs-like particle at the CERN Large Hadron
Collider (LHC) in 2012, there have been many theoretical and
phenomenological studies dedicated to non-minimal Higgs sector models that
can explain the observed Higgs-like particle and account for some weakness
of the Standard Model (SM).  One common feature of many extensions of the minimal
Higgs sector is the presence of extra neutral Higgs bosons
as well as singly-charged Higgs bosons in the physical spectrum.
Therefore, the discovery of charged Higgs bosons would be an
unambiguous sign of physics beyond the SM. One of the most popular
models with extended Higgs sector is the two-Higgs-Doublet Model
(2HDM) \cite{Lee:1973iz,Branco:2011iw} in which one introduces two Higgs doublet
fields to break the $SU_L(2)\times U_Y(1)$ symmetry down to the
$U(1)_{\rm em}$ symmetry.  In order to avoid tree-level
flavor-changing neutral
current couplings, one can advocate a natural flavor
conservation imposed by a discrete  $Z_2$ symmetry
\cite{Glashow:1976nt}.
Depending on the Higgs and fermion field transformations under the
$Z_2$, one can have a number of Yukawa textures for the fermion
sector, denoted by Type I, II, X, and Y 2HDM's.  After electroweak symmetry
breaking driven by the two Higgs fields takes place, the physical
spectrum of the model consists of 2 CP-even Higgs bosons $h^0$, $H^0$
(one of them could be identified as the observed 125 GeV Higgs-like
particle), a CP-odd Higgs boson $A^0$ and a pair of charged Higgs
bosons $H^\pm$.

At hadron colliders, charged Higgs bosons can be produced in a number of 
channels.  An important source of light charged Higgs bosons is from 
$t\bar{t}$ production, followed by the top decay into a charged
Higgs boson and a bottom quark if kinematically allowed.
Other important mechanisms for singly-charged Higgs production are the QCD processes
$gb \to t H^-$ and $gg\to t\bar{b}H^-$ \cite{Barger:1993th}. We refer
to Ref.~\cite{Akeroyd:2016ymd} for an extensive review on charged Higgs
phenomenology.   Charged Higgs bosons have been searched for in the past
at both LEP \cite{Abbiendi:2013hk} and Tevatron
\cite{Abazov:2009wy}. An upper limit of the order of 80 GeV has been
set at LEP experiments both from fermionic and bosonic decays 
$H^\pm\to W^\pm A^0$ \cite{Abbiendi:2013hk}.
While at the Tevatron a search for
the charged Higgs from top decay had been performed in various decay
channels of $H^\pm$ and limits on $B (t\to H^+b)$ have been set
\cite{Abazov:2009wy}.
At the LHC, one can search for 
light $H^\pm$ from top decay and for heavy $H^\pm$ from $gb \to t H^-$
and $gg\to t\bar{b}H^-$.   Light charged Higgs boson ($\leq m_t - m_b$)
would decay dominantly into $\tau \nu_\tau$, $c\bar{s}$ or $c\bar{b}$ final
states. In case of light pseudoscalar boson $A^0$,
$H^\pm$ can also decay into $W^\pm A^0$.  
However, a heavy $H^+$ can also decay into $t\bar{b}$, $W^+ h^0$, $W^+A^0$,
or $W^+ H^0$ if kinematically open.
Both at the LHC Run-1 and Run-2, ATLAS and
CMS had already set exclusion limits on
$B(t \to bH^+) \times B(H^+\to \tau^+ \nu_\tau)$
\cite{ATLAS1,ATLAS2,ATLAS3,CMS1,CMS2,CMS3}, which can be used to set limits on
$\tan\beta$ for a given charged Higgs mass less than
$\leq m_t - m_b$.  Moreover, 
from $t\to b H^+$ there has been also a search for $H^+\to c\bar s$
channel
both by ATLAS and CMS
\cite{Aad:2013hla} at 7 TeV and 8 TeV. The limit obtained on $B(t \to bH^+)$ is rather weak 
compared to the $ \tau \nu_{\tau}$ mode.
Both ATLAS and CMS also searched for $H^\pm \to tb$ decay, in which no $H^\pm$ signal was
observed and upper limits on the $\sigma(pp\to tbH^\pm) \times
B(H^\pm \to tb)$ are set
\cite{ATLAS:2016qiq,Aaboud:2018cwk,CMS:1900zym,CMS:2019yat}.


In the 2HDM, it has been shown
\cite{Arhrib:2016wpw,Kling:2015uba,Coleppa:2014cca}
that the charged Higgs boson
can decay dominantly into the
bosonic final state $H^\pm \to W^\pm A^0$ when kinematically open. 
Other models
beyond SM could also have similar features such as 2HDM with 
singlet scalars \cite{Dermisek:2012cn} and also the next-to-minimal
supersymmetric standard model \cite{Akeroyd:2007yj}.
  At LEPII \cite{Abdallah:2003wd}, pair-produced charged Higgs bosons
  have been searched in various final states, including 
  $\tau^+ \nu_{\tau} \tau^- \bar{\nu}_{\tau}$, $c\bar{s} \bar{c} s$,
  $c\bar{s} \tau^- \bar \nu_{\tau}$, $W^* A W^* A$ and
  $W^* A \tau^- \bar \nu_\tau$,
  and an upper limit of the order 80 GeV was set on the charged Higgs mass.
Recently, CMS also performed
a search for such bosonic decays of the charged Higgs
\cite{Sirunyan:2019zdq}. The study was only dedicated to light charged
Higgs produced from top decay followed by $H^\pm \to W^\pm A^0$, where
$A^0$ decays into a pair of muons and $W^\pm$ decays into a
charged lepton $(e,\mu)$ and a neutrino.
Assuming that $H^\pm$
decays 100\% into $W^\pm A^0$ and $B(A^0 \to \mu\mu)=3\times 10^{-4}$,
CMS set a new and first limit from bosonic decay of $H^\pm$ on
$B(t\to bH^+)$.

Recently, Ref.~\cite{Aiko:2019mww} proposed a new mechanism where
a pair of same-sign charged Higgs bosons are produced via vector boson 
fusion (VBF) at hadron colliders.  Such a
process can shade some light on the 
global symmetry of the underlying scalar potential.
Assuming that the charged Higgs bosons decay into
$\tau \nu_{\tau}$ or $tb$, Ref.\cite{Aiko:2019mww} evaluated the
signal and the SM backgrounds, and discussed the feasibility of the new
process both for the high-luminosity LHC (HL-LHC)
with 14 TeV center of mass energy and also for
the future high-energy LHC (HE-LHC) 27 TeV.

In this work, motivated by the recent CMS search for the bosonic decay
$H^\pm\to W^\pm A^0$, we investigate same-sign charged Higgs production
from VBF, followed by bosonic decays of the charged Higgs boson:
\begin{eqnarray}
pp \to j j W^{\pm *} W^{\pm *} \to j j H^\pm H^\pm 
\to jj (W^\pm A^0) (W^\pm A^0) \;.
\label{eq:2W2A2j}
\end{eqnarray}
in Type-I and X 2HDM's.
We calculate the signal and various SM backgrounds, and estimate the sensitivity
at the HL-LHC  as well as for the future hadron collider HE-LHC with 27 TeV
center of mass energy.
  Another important observation that motivates our work is because
  the fermionic production modes for these new scalars are highly
  suppressed by large $\tan\beta$ in both Type-I and Type-X 2HDMs,
  the discovery of these new scalars via fermionic modes is indeed
  challenging at the LHC. Therefore, we are exploring the
  bosonic decay mode of the charged scalar $H^\pm \to W^\pm A^0$, which dominates
  for $\tan\beta > 5$.

We should emphasize, instead of studying each new scalar (or two of
them) in different processes separately, the novel process we consider
here involves the effects of all new scalar masses. It means that we
have the chance to simultaneously test the whole mass spectrum in the
2HDM for some specific mass relations via a single process.  Finally,
we show that the mass spectrum of $ m_{A^0} = 63-100 $ GeV and
$\Delta m \equiv m_{H^0} - m_{A^0} = 200-250 $ GeV in Type-I and X 2HDM's
can be explored at the HE-LHC in the future 
after the accelerator and detector are further upgraded.
  
The strategy in this work is two-fold.  If any new scalar has been
discovered in the future, the signal process considered here serves
to confirm or rule out some specific mass spectra in the 2HDMs.
On the other hand, if we do not find any positive evidence of new
scalars in the future, the signal process in this study can also help
to clarify which kind of mass spectra in 2HDMs is not preferred.
  
The organization is as follows. In the next section, we describe
briefly the 2HDM's and relevant interactions. In Sec. III, we discuss
the constraints on the model from theoretical requirements,
electroweak precision test measurements,
$B$ decays, and direct searches. In Sec. IV, we calculate the
same-sign charged Higgs production cross sections, and perform the
signal-background analysis. We conclude in Sec. V.

\section{Brief review of two-Higgs-doublet models}
Many beyond Standard Model process an extended Higgs sector with more Higgs doublets, Higgs singlets , Higgs triplets or a mixture of all. One of the simplest, popular and well motivated extension of the SM is the two Higgs doublets model. A variety of which can be used in the minimal supersymmetric SM.
In the two-Higgs-Doublet Model (2HDM), two Higgs doublet
fields $\Phi_{1,2}$ with hypercharge $Y_{\Phi_{1,2}}=1/2$ are introduced.  
The most general renormalizable scalar potential, which respects the 
$SU_L(2)\otimes U_Y(1)$ gauge symmetry, has the following form:
\begin{eqnarray}
V(\Phi_1,\Phi_2) &=&
m^2_{11}\, \Phi_1^\dagger \Phi_1
+ m^2_{22}\, \Phi_2^\dagger \Phi_2 + 
\left(m_{12}^2\, \Phi_1^\dagger \Phi_2 + h.c\right) +\frac{\lambda_1}{2} \left( \Phi_1^\dagger \Phi_1 \right)^2
+ \frac{\lambda_2}{2} \left( \Phi_2^\dagger \Phi_2 \right)^2 \nonumber \\
&+& \lambda_3\, \Phi_1^\dagger \Phi_1\, \Phi_2^\dagger \Phi_2 + \lambda_4\, \Phi_1^\dagger \Phi_2\, \Phi_2^\dagger \Phi_1 +
\left[\frac{\lambda_5 }{2} \left( \Phi_1^\dagger \Phi_2 \right)^2+ h.c\right] 
\nonumber \\ 
&+&
\left[ \left( \lambda_6 \left( \Phi_1^\dagger \Phi_1 \right)  +
 \lambda_7 \left( \Phi_2^\dagger \Phi_2 \right) \right)  \left( \Phi_1^\dagger \Phi_2 \right) + h.c \right] 
\label{eq:pot}
\end{eqnarray}
where $m_{11}^2$, $m_{22}^2$ and $\lambda_{1,2,3,4}$ are real, while
$m_{12}^2$ and $\lambda_{5,6,7}$ could be complex for CP violation
purpose.
If we require that the potential to be invariant under a discrete $Z_2$ symmetry
$\Phi_1\to \Phi_1$, $\Phi_2\to -\Phi_2$
which is needed for natural flavor conservation in the Yukawa sector
(see discussion below), this would lead to $\lambda_{6,7}=0$. One
can still allow a soft violation of the discrete symmetry by a
dimension two terms $m_{12}^2$. In what follow, we assume that
$\lambda_{6,7}=0$ and $m_{12}^2\neq 0$.

Assuming that both $\Phi_1$ and $\Phi_2$ acquire a
vacuum expectation value (VEV) $v_{1,2}$
that can induce electroweak symmetry breaking,
the  two complex scalar $SU_L(2)$ doublets can be decomposed according to
\beqa
\Phi_i = \left( \begin{array}{c} \phi_i^+ \\
\left(v_i +  \rho_i + i \eta_i \right) \left/ \sqrt{2} \right.
\end{array} \right),
\quad i= 1, 2 \;.
\label{eq:doubletcomponents}
\eeqa
The mass eigenstates for the Higgs sector are obtained by  orthogonal transformations,
\beqa
\left(\begin{array}{c}
\phi_1^\pm\\
\phi_2^\pm
\end{array}\right)&=R_\beta
\left(\begin{array}{c}
G^\pm\\
H^\pm
\end{array}\right),\quad 
\left(\begin{array}{c}
\rho_1\\
\rho_2
\end{array}\right)=R_\alpha
\left(\begin{array}{c}
H^0\\
h^0
\end{array}\right) ,\quad 
\left(\begin{array}{c}
\eta_1\\
\eta_2
\end{array}\right)
=R_\beta\left(\begin{array}{c}
G^0\\
A^0
\end{array}\right) ,
\eeqa
with the generic form $(\theta = \alpha,\beta)$
$$R_\theta = 
\left(
\begin{array}{cc}
\cos\theta & -\sin\theta\\
\sin\theta & \cos\theta
\end{array}\right). $$

From the eight degrees of freedom initially present in the two scalar doublets, 
three of them, namely the Goldstone bosons 
$G^\pm$ and $G^0$, are eaten by the longitudinal component
of $W^\pm$ and $Z^0$, respectively. 
The remaining five degrees of freedom should 
manifest as physical Higgs bosons: two CP-even $H^0$ and $h^0$, one CP-odd $A^0$, 
and a pair of charged scalars $H^\pm$.
In the CP conserving case, the above potential contains 10 parameters
(including the VEV's of the Higgs doublets). $m_{11}^2$ and
$m_{22}^2$ can be eliminated by the use of the 2 minimization
conditions. One of the VEV's can be traded from the $W$
mass as a function of
the ratio $\tan\beta \equiv v_2/v_1$. We are
then left with seven independent
parameters which can be taken as: the four physical masses $m_h$, $m_H$, $m_A$ and $m_{H\pm}$, 
CP-even mixing angle $\alpha$, $\tan\beta$ and $m_{12}^2$.

In the Yukawa sector, it is well known that if we assume that
both Higgs doublets couple to all fermions we will end up with
large Flavor-Changing Neutral Currents (FCNC) mediated by the neutral
Higgs scalars at tree level.
In order to avoid such FCNC's, a discrete $Z_2$ symmetry
(where $\Phi_1 \to \Phi_1$ and $\Phi_2 \to -\Phi_2$) is 
imposed \cite{Glashow:1976nt}. Note that in the above potential,
the $Z_2$ symmetry is only violated by the dimension-two
term involving $m_{12}^2$. Depending on the $Z_2$
charge assignment to the lepton and quark fields 
\cite{Aoki:2009ha,Branco:2011iw,Barger:1989fj}, one can have $4$ different types of 
Yukawa textures.
\footnote{
  Here we follow the same notation as in Ref.~\cite{Aoki:2009ha}.
  }  
In the type-I model, only the second doublet
$\Phi_2$ interacts with all the fermions like in the SM 
 while in the type-II model $\Phi_1$ interacts
with the charged leptons and down-type quarks and 
$\Phi_2$ interacts with up-type quarks.
In the type-X (lepton-specific) model, charged leptons couple
to $\Phi_1$ while all the quarks 
couple to $\Phi_2$. Finally, in the type-Y (flipped) model 
down-type quarks acquire masses from their couplings to $\Phi_1$ while
charged leptons and up-type quarks couple to $\Phi_2$. 
The most general Yukawa interaction can be written as follows 
\cite{Branco:2011iw},
\beqa
-{\mathcal L}_\text{Yukawa}^\text{2HDM} =
&{\overline Q}_LY_u\widetilde{\Phi}_2 u_R^{}
+{\overline Q}_LY_d\Phi_dd_R^{}
+{\overline L}_LY_\ell\Phi_\ell \ell_R^{}+\text{h.c},
\label{eq:yukawah}
\eeqa
where $\Phi_{d,l}$ ($d,l=1,2$) represent $\Phi_1$ or $\Phi_2$, 
$Y_f$ ($f=u,d$ or $\ell$) stand for $3\times 3$ Yukawa matrices and 
$\widetilde{\Phi}_2=i \sigma_2 {\Phi}_2^{\star}$.

Writing the Yukawa interactions Eq.~(\ref{eq:yukawah}) in terms of 
 the mass eigenstates of the neutral and charged Higgs bosons yields 
\beqa
-{\mathcal L}_\text{Yukawa}^\text{2HDM} &=&
 \sum_{f=u,d,\ell} \frac{m_f}{v} \left(
\xi_f^{h^0} {\overline f}f h^0
+ \xi_f^{H^0}{\overline f}f H^0
- i \xi_f^{A^0} {\overline f}\gamma_5f A^0
\right)    \\ & &
+\left\{\frac{\sqrt2V_{ud}}{v}\,
\overline{u} \left( m_u \xi_u^{A^0} \text{P}_L
+ m_d \xi_d^{A^0} \text{P}_R \right)
d H^+
+\frac{\sqrt2m_\ell\xi_\ell^{A^0}}{v}\,
\overline{\nu_L^{}}\ell_R^{}H^+
+\text{h.c}\right\},    \nonumber
\label{Eq:Yukawa}
\eeqa
where $v^2=v_1^2+v_2^2=(2\sqrt{2} G_F)^{-1}$; 
$P_{R}$ and $P_{L}$ are the 
right- and left-handed  projection operators, respectively. 
The coefficients for $\xi^{A^0}_f$ ($f=u, d, l$)
in the four 2HDM types, which are relevant to this work,
are given in the Table~\ref{Ycoupl}.

\begin{table}[t!]
  \begin{tabular}{l @{\extracolsep{0.2in}} r r r }
   \hline \hline
type    & $\xi_u^{A^0}$ & $\xi_d^{A^0}$ & $\xi_l^{A^0}$ \\ \hline
    I  & $\cot\beta$ & $-\cot\beta$ & $-\cot\beta$ \\ \hline
    II & $\cot\beta$ &  $\tan\beta$ & $\tan\beta$ \\ \hline 
     X & $\cot\beta$ & $-\cot\beta$ & $\tan\beta$ \\ \hline
     Y & $\cot\beta$ & $\tan\beta$ & $-\cot\beta$ \\ \hline \hline
    \end{tabular}
    \caption{\small
 Yukawa coupling coefficients $\xi_f^{A^0}$
to the up-quarks, down-quarks and the charged leptons ($f=u, d, l$) 
in the four 2HDM types.}
\label{Ycoupl}
\end{table}

\section{Constraints}

 We consider both theoretical and experimental constraints on 2HDM's.

\subsection{Theoretical and electroweak precision constraints}
For theoretical constraints we take into account all set of tree-level
perturbative unitarity conditions
\cite{Lee:1977eg,Kanemura:1993hm,Kanemura:2015ska}. We use the unitarity constraints from  Ref.\cite{Kanemura:2015ska}
   and require that the eigenvalues of the scattering amplitudes
   satisfy the original Lee-Quigg-Thacker bound \cite{Lee:1977eg}.
We also require that all $\lambda_i$'s remain perturbative.  
Moreover, we demand that the
potential remains bounded from below when the Higgs fields become
large in any direction of the field space \cite{Branco:2011iw}, which
results in the following set of constraints:
\begin{eqnarray}
  \lambda_1>0 \ , \ \lambda_2>0 \ , \ \sqrt{\lambda_1\lambda_2}+\lambda_3+
  {\rm min}(0,\lambda_4+\lambda_5 ,
\lambda_4-\lambda_5) >0 \;.
\label{eq:bfb}
\end{eqnarray}
 
%
For experimental constraints we can further divide them into indirect
and direct searches. The indirect searches mainly arise from
Electro-Weak Precision Observables (EWPOs) and flavor physics.  The
EWPOs can be represented by a set of oblique parameters $ S, T$ and $U
$. From 2018 Particles Data Group (PDG) review
\cite{Tanabashi:2018oca} with a fixed $ U=0 $, the best fit of $ S, T
$ parameters can be represented as $ S=0.02\pm 0.07 $ and $ T=0.06\pm
0.06 $.  We emphasize that the $ T $ parameter, which is related to
the amount of isospin violation, is sensitive to the mass splitting
among $H^{\pm}$, $H^0$, and $A^0$.  It will restrict the allowed mass
spectrum for the scalars in our analysis below.  In order to fulfill
the $T$ constraint in the 2HDM, the spectrum should be chosen close to
the approximate custodial symmetry \cite{Haber:1992py}, which is
satisfied in one of the following limits: i) $m_{H^\pm}=m_{A^0}$, ii)
$m_{H^\pm}=m_{H^0}$ together with $\sin(\beta-\alpha)=1$, or iii)
$m_{H^\pm}=m_{H^0}$ together with $\cos(\beta-\alpha)=1$.

\begin{figure}[t!]
 \centering
\includegraphics[width=0.32\textwidth,clip]{{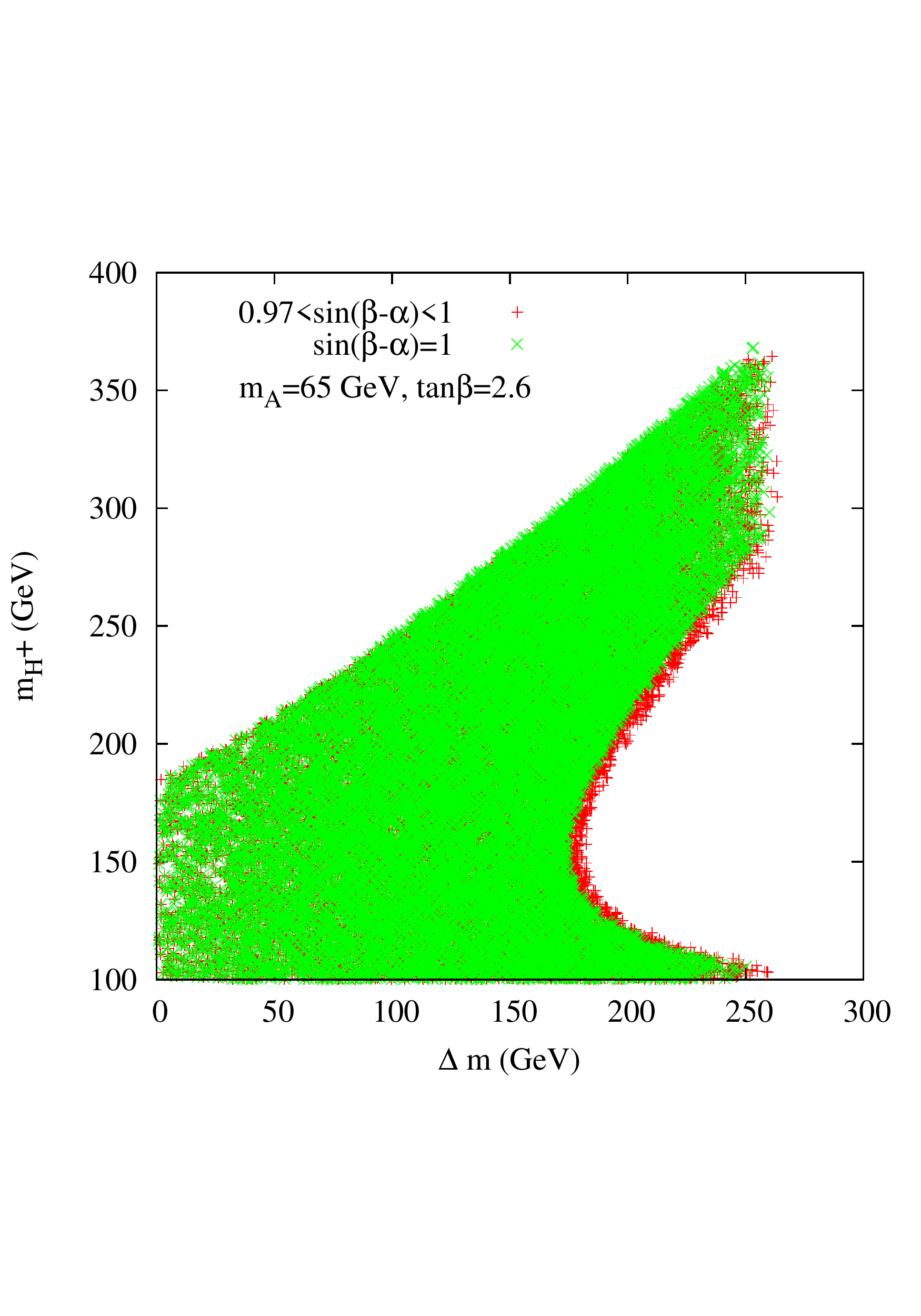}}
\includegraphics[width=0.32\textwidth,clip]{{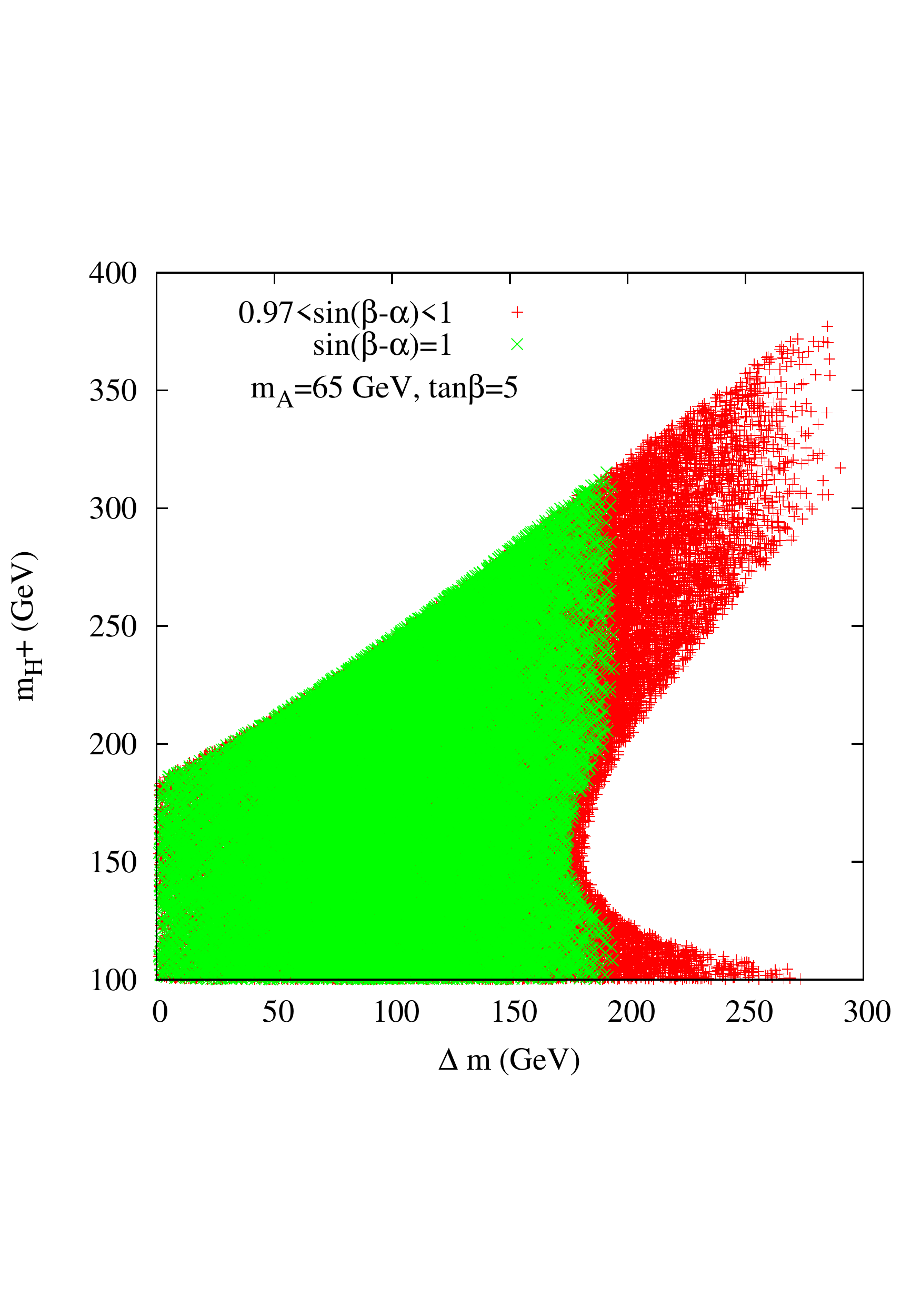}}
\includegraphics[width=0.32\textwidth,clip]{{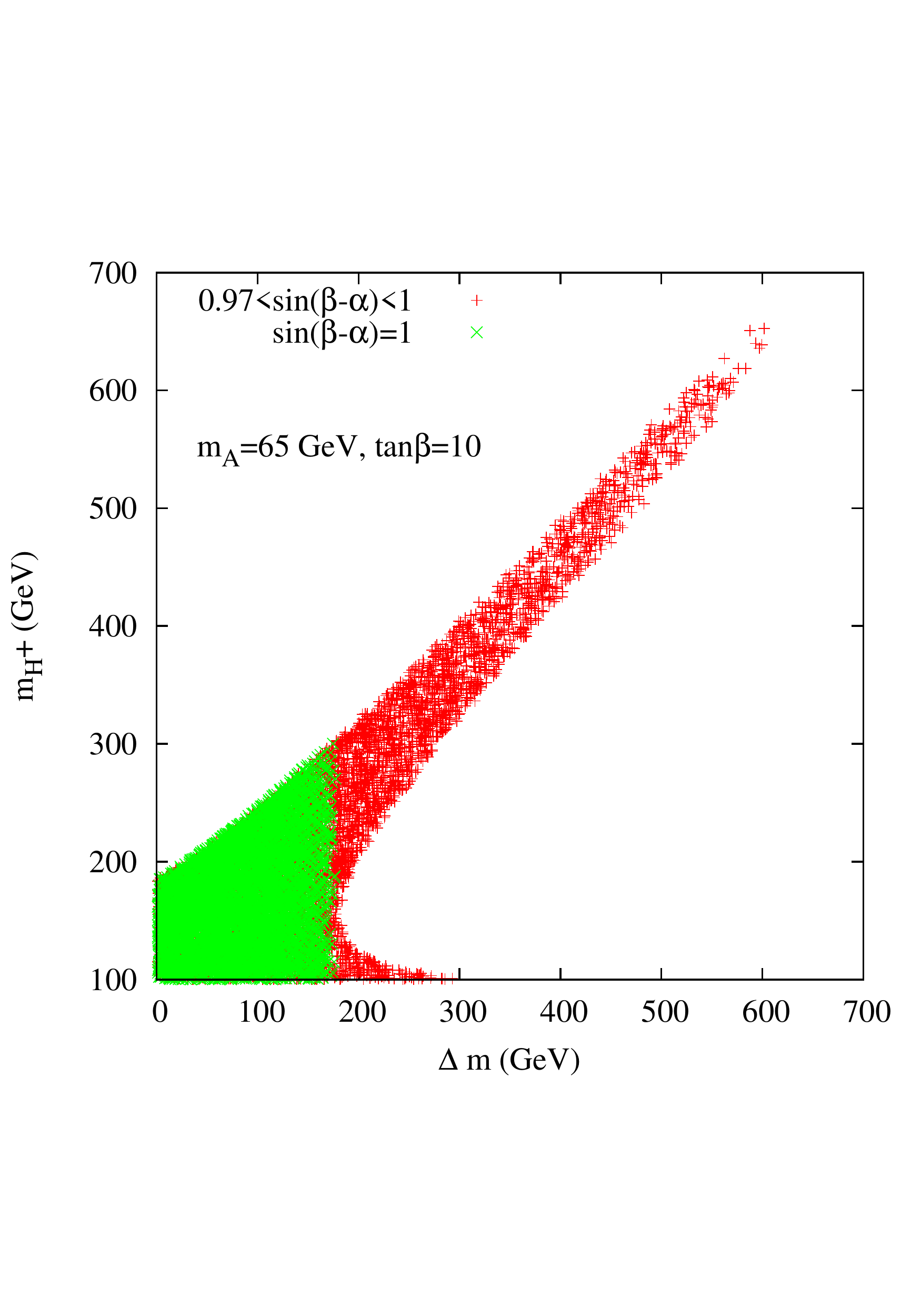}}

\vspace{-0.5in}
\includegraphics[width=0.32\textwidth,clip]{{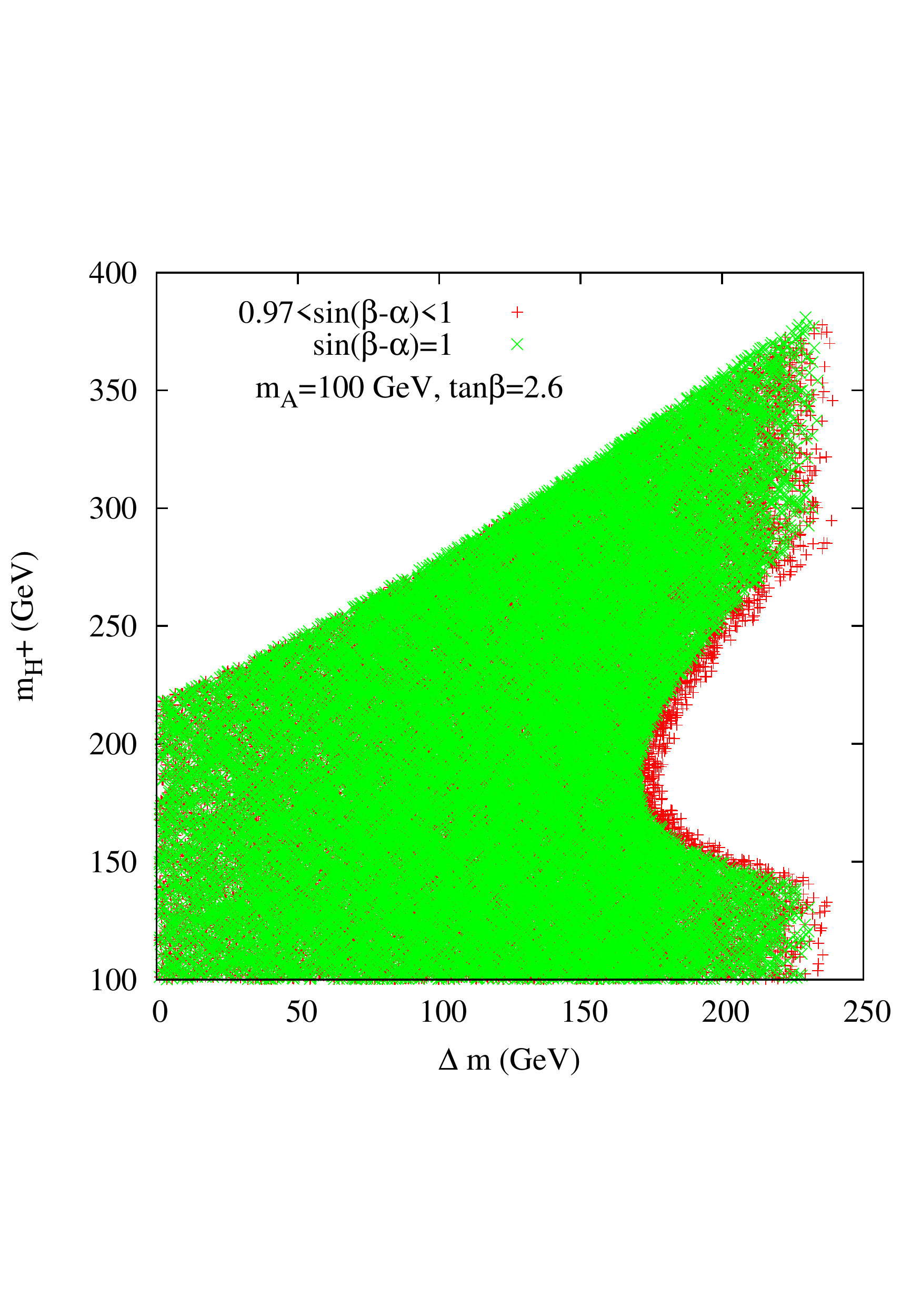}}
\includegraphics[width=0.32\textwidth,clip]{{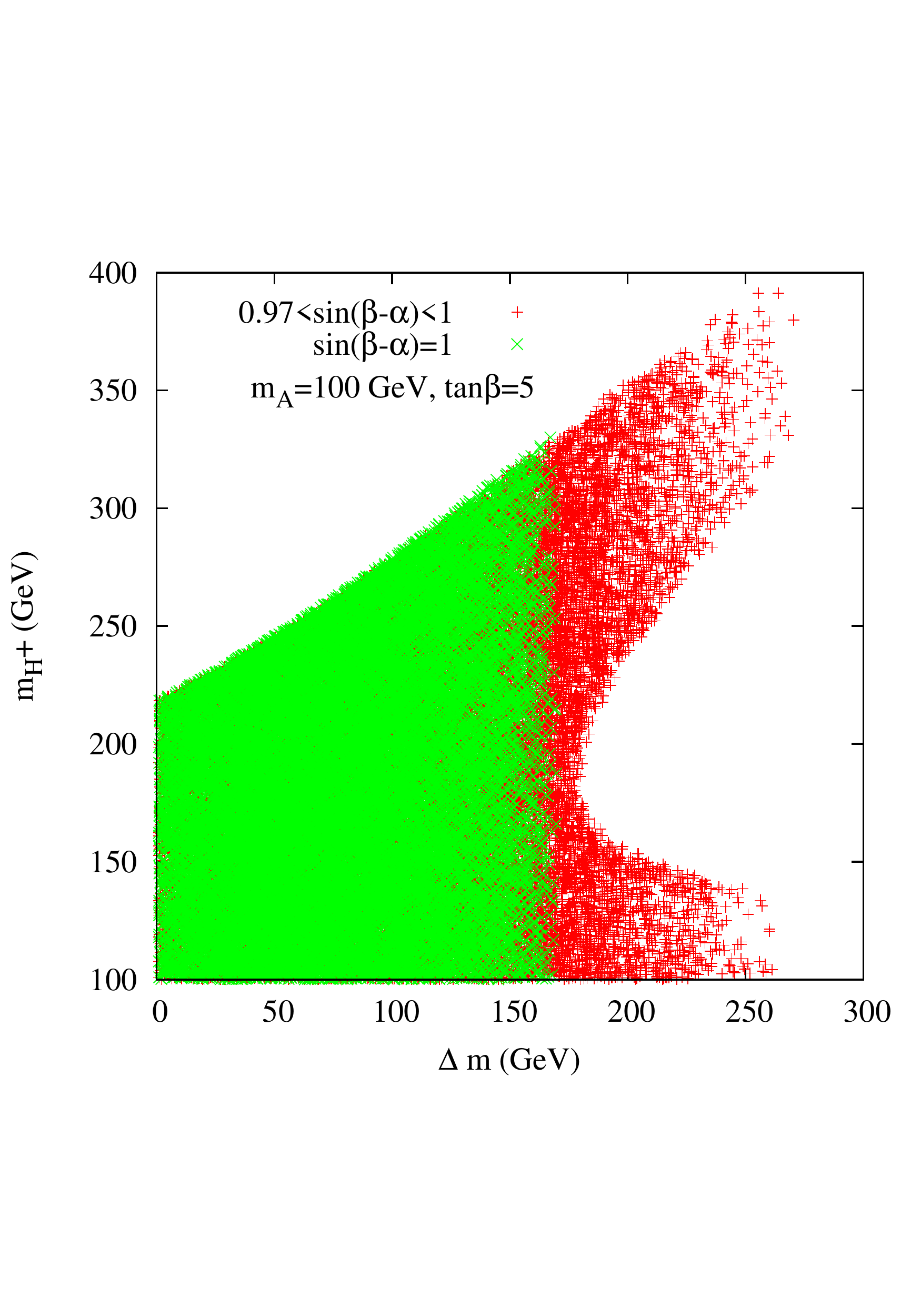}}
\includegraphics[width=0.32\textwidth,clip]{{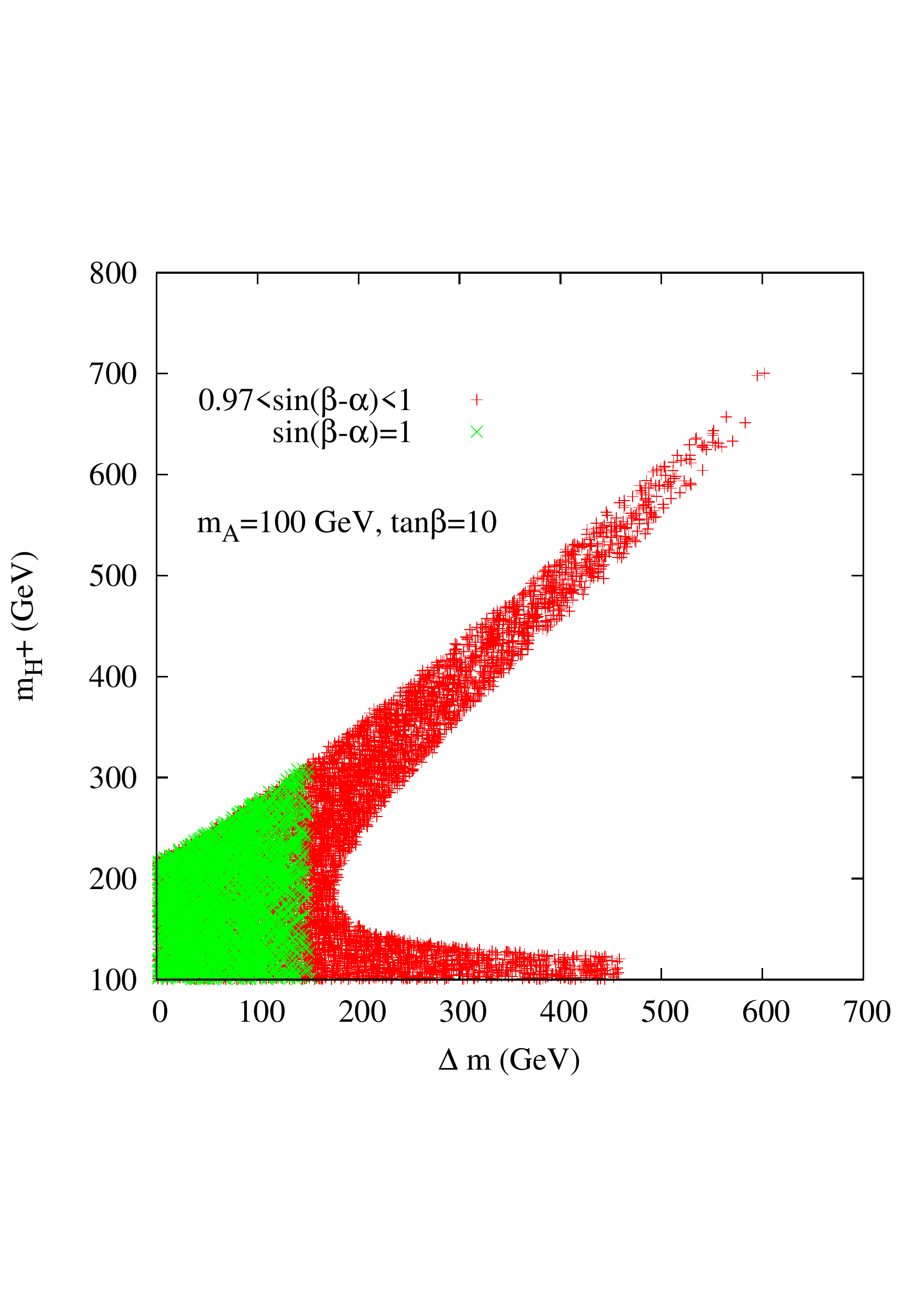}}
\vspace{-0.5in}
\caption{ The allowed parameter space in the plane of
  $( \Delta m \equiv m_{H^0} - m_{A^0}$, $  m_{H^{\pm}} )$
  due to the constraints from the oblique $S$ and $T$
  parameters, and all theoretical constraints. The upper panels are for
  $ m_{A^0} =65 $ GeV while the lowers for $m_{A^0} =100$ GeV, in
  which $ \tan\beta =2.6$ (left), 5 (middle), and 10 (right) are
  shown.  The green points are right at the alignment limit $
  \sin(\beta-\alpha)=1 $ while the red points satisfy $ 0.97 <
  \sin(\beta-\alpha) < 1 $ (near alignment limit).  }
\label{fig:tbdm}
\end{figure}

As mentioned before, the oblique parameter $T$ is highly sensitive to
the mass splitting among $H^{\pm}$, $H^0$, and $A^0$. 
\footnote{
Here we assume the SM-like Higgs boson is the lightest 
CP-even scalar ($m_{H^0} > m_{h^0}$).
  For the reversed case $m_{H^0} = 125 $ GeV and $m_{h^0} < m_{H^0}$, with
  another near alignment limit of 
$ \cos(\beta-\alpha) $ one can also consider another process
\[
pp \to j j W^{\pm *} W^{\pm *} \to j j H^\pm H^\pm 
\to jj (W^\pm h^0) (W^\pm h^0) \;,
\]
which is similar to the process considered in this work.
}
In order to obtain
the allowed parameter space for the mass of charged Higgs boson and
the mass splitting $\Delta m=m_{H^0}-m_{A^0}$, we consider all the
above theoretical constraints and $ 3\sigma $ allowed regions of the
$S$ and $T$ parameters in Fig.~\ref{fig:tbdm} for $ \tan\beta = $ 2.6,
5, and 10 with $ m_{A^0} = $ 65 and 100 GeV, respectively.  We also
scan on $m_{12}^2$ in the following range $[0,10^6]$ GeV$^2$ in order
to satisfy the perturbative unitarity and vacuum stability constraints
for a fixed set of physical masses and mixings.  We notice that, in
our parameter space, the $S$ parameter is always within the best-fit
range while the $T$ parameter severely constrains the splitting
between $m_{A^0}$ and $m_{H^\pm}$, and also $\Delta m$.

For $ \tan\beta =2.6 $, there is no significant difference in the
allowed region between the alignment limit $ \sin(\beta-\alpha)=1 $
and the near-alignment limit $0.97 < \sin(\beta-\alpha) < 1$.  In the
case where $ \tan\beta =5$, one can see that $\Delta m$ is constrained
to be less than about 200 GeV in the exact alignment limit. This cut
on $\Delta m$ is in fact due to the vacuum stability constraints in
Eq.(\ref{eq:bfb}), where either $\lambda_1$ or the third constraint in
Eq.(\ref{eq:bfb}) becomes quickly negative.  While in the case
near-alignment limit $0.97 < \sin(\beta-\alpha) < 1$, which allows the
vacuum stability to be fulfilled and $\Delta m$ can reach up to 280
GeV.  This correlation between vacuum stability and
$\sin(\beta-\alpha)\in [0.97,1]$ is also observed in the case
$\tan\beta=10$ and is even more pronounced where one can see that
$\Delta m$ can reach up to 600 GeV.
The parameter space can be divided into two parts.
The first region of parameter space
is for light $ H^{\pm} $. Once $m_{H^{\pm}}\sim m_{A^0} $,
the mass splitting $\Delta m$ can be as large as $ 300-450 $ GeV.
The second region is for heavy $ H^{\pm} $. When $m_{H^{\pm}}\sim
m_{H^0} $, the mass splitting $\Delta m$ can be extended to about $
600 $ GeV for $\tan\beta=10$. While in the case $ \tan\beta =5$, the
maximum mass splitting $\Delta m$ is less than $ 200 $ GeV in the
alignment limit $\sin(\beta-\alpha)=1 $, and could be extended to more
than $ 250 $ GeV for $ 0.97 < \sin(\beta-\alpha) < 1$.  We stress that
even in the case where $\Delta m$ is rather small, the $T$ parameter
severely constrains the charged Higgs mass to be less than about 200
GeV for $\tan\beta=2.6, 5$ and $10$.
\begin{figure}[t!]\centering
\includegraphics[width=0.48\textwidth]{{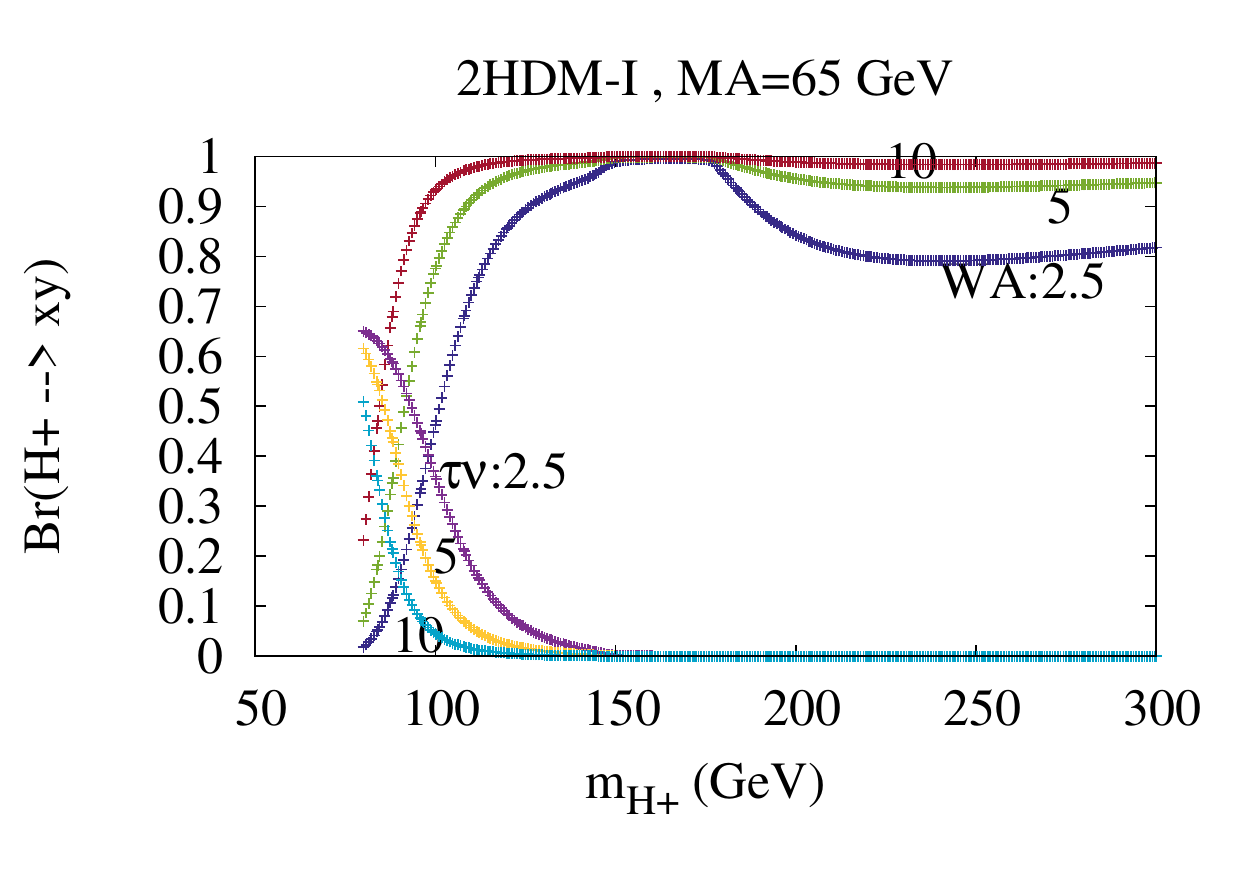}}
\includegraphics[width=0.48\textwidth]{{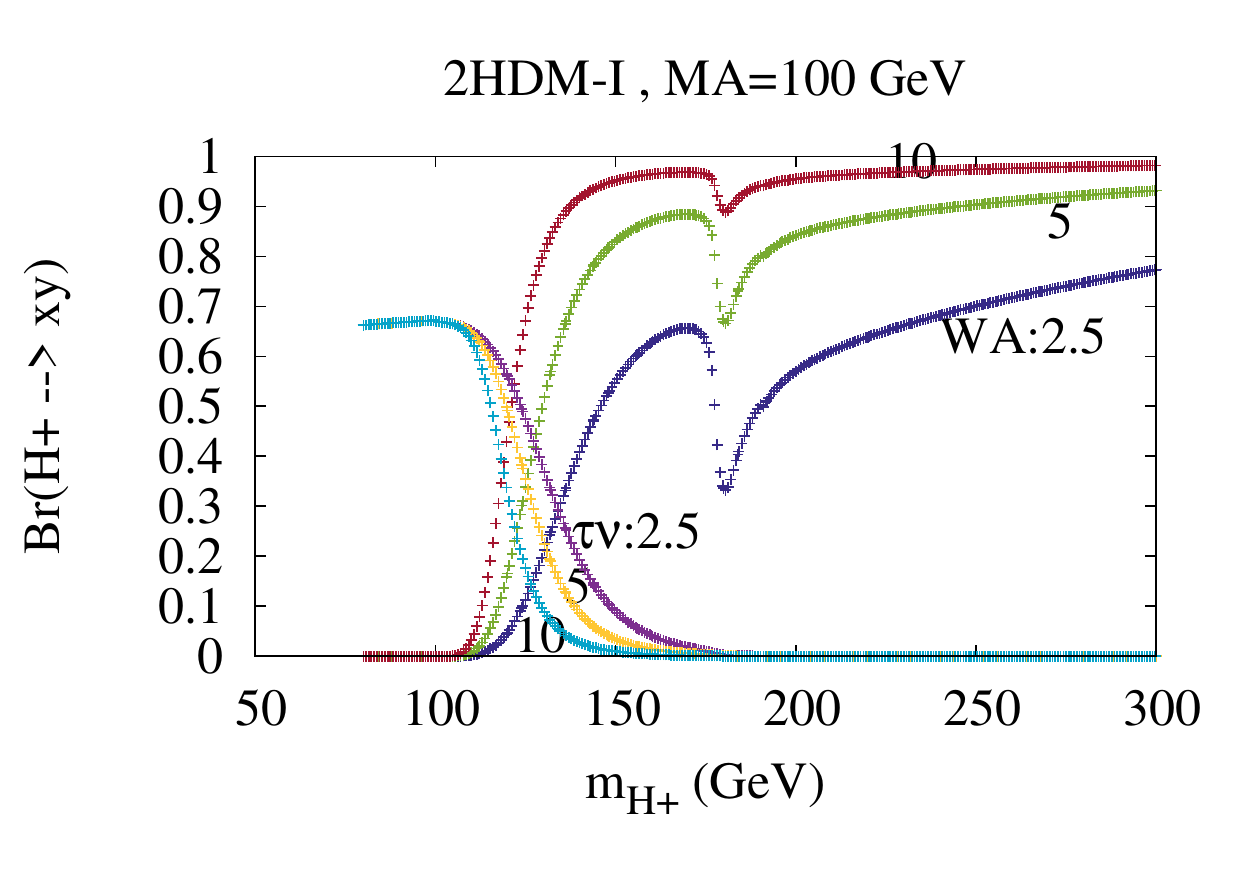}}
\includegraphics[width=0.48\textwidth]{{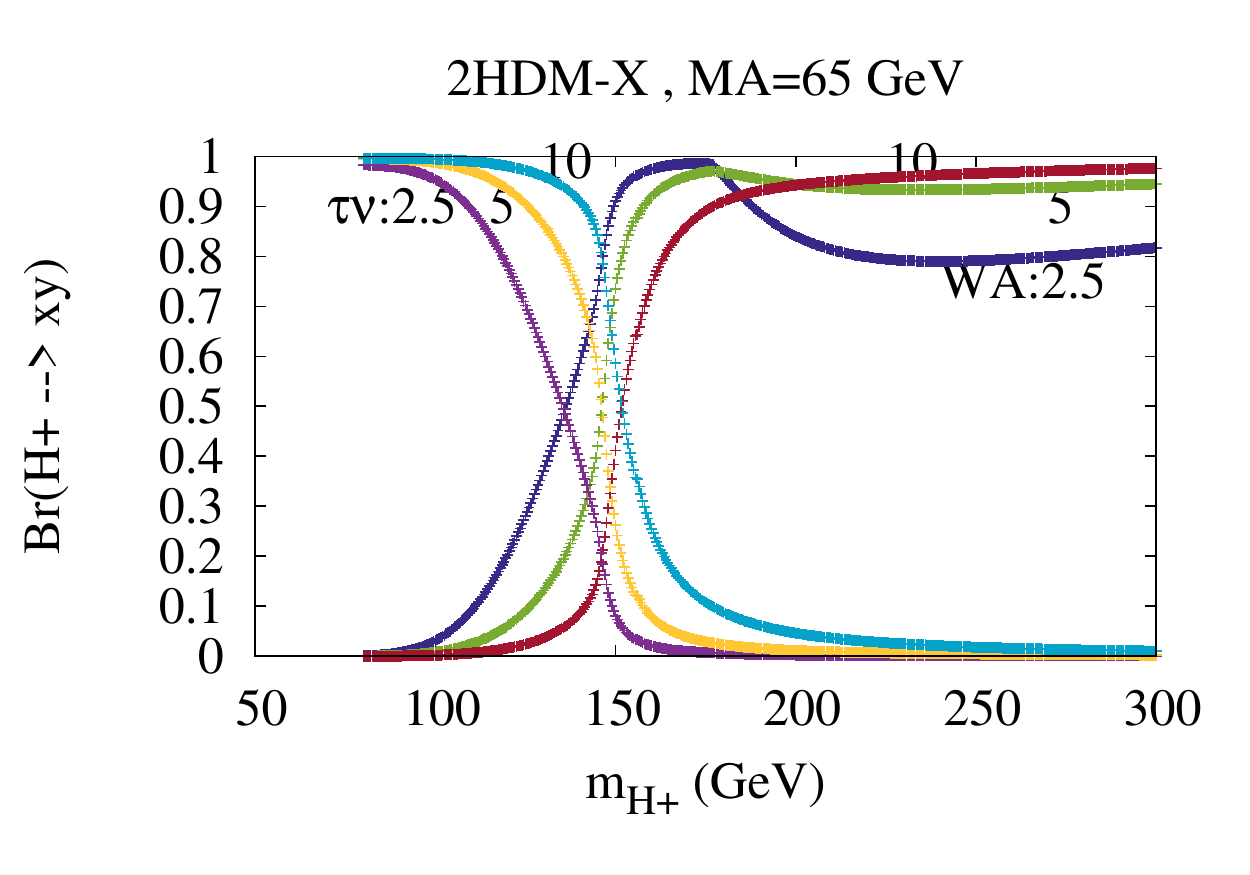}}
\includegraphics[width=0.48\textwidth]{{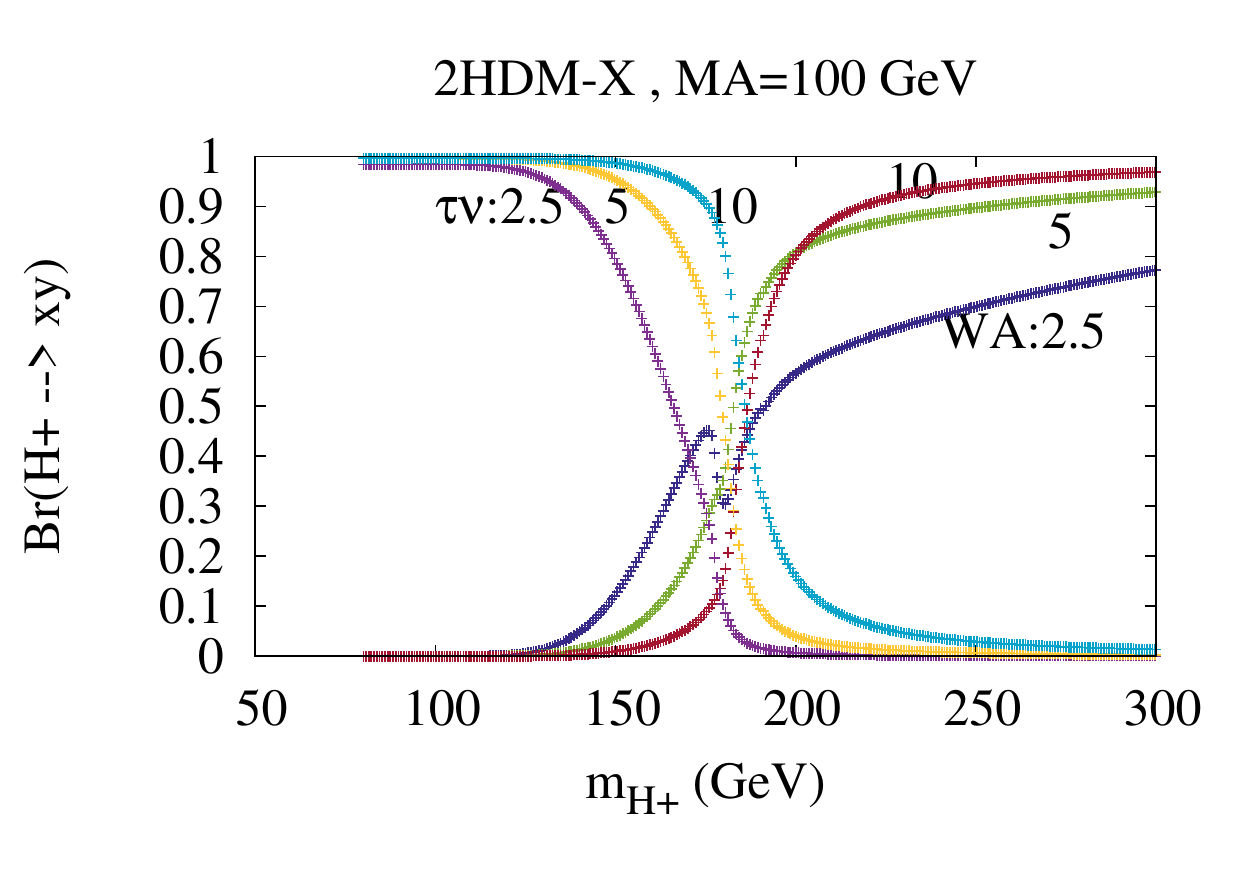}}
\caption{
  Branching fractions of the charged Higgs boson versus  $ m_{H^{\pm}} $
  for type-I 2HDM with $ m_{A^0} =65 $ GeV (upper-left panel),
  $ m_{A^0} =100 $ GeV (upper-right panel) and for type-X 2HDM with
  $ m_{A^0} =65 $ GeV (lower-left panel), $ m_{A^0} =100 $ GeV (lower-right panel).
  The alignment limit $ \sin(\beta-\alpha)=1 $ is assumed.
}
\label{fig:HCHBR}
\end{figure}

\subsection{B physics constraints}
The most severe constraints in flavor physics are due to the
measurements of $B(B\rightarrow X_s \gamma)$,
$B(B_{d,s}\rightarrow\mu^+\mu^-) $ and $\Delta m_s$ of $B$ mesons.
For $B(B\rightarrow X_s \gamma)$, according to the latest analysis by
\cite{Misiak:2017bgg}, we have:
\begin{itemize}
\item In 2HDM type-II and Y, the $b\to s\gamma$ constraint forces
the charged Higgs mass to be  heavier than 
580~GeV~\cite{Misiak:2017bgg,Misiak:2015xwa} 
for any value of $\tan\beta \geq 1$. 
\item In 2HDM-I and X, charged Higgs with mass as low as
  $\sim 100-200$\ GeV ~\cite{Misiak:2017bgg,Enomoto:2015wbn} is still allowed as  long as $\tan\beta\geq 2$. 
\end{itemize}
For other B-physics observables we refer to the recent analysis
\cite{Haller:2018nnx}, in which they also included $\Delta m_s$ and
$B_{d,s}\to \mu^+ \mu^-$. For a light charged Higgs boson, $100<
m_{H^\pm}< 200$ GeV, of interest in this study, one can conclude from
\cite{Haller:2018nnx} that $\tan\beta \geq 3$ is allowed for 2HDM type
I and X.

\subsection{$H^\pm$ and $A^0$ branching ratios and Direct searches}
Before discussing the constraints coming from direct searches, we first 
show the branching ratios of $H^\pm$ and $A^0$ in both 2HDM type I and X
in the following subsection. Calculations of these branching ratios
are performed using the public code 
\textbf{2HDMC} \cite{Eriksson:2009ws}.

\begin{figure}[t!]\centering
\includegraphics[width=0.48\textwidth]{{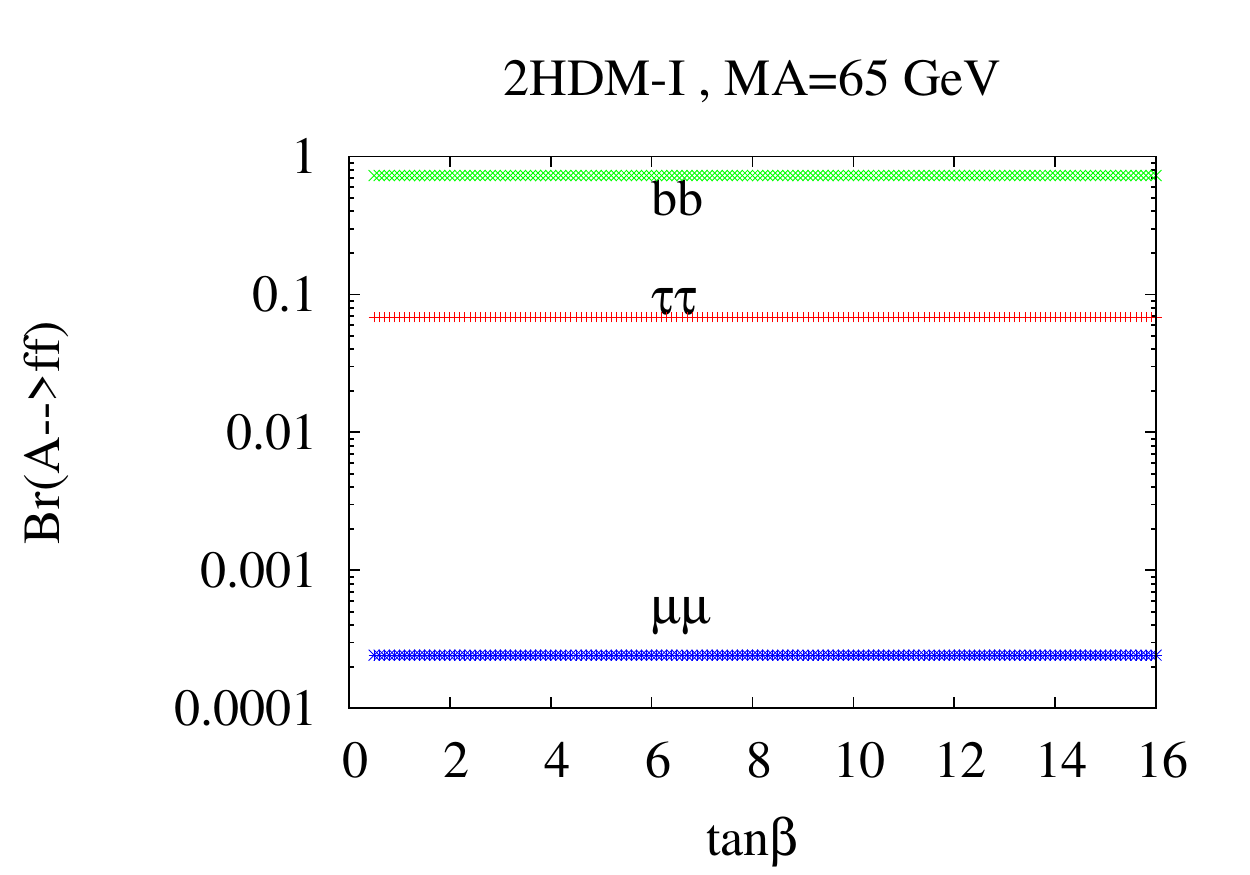}}
\includegraphics[width=0.48\textwidth]{{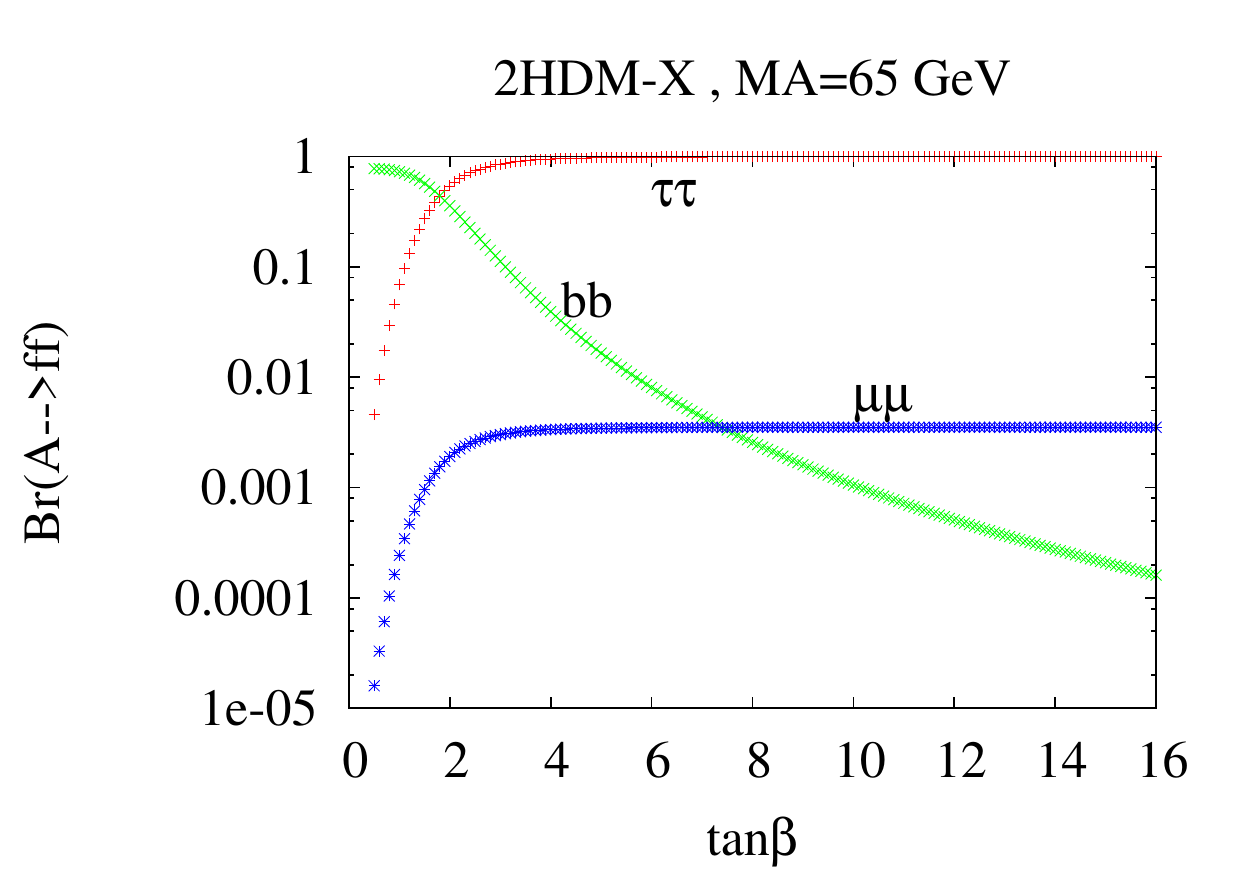}}
\caption{
  Branching fractions of the CP-odd Higgs boson $A^0$
  versus $ \tan\beta $
  for $ m_{A^0} =65 $ GeV in type-I 2HDM (left panel) and
  type-X 2HDM (right panel).
}
\label{fig:ABR}
\end{figure}

\subsubsection{Branching ratios of $H^\pm$ and $A^0$}
We illustrate in Fig.~\ref{fig:HCHBR} the
branching ratios of the charged Higgs boson
for both 2HDM type I and X.
It is clear that once the bosonic decay 
mode $ H^{\pm}\rightarrow W^{\pm}A^0 $ is open,
it can be the dominant decay 
 mode and both $ B(H^{+}\rightarrow t\overline{b}) $ and
$B(H^{\pm}\rightarrow\tau^{\pm}\nu_{\tau})$ are highly suppressed. 

In type I, one can see that the full dominance of the bosonic decay
needs $\tan\beta>5$ which reduces the $H^\pm\to\tau^\pm \nu_{\tau}$ and
$H^\pm \to tb$ channels.  The decay channel $H^\pm\to W^\pm h^0$ is
vanishing because $H^\pm W^\mp h^0$ coupling is proportional to
$\cos(\beta-\alpha)\approx 0$.  In 2HDM type X, the coupling
$H^\pm \tau^\mp \nu_{\tau}$ is proportional to $\tan\beta$
and since we assume
that $\tan\beta\geq 2.5$, the $\tau\nu_{\tau}$ channel is slightly
larger than in the 2HDM type I. It is clear from the lower panels of
Fig.~\ref{fig:HCHBR} that before the $W^\pm A^0$ threshold, $H^\pm \to
\tau^\pm \nu_{\tau}$ is the dominant decay mode and it is amplified by
taking large $\tan\beta$. In fact, such a large $\tan\beta$ not
only enhances the $\tau\nu_{\tau}$ channel but also reduces
$H^\pm \to cb, cs, tb$ modes,
which are all proportional to $\cot\beta$.  After
crossing $W^\pm A^0$ threshold, $H^\pm \to W^\pm A^0$ becomes the
dominant decay mode and taking large $\tan\beta$ can further
suppress $H^\pm \to tb$ and makes $H^\pm \to W^\pm A^0$ even larger.
Note that in the
alignment limit $\cos(\beta-\alpha)=0$, the coupling $H^\pm W^\mp h^0$
vanishes while $H^\pm W^\mp H^0$ is maximal and becomes similar to
$H^\pm W^\mp A^0=g/2$.  Therefore, if $H^\pm \to W^\mp H^0$ is
kinematically open it will compete on equal footing with $H^\pm \to
W^\pm A^0$.

If $\tan\beta$ increases beyond 20 (45), the $\tau\nu$ mode could
become comparable to the $WA$ mode for $m_{H^\pm}\agt 200$ GeV and
$m_{A^0} = 100 \; (65)$ GeV in type-X. In such a case, the model
would be subject to the current charged Higgs searches via the $\tau
\nu$ mode. In the following, we will concentrate on a  scenario in
which the $WA$ is the dominant mode.

The branching fractions for $A^0$ are depicted in Fig.\ref{fig:ABR} as
a function of $\tan\beta$ for $m_{A^0}=65$ GeV
in 2HDM type-I (left panel)
and type-X (right panel). In 2HDM type I, all couplings $A^0ff$
are proportional to $\cot\beta$. Therefore, the $\tan\beta$ factorizes
out in the branching ratio calculation leading to
constant $B(A^0\to b\bar{b}, \tau^+ \tau^- , \mu^+\mu^-)$
as a function of $\tan\beta$.
In the case of type X, the branching ratios $B(A^0\to \tau^+ \tau^-,
\mu^+ \mu^-)$ are enhanced for large $\tan\beta$ while $B(A^0\to
b\bar{b})$ is suppressed. Note for $m_{A^0}=100$ GeV, none of $A^0\to
Z^*h^0$ and $A^0\to W^{\mp *}H^\pm$ are open, we observe similar
behavior for $B(A^0\to f\bar{f})$ in both type I and X.

\subsubsection{LHC Constraint from $t\to bH^+ \to b \tau \nu_{\tau}$}

For direct searches the LEP collaborations \cite{Abbiendi:2013hk} had
searched for charged Higgs pair production via the Drell-Yan process $
e^+ e^-\rightarrow Z/\gamma\rightarrow H^+ H^-$, excluding $
M_{H^{\pm}} < 80 $ GeV (Type-II) and $ M_{H^{\pm}} < 72.5 $ GeV
(Type-I) at 95\% confidence level.
The LHC collaborations also reported their charged Higgs search
results for various mass regions. In the low mass region, the main
decay mode is via $t \to b H^+$ followed by
$H^{\pm}\rightarrow\tau^{\pm}\nu_{\tau} $ from CMS \cite{CMS1,CMS2}
and ATLAS \cite{ATLAS1,ATLAS2}.  In the high mass region, the main
decay mode is $ H^{+}\rightarrow t\overline{b} $ from
CMS \cite{CMS1,CMS:1900zym} and ATLAS \cite{Aad:2015typ}.
  
\begin{figure}[t!]
  \centering
  \includegraphics[width=3.2in]{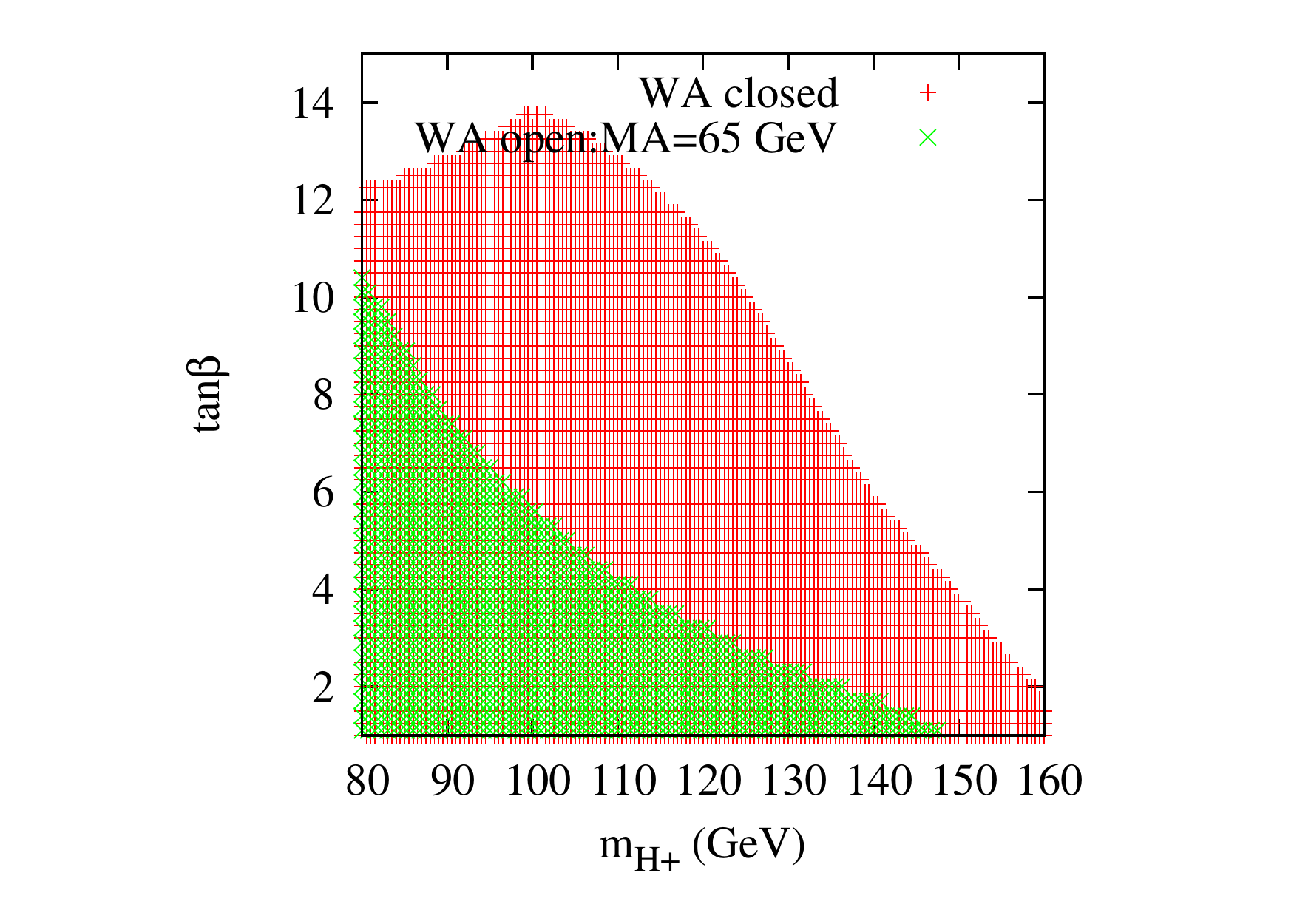}
  \includegraphics[width=3.2in]{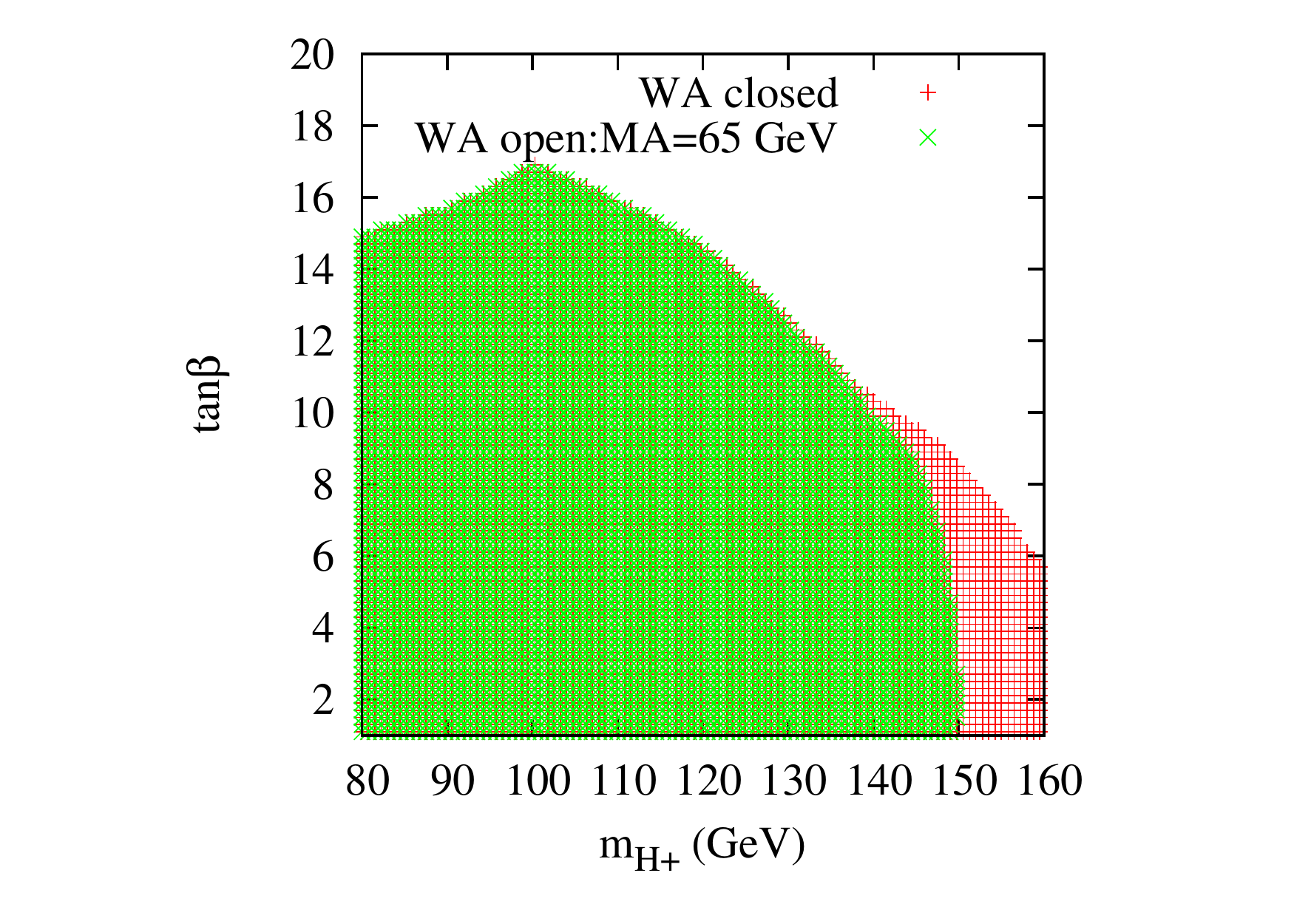}
  \caption{\small \label{fig:cms-limit-tau}
    Interpretation of CMS exclusion regions
      \cite{CMS1,CMS2,CMS3}
    in the 2HDM type I
    (left panel) and  type X (right panel) projected on
    the plane of $(m_{H^\pm},\, \tan\beta)$.
    The red points stand for the case the $WA$ mode is closed while
    the green points are for the case that the $WA$ mode
    is open, with $m_{A^0}= 65$ GeV.
  }
  \end{figure}

When the charged Higgs mass is below $m_t - m_b$, it can be abundantly
produced in top-quark decays, $t\to b H^+$, followed by charged Higgs
decay $H^+ \to \tau^+ \nu_{\tau}$ or $H^+ \to W^+ A^0$.
The CMS search for $t \to b H^+ \to b (\tau^+ \nu_{\tau})$
\cite{CMS1,CMS2,CMS3} set limits
on $B(t \to b H^+)\times B(H^+ \to \tau^+ \nu_{\tau})$. 
We rescale their limits to the type I and X 2HDM's and
show the exclusions in $(m_{H^\pm},\, \tan\beta)$ plane. 
We note that in type I and X the decay width of
$t \to b H^+$ scales as $\cot^2\beta$:
\begin{eqnarray}
    \Gamma( t\to b H^+) &=& \frac{G_F}{8\sqrt{2} \pi}  \frac{|V_{tb}|^2}{m_t}
    \lambda^{1/2} \left( 1, \;\frac{m_b^2}{ m_t^2},\; \frac{m_{H^\pm}^2}{m_t^2} \right)
    \nonumber \\
    & \times &
    \left[ (m_t^2 + m_b^2) \cot^2\beta (m_t^2 + m_b^2 - m_{H^\pm}^2 )
      - 4 m_t^2 m_b^2  \cot^2\beta \right ] \;.
      \label{eq:tbhc}
\end{eqnarray}
where $ \lambda^{1/2}(1,x^2,y^2)\equiv\sqrt{[1-(x+y)^2][1-(x-y)^2]} $.

Interpretation of the 
  CMS exclusion region \cite{CMS1,CMS2,CMS3}
in the framework of 2HDM type I and X in
$(\tan\beta, m_{H^\pm})$ plane is illustrated in
Fig.~\ref{fig:cms-limit-tau} for both cases: $H^\pm \to W^\pm A^0$
closed and $H^\pm \to W^\pm A^0$ open.\footnote{ 
Here the results presented in Fig.~\ref{fig:cms-limit-tau} are consistent with
the Fig.3 of Ref.~\cite{Sanyal:2019xcp}, in which the $WA$ mode was not considered. }
It is clear that for charged
Higgs mass $\leq 120$ GeV with the $W^\pm A^0$ channel closed,
$\tan\beta \leq 12$ is excluded. This exclusion is reduced for
$m_{H^\pm}\geq 120$ GeV due to the fact that
$B(H^+\to \tau^+ \nu_{\tau})$
is highly suppressed for 2HDM type I as $\tan\beta$ increases.
On the other hand, when the $WA$ mode is open, the
exclusion region in $(\tan\beta, m_{H^\pm})$ plane is significantly
reduced in 2HDM type I.
In the case of 2HDM type X, one can see
from the right panel that $\tan\beta\leq 6$ is excluded for any value
of charged Higgs mass provided that $H^\pm \to W^\pm A^0$ is
closed.
This limit on $\tan\beta$ is slightly more severe than what we can
get from flavor physics (see the above discussion).
When $H^\pm \to W^\pm A^0$ is open,
starting from $m_{H^\pm}\geq 145$ GeV for $m_{A^0}=65$ GeV,
$H^\pm \to \tau \nu_{\tau}$ mode is suppressed leading to no
exclusion for any $\tan\beta$.
Below the $W^\pm A^0$ threshold, $H^\pm \to \tau \nu_{\tau}$
channel is still the dominant one, one can see that the
green exclusion completely overlaps with the red one
in 2HDM X.

\subsubsection{LHC Constraint from $t\to bH^+ \to bW^+A^0\to bW^+ \mu^+ \mu^-$ }

\begin{figure}[t!]
  \centering
  \includegraphics[width=3.2in]{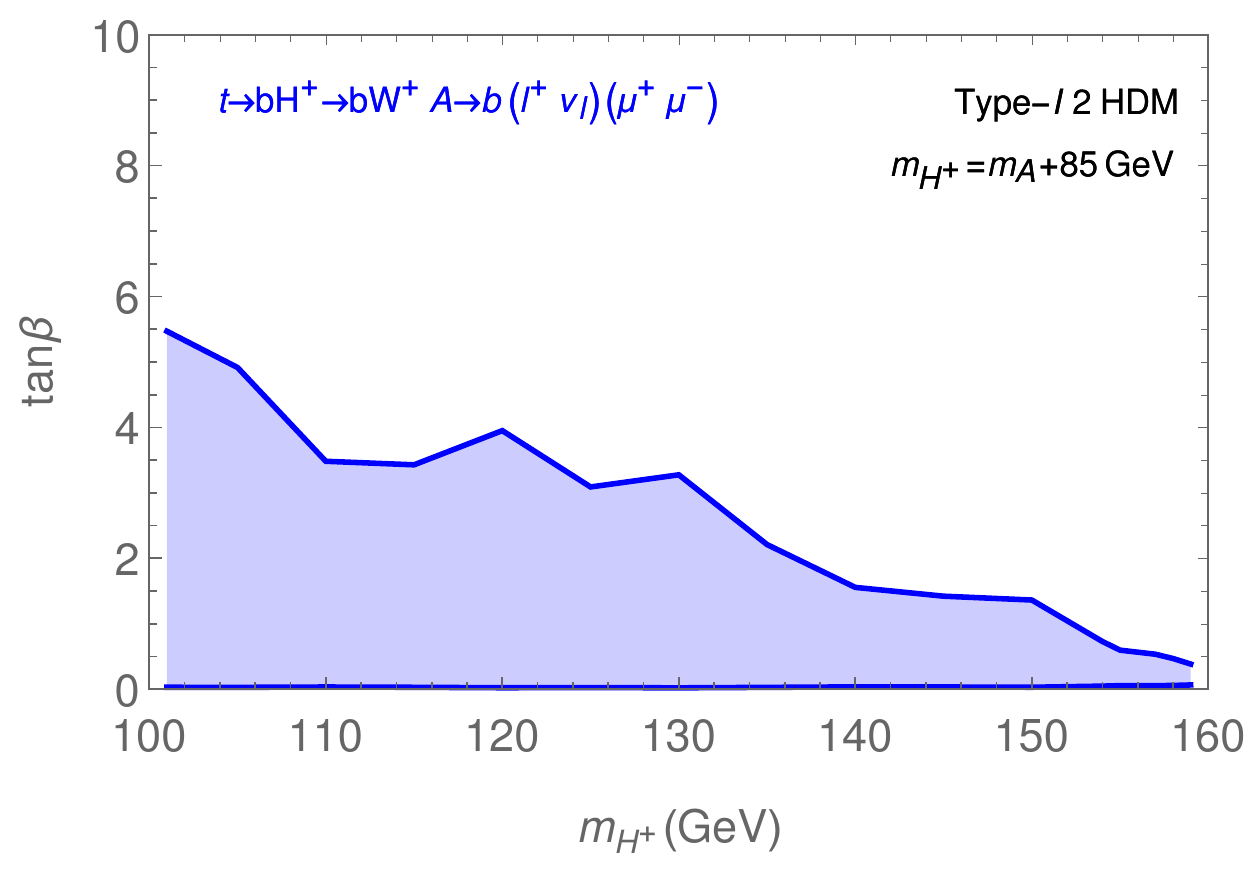}
  \includegraphics[width=3.2in]{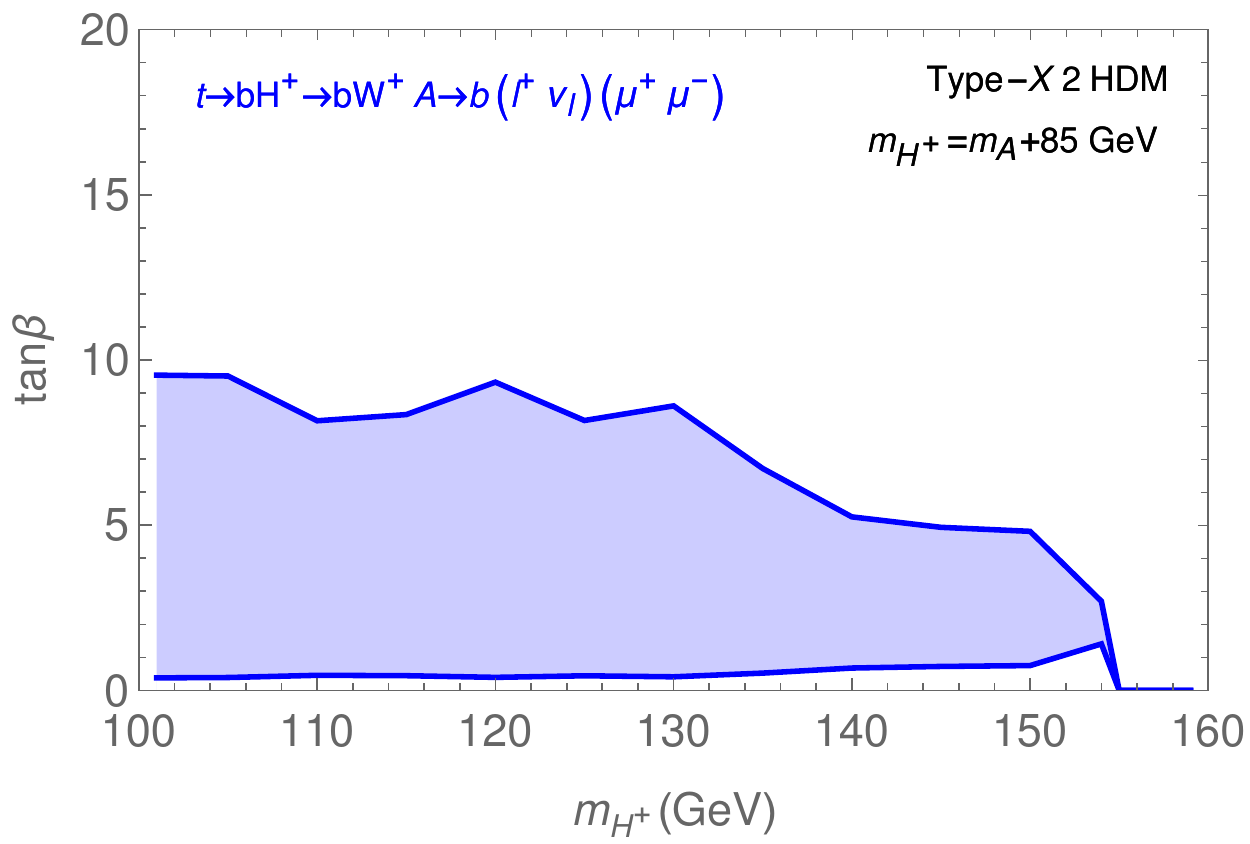}
  \caption{\small \label{fig:cms-limit-wa}
    Exclusions in the parameter space of
    $(m_{H^\pm},\, \tan\beta)$ for type I (left panel)
    and for type X (right panel) 2HDM's
    obtained by rescaling the observed limits of the CMS results
    in Refs.~\cite{Sirunyan:2019zdq}
    based on $t \to b H^+ \to b W^+ A \to b (l^+ \nu_l) (\mu^+ \mu^-)$.
  }
  \end{figure}

Recently, the CMS collaboration \cite{Sirunyan:2019zdq} also
reported the
direct search for light charged Higgs via $t \to b H^+ \to b (W^+ A^0)
\to b (l^+ \nu_l) (\mu^+ \mu^-) $ with $l=e,\mu$ \cite{Sirunyan:2019zdq}
assuming that $H^\pm$ decays 100\% into $W^\pm A^0$ and $B(A^0 \to
\mu^+\mu^-)=3\times 10^{-4}$ and set a limit on $B(t\to bH^+)$. We
rescale the CMS limit and interpret it for 2HDM type I and X, which
are depicted in Fig. \ref{fig:cms-limit-wa}.  It is clear that
the exclusion based on $A^0 \to \mu^+ \mu^-$ also shows some
differences between type I and X. It is easy to see from
Fig.~\ref{fig:ABR} that $B(A^0 \to \mu^+\mu^-)$ is only about $2\times
10^{-4}$ in type I but is as large as $3 \times 10^{-3}$ in type X
for $\tan\beta > 3$.  Therefore, the excluded region (blue shaded) in
Fig.~\ref{fig:cms-limit-wa} for type X is much larger than that of
type I.
  
In the rest of this work, we focus on type I and X 2HDM's, in which
the charged Higgs mass is much less restricted.  In addition, we also
focus on the currently-allowed parameter space region where $H^\pm$
decays dominantly into $W^\pm A^0$ via VBF production of same-sign charged Higgs
boson pair. This is complementary to the study in
Ref.~\cite{Aiko:2019mww}.

Before moving to the next section, we make some comments for direct
searches of light $H^0$ and $A^0$ at the LHC. In the (near) alignment
limit, only fermionic production channels $ gg\rightarrow H^0/A^0 $, $
pp\rightarrow t\overline{t}H^0/A^0 $ and $ pp\rightarrow
b\overline{b}H^0/A^0 $ with decay modes $ H^0/A^0\rightarrow
b\overline{b},\tau^+\tau^-,\mu^+\mu^- $ and $ \gamma\gamma $ are
possible to directly detect light $H^0$ and $A^0$
\cite{Sirunyan:2018wim,CMS:2019hvr,Sirunyan:2018aui,ATLAS:2018xad}.
\footnote{
    The most stringent constraint from the direct search of
    light pseudoscalar $A^0\rightarrow\tau^+\tau^-$ at the LHC comes
    from Ref.~\cite{CMS:2019hvr}. If we take $\tan\beta=3$ for Type-I
    and X 2HDMs, and compare the constraints from
    Ref.~\cite{CMS:2019hvr} for the process $pp\rightarrow
    b\overline{b}A^0$ with $A^0\rightarrow\tau^+\tau^-$, then the cross
    sections for Type-I (Type-X) are about 3 (2) orders smaller than the
    current constraints for $25 < m_{A^0} < 70$ GeV. Therefore, we
    will ignore these constraints in our study. }
In type I and X 2HDM's, all of these production channels
are proportional to $\cot^2\beta $.  Therefore, it is rather challenging
to detect both of them for large $\tan\beta $.  Besides, it is also
hard to distinguish between the CP properties of light $H^0$ and $A^0$ at the
LHC, even if we already observe two different resonance peaks from
their fermionic channels.  Based on these difficulties, we argue that
the process in Eq.~\ref{eq:2W2A2j} can be another way to double check
the mass splitting $\Delta m $ between $H^0$ and $A^0$.

Note that the case of relatively light CP-odd ($m_A<60$ GeV) is 
now rather severely constrained by LHC searches. Several dedicated searches can be used to constraint such scenario. 
The first search is $pp\to h^0\to A^0A^0\to 4f$ 
\cite{Aaboud:2018esj,Sirunyan:2018pzn,Aaboud:2018iil,Sirunyan:2018mot,Sirunyan:2018mbx}  which is  performed both by ATLAS and CMS and the second one is 
 $pp\to H^0\to ZA^0\to 2b l^+ l^-$ \cite{Aaboud:2018eoy}.
Even though the $h^0 A^0 A^0$ coupling can be adjusted to be very small by tuning the parameter $m^2_{12}$, this $m^2_{12}$ may also violate theoretical and EWPT constraints as well, especially for large mass splitting between $H^0$ and $A^0$.  
In this regards, we perform a global scan
for the benchmark point $m_{A^0}\in [15-60] $ GeV with $0.97\leq \sin(\beta -\alpha)\leq 1$ by using the public softwares 2HDMC, 
\textbf{HiggsBounds} \cite{Bechtle:2013wla} and \textbf{Higgssignal} 
\cite{Bechtle:2013xfa}. For such light CP-odd, $h^0$ can decay 
with a significant branching ratio into $A^0A^0$. In addition, 
the heavy CP-even $H^0$ can also decay dominantly into $A^0Z$
because $H^0A^0Z$ coupling being proportional to $\sin(\beta-\alpha)\approx 1$.
However, we found that the allowed parameter space that survive to the theoretical and EWPT constraints is now almost 
excluded either  by  $pp\rightarrow h^0\rightarrow A^0 A^0 
\rightarrow \{ b\overline{b} b\overline{b}, \mu^+ \mu^- b\overline{b}, \mu^+ \mu^- \tau^+ \tau^-, b\overline{b} \tau^+\tau^-\}$ \cite{Aaboud:2018esj,Sirunyan:2018pzn,Aaboud:2018iil,Sirunyan:2018mot,Sirunyan:2018mbx} or by $pp\rightarrow H^0\rightarrow A^0Z\rightarrow b\overline{b} l^+l^- $ \cite{Aaboud:2018eoy} searches.

%
%

\section{Same-sign charged Higgs pair production}

\subsection{The behavior of $ p p\rightarrow H^{\pm}H^{\pm} j_F j_F $
  process}

Recently, the novel process of same-sign charged Higgs pair production
was proposed in Ref.~\cite{Aiko:2019mww}, and especially this process
is very sensitive to the mass splitting
$ \Delta m \equiv m_{H^0} - m_{A^0}$
in the 2HDMs as it will be shown below. The cross section is enhanced
according to the large mass splitting $ \Delta m $.  This process can
be generated via the same-sign $W$ boson fusion, $ p p \rightarrow
W^{\pm\ast}W^{\pm\ast} j_F j_F \rightarrow H^{\pm}H^{\pm} j_F j_F $ at
hadron colliders, where $ j_F $ denotes the forward and energetic jet
directly from the initial parton.

The relation between the mass splitting $ \Delta m $ and same-sign
charged Higgs pair production can be understood in the $ 2\rightarrow
2 $ subprocess $ W^+ W^+ \rightarrow H^+ H^+ $ at amplitude level.
This subprocess is induced by three t-channel diagrams with
$h^0$, $H^0$ and $A^0$ exchange. 
In the alignment limit, $\cos(\beta - \alpha) = 0$,
which is favored by the current Higgs data, the scattering amplitude  for
\[
W^+ ( p_1) W^+ (p_2) \to H^+ (q_1) H^+ (q_2)
\]
is only mediated by $H^0$ and $A^0$ and is given by
\beqa
i {\cal M}^{H^0 +A^0} & =
i g^2 q_1 \cdot \epsilon (p_1) \; q_2 \cdot \epsilon(p_2) \;
\left[ \frac{1}{ t - m_{A^0}^2} - \frac{1}{t - m_{H^0}^2} \right]  \;\; + \;\;
\left (q_1 \leftrightarrow q_2, t \leftrightarrow u \right ) \nonumber \\ & 
\propto \Delta m \times \frac{m_{H^0}+m_{A^0}}{(t - m_{H^0}^2)(t - m_{A^0}^2)} \;
q_1 \cdot \epsilon (p_1) \; q_2 \cdot \epsilon(p_2) \;\; + \;\;
\left (q_1 \leftrightarrow q_2, t \leftrightarrow u \right )
\label{amp}
\eeqa
where $t = (p_1 - q_1)^2$ and $u = (p_1 - q_2)^2$, and $\epsilon(p_{1,2})$
are the polarization 4-vectors of the incoming $W^+$ bosons. As it can
be seen, the above amplitude is proportional to $ \Delta m $.

\begin{figure}[t!]
\centering
\includegraphics[width=0.45\textwidth]{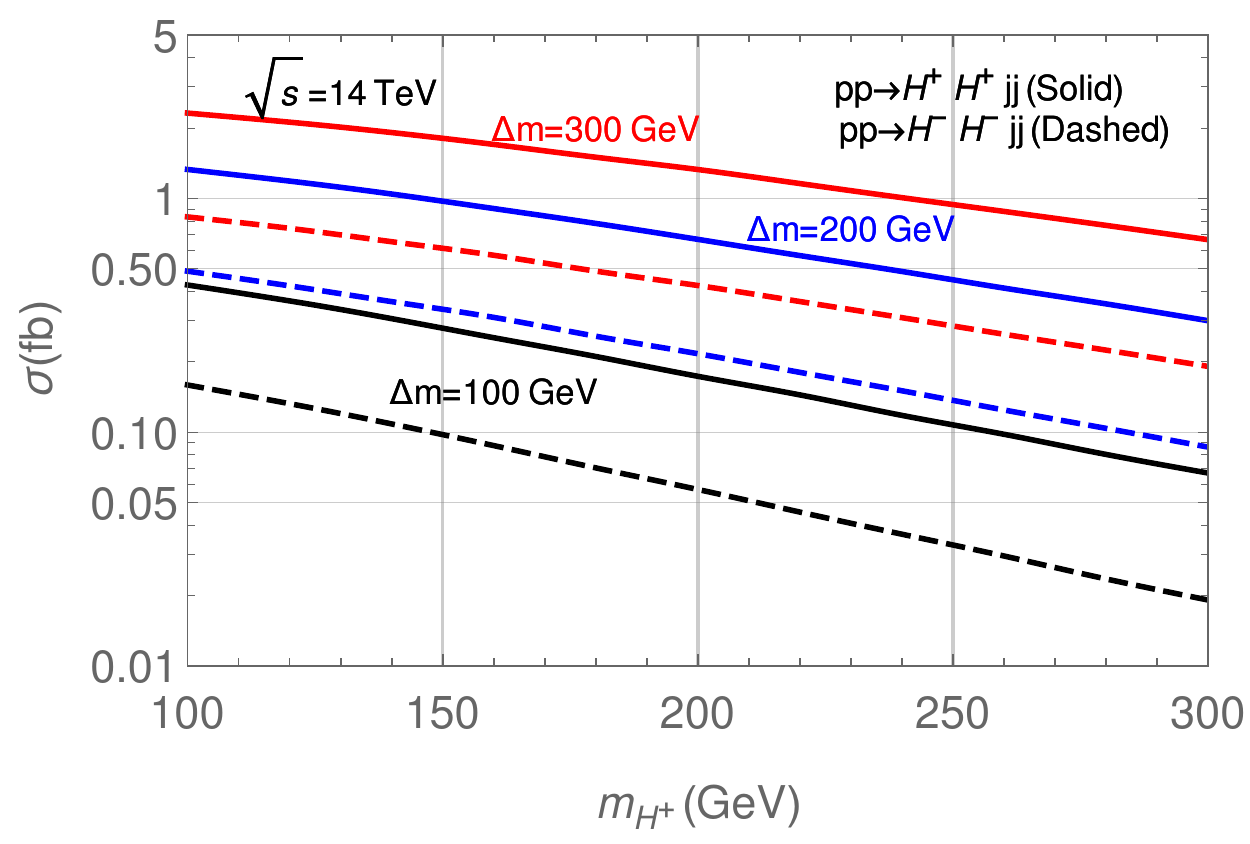}
\includegraphics[width=0.45\textwidth]{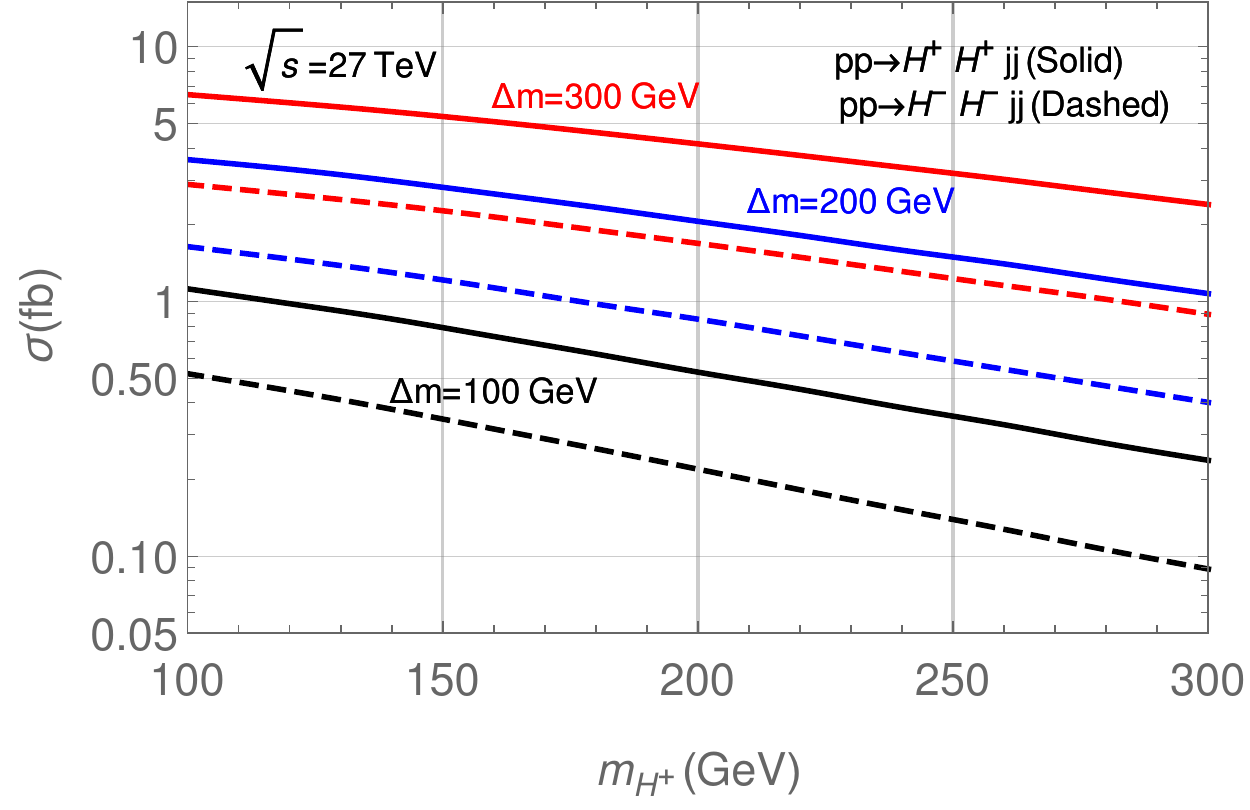}
\caption{ The production cross sections of
  $ p p \rightarrow  H^{+}H^{+} j_F j_F $ (solid line) and
  $ p p \rightarrow  H^{-} H^{- } j_F j_F $ (dashed line)  versus $ m_{H^{\pm}} $
  at $\sqrt{s}= 14 $ TeV (left panel) and $\sqrt{s}= 27 $ TeV (right panel), for $ \Delta m = $100 GeV (black),
  200 GeV  (blue), and 300 GeV (red).
  Notice the VBF cut $ \eta_{j_1}\times\eta_{j_2} < 0 $ and
  $|\Delta\eta_{jj}| > 2.5 $ for the minimum rapidity
  difference between the forward jet pair are applied.
  }
\label{fig:Xsec1}
\end{figure}

\begin{table}[th!]
\centering     
\caption{\small \label{tab:sinba14}
  Sum of cross sections for $\sigma (p p\rightarrow H^+ H^+ j_F j_F)$ and
  $ \sigma (p p\rightarrow H^- H^- j_F  j_F)$ [fb] at $\sqrt{s}= 14 $ TeV
  for $ \sin(\beta-\alpha)=1, 0.95, 0.9 $ with the
  benchmark points $ \Delta m =100, 200, 300 $ GeV and $ m_{H^{\pm}}=
  100, 200, 300 $ GeV.
  Notice the VBF cut $ \eta_{j_1}\times\eta_{j_2} < 0 $ and $ |\Delta\eta_{jj}| > 2.5 $
  for the minimum rapidity difference between the forward jet pair have been applied. }
\begin{ruledtabular}
\begin{tabular}{c c c c c}
  & & \multicolumn{3}{c}{$\sigma (p p\rightarrow H^\pm H^\pm  j_F j_F)$ [fb] } \\
\hline
$ \Delta m $ (GeV) & $ m_{H^{\pm}} $ (GeV) & $\sin(\beta-\alpha)=1 $
            & $ \sin(\beta-\alpha)=0.95 $ & $\sin(\beta-\alpha)=0.9 $ \\ \hline
 & 100 & $ 5.84\times 10^{-1} $ & $ 5.43\times 10^{-1}$ & $ 5.03\times 10^{-1} $ \\
100 & 200 & $ 2.30\times 10^{-1} $ & $ 2.11\times 10^{-1}$ & $1.95\times 10^{-1} $ \\
& 300 & $ 8.57\times 10^{-2}$ & $ 7.86\times 10^{-2}$ & $ 7.21\times 10^{-2} $\\
\hline
 & 100 & $ 1.81 $ & $ 1.59 $ & $ 1.39 $ \\
200 & 200 & $ 8.82\times 10^{-1} $ & $ 7.66\times 10^{-1}$ & $ 6.62\times 10^{-1} $ \\
& 300 & $ 3.85\times 10^{-1} $ & $ 3.33\times 10^{-1}$ & $2.85\times 10^{-1}$ \\
\hline
 & 100 & $3.14$ & $2.70$ & $2.32$ \\
300 & 200 & $ 1.75$ & $ 1.49 $ & $ 1.26 $ \\
 & 300 & $ 8.54\times 10^{-1} $ & $ 7.21\times 10^{-1} $ & $ 6.05\times 10^{-1} $ \\ 
\end{tabular}
\end{ruledtabular}
\end{table}

\begin{table}[th!]
\centering     
\caption{\small \label{tab:sinba27}
  Sum of cross sections for $\sigma (p p\rightarrow H^+ H^+ j_F j_F)$ and
  $ \sigma (p p\rightarrow H^- H^- j_F  j_F)$ [fb] at $\sqrt{s}= 27 $ TeV
  for $ \sin(\beta-\alpha)=1, 0.95, 0.9 $ with the
  benchmark points $ \Delta m =100, 200, 300 $ GeV and $ m_{H^{\pm}}=
  100, 200, 300 $ GeV.
  Notice the VBF cut $ \eta_{j_1}\times\eta_{j_2} < 0 $ and $ |\Delta\eta_{jj}| > 2.5 $ for the minimum rapidity difference between the forward jet pair have been applied.
  }
\begin{ruledtabular}
\begin{tabular}{c c c c c}
  & & \multicolumn{3}{c}{$\sigma (p p\rightarrow H^\pm H^\pm  j_F j_F)$ [fb] } \\
\hline
$ \Delta m $ (GeV) & $ m_{H^{\pm}} $ (GeV) & $\sin(\beta-\alpha)=1 $
            & $ \sin(\beta-\alpha)=0.95 $ & $\sin(\beta-\alpha)=0.9 $ \\ \hline
 & 100 & $ 1.64 $ & $ 1.52 $ & $ 1.41 $ \\
100 & 200 & $ 7.46\times 10^{-1} $ & $ 6.87\times 10^{-1}$ & $6.34\times 10^{-1} $ \\
& 300 & $ 3.26\times 10^{-1}$ & $ 2.99\times 10^{-1}$ & $ 2.75\times 10^{-1} $\\
\hline
 & 100 & $ 5.24 $ & $ 4.59 $ & $ 4.00 $ \\
200 & 200 & $ 2.91 $ & $ 2.53 $ & $ 2.18 $ \\
& 300 & $ 1.47 $ & $ 1.27 $ & $ 1.09 $ \\
\hline
 & 100 & $9.35$ & $8.04$ & $6.87$ \\
300 & 200 & $ 5.84 $ & $ 4.97 $ & $ 4.20 $ \\
 & 300 & $ 3.29 $ & $ 2.77 $ & $ 2.33 $ \\ 
\end{tabular}
\end{ruledtabular}
\end{table}

As indicated by Eq.~(\ref{amp}) the production cross section of
$ p p\rightarrow H^{\pm}H^{\pm} j_F j_F $ scales as the square of the mass
splitting $\Delta m$.  We quantitatively show this relation by plotting the
production cross sections versus $m_{H^\pm}$ in Fig.~\ref{fig:Xsec1}
with $\Delta m =$ 100, 200, and 300 GeV at $\sqrt{s}= 14 $ TeV (left panel)
and $\sqrt{s}= 27 $ TeV (right panel).
It is clear to observe that
the cross section is enhanced according to the large
mass splitting $ \Delta m $.  Note that we have used the general
Two-Higgs-Doublet Model UFO model file \cite{Degrande:2014vpa} and
employ \textbf{Madgraph5 aMC@NLO} \cite{Alwall:2014hca} with VBF cut
$ \eta_{j_1}\times\eta_{j_2} < 0 $ and
$|\Delta\eta_{jj}| > 2.5 $ for the minimum rapidity difference between
the forward jet pair to evaluate the cross sections.
Furthermore, in order to study the effects of the near-alignment limit
on the production cross sections, we list some benchmark points for
the relation of cross sections with $\sin(\beta-\alpha) =1, 0.95, 0.9$
in Table~\ref{tab:sinba14} at $\sqrt{s}= 14 $ TeV and
Table~\ref{tab:sinba27} at $\sqrt{s}= 27 $ TeV, respectively.

We stress first that the production cross section
$pp\to H^\pm H^\pm j_Fj_F$ is the same for 
both 2HDM type I and X. Only the decay of the charged Higgs bosons
that will make the process model dependent.
The full signal process including decays of $H^\pm$, $W^\pm$, $A^0$ is given by
\beqa
p p\rightarrow W^{\pm\ast}W^{\pm\ast} j_F j_F\rightarrow H^{\pm}H^{\pm} j_F j_F
\rightarrow (W^{\pm}A^0)(W^{\pm}A^0) j_F j_F\rightarrow l^{\pm}
\nu_l(b\overline{b})l^{\pm}\nu_l(b\overline{b}) j_F j_F \nonumber \\
\label{eq:VBF-1}
\eeqa
in  type-I 2HDM, and
\beqa
p p\rightarrow W^{\pm\ast}W^{\pm\ast} j_F j_F\rightarrow H^{\pm}H^{\pm} j_F j_F
\rightarrow (W^{\pm}A^0)(W^{\pm}A^0) j_F j_F\rightarrow l^{\pm}
\nu_l(\tau^+\tau^-)l^{\pm}\nu_l(\tau^+\tau^-) j_F j_F \nonumber \\
\label{eq:VBF-2}
\eeqa
in type-X 2HDM. We advocate that the novel signatures including
the combination of a pair of same-sign dileptons ($ l^{\pm} l^{\pm} $), a forward
and energetic jet pair ($ j_F j_F $), and two pairs of bottom quarks
($ b\overline{b} $) or tau leptons ($ \tau^+\tau^- $) coming from two
light pseudoscalars $A^0$ can  largely reduce the possible SM backgrounds.

Even in case that the masses of $H^0$ and $A^0$ are separated wide enough and they can be 
directly measured from other production channels, the current VBF process is still worthwhile to search for. 
First, the advantage of this process is that it does not
depend on Yukawa couplings, in contrast to direct searches of $H^0$,
$A^0$, or $H^{\pm}$.  The cross section of the this process is a
function of mainly $m_{H^{\pm}}$, $\Delta M$ in the limit $\sin(\beta
-\alpha)=1$. Therefore, if no such process is observed, it can exclude
the charged Higgs mass or mass correlation between $m_{H^0}$ and
$m_{A^0}$.
Second, since $H^0$ is difficult to be discovered in the
(near) alignment limit $(\sin(\beta - \alpha)\approx 1)$ in type-I or
type-X 2HDM, this process can imply the mass of $H^0$. Nevertheless,
this is only true in 2HDMs.  
If the light boson $A^0$ can be discovered in the near future, the usefulness of this process is to tell the mass difference between $H^0$ and $A^0$ even we do not find the heavier boson $H^0$.
On the other hand, if both $H^0$ and $A^0$ have been discovered, the usefulness of this process is to tell if the cross section matches the prediction in 2HDM.



\subsection{Signal-background analysis for Type-I 2HDM}

The signal process in Eq.~(\ref{eq:VBF-1}) is unique with a signature
including the combination of a pair of same-sign dileptons ($
l^{\pm}l^{\pm} $), a pair of forward and energetic jets ($ j_F j_F $),
and two pairs of bottom quarks ($ b\overline{b} $) coming from two
light pseudoscalar $A^0$.
There are a few SM backgrounds that can mimic this kind of
final states. We
consider the following four processes as the main SM backgrounds,
\beqa pp\rightarrow t\overline{t}t\overline{t}\rightarrow
(bW^+)(\overline{b}W^-)(bW^+)(\overline{b}W^-)\rightarrow
l^{\pm}l^{\pm}4b4j,
\label{eq:4tBG}
\eeqa
\beqa
pp\rightarrow t\overline{b}\overline{t}b l^+ l^-\rightarrow (bW^+)\overline{b}(\overline{b}W^-)b l^+ l^-\rightarrow l^{\pm}l^+ l^- 4b2j,
\label{eq:2t2b2lBG}
\eeqa
\beqa
pp\rightarrow t\overline{t}t\overline{b}\rightarrow (bW^+)(\overline{b}W^-)(bW^+)\overline{b}\rightarrow l^+ l^+ 4b2j \nonumber \\
or\quad pp\rightarrow t\overline{t}\overline{t}b\rightarrow (bW^+)(\overline{b}W^-)(\overline{b}W^-)b\rightarrow l^- l^- 4b2j,
\label{eq:3tlbBG}
\eeqa
\beqa
pp\rightarrow tt\overline{b}\overline{b}jj\rightarrow (bW^+)(bW^+)\overline{b}\overline{b}jj\rightarrow l^+ l^+ 4b2j \nonumber \\
or\quad pp\rightarrow \overline{t}\overline{t}bbjj\rightarrow (\overline{b}W^-)(\overline{b}W^-)bbjj\rightarrow l^- l^- 4b2j \,.
\label{eq:2t2b2jBG}
\eeqa

\begin{table}[t]
\centering     
\caption{ \small \label{tab:SSCF-1}
  Cut flow table for the Type-I 2HDM signal
  $ p p\rightarrow H^{\pm} H^{\pm} j_F j_F $ with
  $m_{H^{\pm}}=205 $ GeV, $m_{A^0} = 65 $ GeV, $\Delta m = 200 $ GeV,
  $\tan\beta =5 $ and $ \sin(\beta-\alpha)=0.97 $, and various backgrounds
  at $ \sqrt{s}=14 $ TeV.  }
\begin{ruledtabular}
\begin{tabular}{l c c c c c}
Cross section (fb) & signal & $t\bar{t}t\bar{t}$ & $ t\bar{t}b\bar{b}l^+l^- $ & $3t1b$ & $2t2b2j$ \\ \hline
Preselection & $2.07\times 10^{-2}$ & $4.94\times 10^{-2}$ & $1.08\times 10^{-2}$ & $7.74\times 10^{-5}$ & $8.29\times 10^{-5}$ \\ \hline
$N(b,l^{\pm})\geq 3,2$, &&&&& \\
$P^{b,l^{\pm}}_T>20$GeV,$|\eta^{b,l}|<2.5$ & $1.76\times 10^{-3}$ & $6.17\times 10^{-3}$ & $9.56\times 10^{-4}$ & $9.57\times 10^{-6}$ & $9.81\times 10^{-6}$ \\ \hline
$N(j)\geq 2$, &&&&& \\
$P^{j}_T>30$GeV,$M_{jj}>500$GeV & $1.46\times 10^{-3}$ & $5.15\times 10^{-3}$ & $4.18\times 10^{-4}$ & $2.88\times 10^{-6}$ & $4.05\times 10^{-6}$ \\ \hline
$m_{H^{\pm}}$ Cuts &&&&& \\
$M_{bbl^{\pm}} < 250 $GeV & $1.41\times 10^{-3}$ & $3.50\times 10^{-3}$ & $2.71\times 10^{-4}$ & $1.85\times 10^{-6}$ & $2.62\times 10^{-6}$ \\
$m_{A}$ Cuts &&&&& \\
$ 50 < M_{bb} < 90 $GeV & $1.30\times 10^{-3}$ & $1.68\times 10^{-3}$ & $1.61\times 10^{-4}$ & $7.58\times 10^{-7}$ & $1.14\times 10^{-6}$ \\
\end{tabular}
\end{ruledtabular}
\end{table}

All signal and SM
background events are simulated at leading order (LO)
using Madgraph5 aMC@NLO.
\footnote{The NLO QCD corrections for the signal process in
  Eq.~(\ref{eq:VBF-1}) and background processes in Eq.~(\ref{eq:4tBG})
  and (\ref{eq:2t2b2lBG}) have been checked with Madgraph5
    aMC@NLO. We assume that
  the kinematic distributions are only mildly affected by
  these higher order QCD effects.  }
In the following, we choose $m_{H^\pm} = 205$ GeV and $m_{A^0} = 65 $
GeV to illustrate the cut flow under a sequence of selection
cuts at $\sqrt{s}= 14 $ TeV.
\begin{enumerate}
\item 
We first identify the forward jet pair ($ j_F j_F $)  in the VBF-type
process and apply the VBF cut $ \eta_{j_1}\times\eta_{j_2} < 0 $ and
$ |\Delta\eta_{jj}| > 2.5 $ for the minimum
rapidity difference between the forward jet pair
in Madgraph5 aMC@NLO at parton level for all
signal and SM background events.
The cross sections for both signal
and background events after this pre-selection cut are shown in the
first row of Table~\ref{tab:SSCF-1}.

\item Then we employ 
\textbf{Pythia8} \cite{Sjostrand:2007gs} for parton showering and
hadronization. \textbf{Delphes3} \cite{delphes3} with default settings
is used for fast detector simulation.
\footnote{
   Notice that we apply the Delphes 3.4.1 in the
   Madgraph5 aMC@NLO. Comparing with the HL-LHC Delphes card in the
   most current version Delphes 3.4.2, they added the conditions
   $|\eta|<2.5$ and $10 < P_T <1000$ GeV for the same $\tau$-tagging
   efficiency and light jet to tau-jet misidentification rate. On the
   other hand, they also included the $\eta$ dependence with similar
   $P_T$ dependence settings compared with our default version. We
   expect these changes will only make very mild modifications of our
   conclusions.}
Finally, all events are analyzed
with \textbf{MadAnalysis5} \cite{MA5}.  We require to see a pair of
same-sign dileptons ($ l^{\pm}l^{\pm} $) and at least $ 3b $ in the event
as the trigger with the following sequence of event selection cuts
\beqa
N(b,l^{\pm})\geq 3,2,\quad P^{l^{\pm}}_T > 20\;{\rm  GeV},
\quad |\eta^{l^{\pm}}| < 2.5,\quad
P^b_T > 20\; {\rm GeV},\quad |\eta^b|<2.5.
\label{eq:trigger}
\eeqa
  The b-jets are selected with the efficiency as a function of $P_T$
  as, $ \epsilon_b = 0.85\times \tanh(0.0025\times P_T)\times
  (25.0/(1+0.063\times P_T)) $ and the misidentification rate from
  c-jets and light jets to b-jets are $ P(c\rightarrow b)= 0.25\times
  \tanh(0.018\times P_T)\times (1/(1+0.0013\times P_T)) $ and $
  P(j\rightarrow b)= 0.01+0.000038\times P_T $, separately.
The cross sections for both signal and background events are shown in the
second row of Table~\ref{tab:SSCF-1}.

\item 
  The forward jet pair is also  required to be energetic with the
  following selection cuts
\beqa
N(j)\geq 2,\quad p^j_T > 30\;{\rm GeV},\quad |\eta^j| < 5,\quad m_{jj} > 500\;{\rm GeV}.
\label{eq:jf-cut}
\eeqa 
The cross sections after this step for both signal
and background events
are shown in the third  row of Table~\ref{tab:SSCF-1}.

\item
 The kinematical distributions of $ M_{bbl^{\pm}} $ and $ M_{bb} $
with $m_{H^\pm} =205$ GeV and $m_{A^0} = 65$ GeV for the
signal and backgrounds are shown in Fig.~\ref{fig:kin1}.
Note that we have applied all the selection cuts except for
  $m_{H^{\pm}}$ and $ m_{A^0} $ cuts in these two kinematical
  distributions.
The signal distribution of $M_{bbl^\pm}$ tends to concentrate
in the region of $M_{bbl^{\pm}} < 250 $ GeV
and decreases more rapidly toward the higher $M_{bbl^\pm}$.
On the other hand, the background is relatively flat after 150 GeV to
500 GeV.
It is also clear to observe the peak shape at 65 GeV in $ M_{bb} $ distribution
for the signal from the resonance of $A^0 $.
These two behaviors can help us to distinguish between the signal 
and the background.

\begin{figure}[t!]
  \centering
\includegraphics[width=0.48\textwidth]{{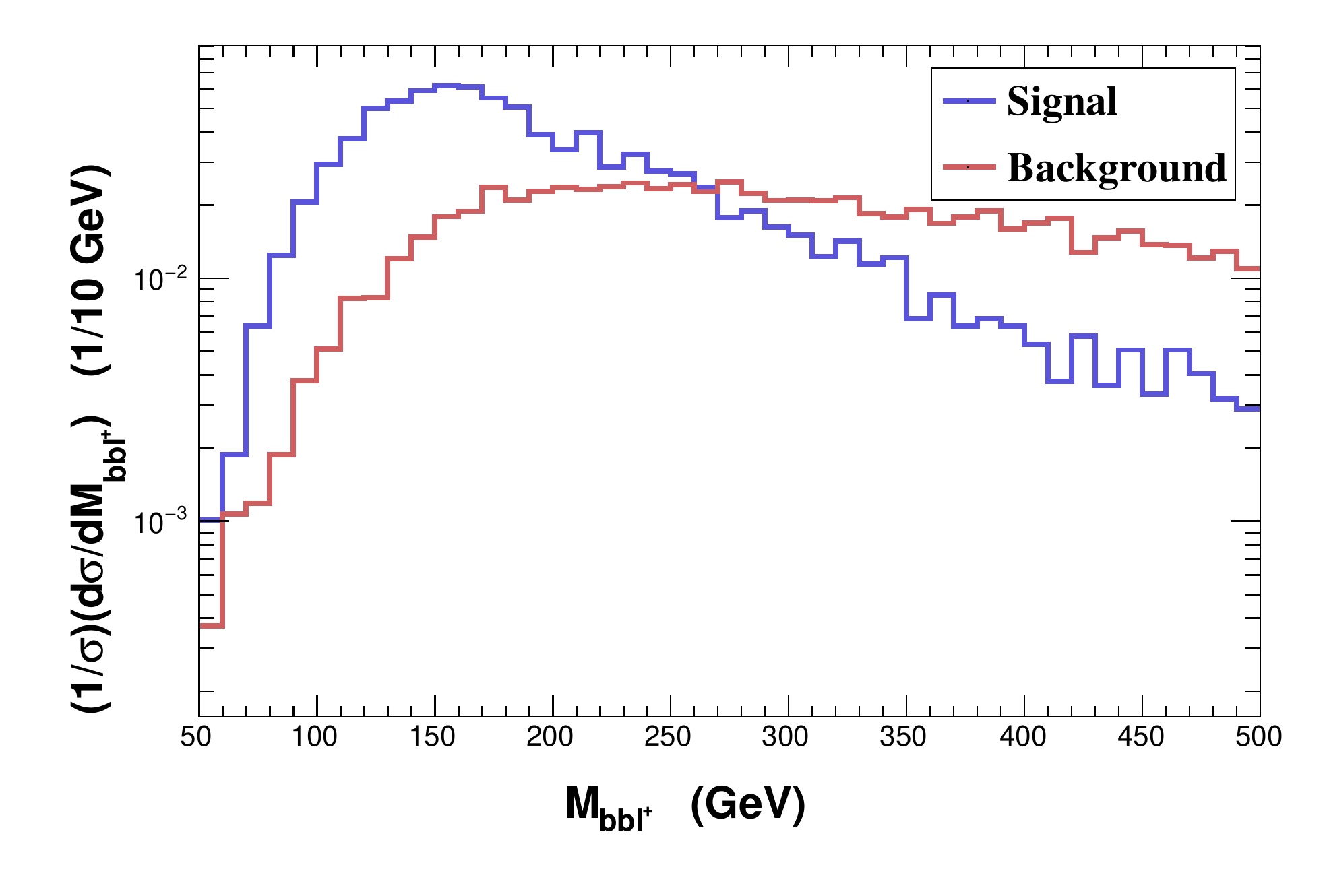}}
\includegraphics[width=0.48\textwidth]{{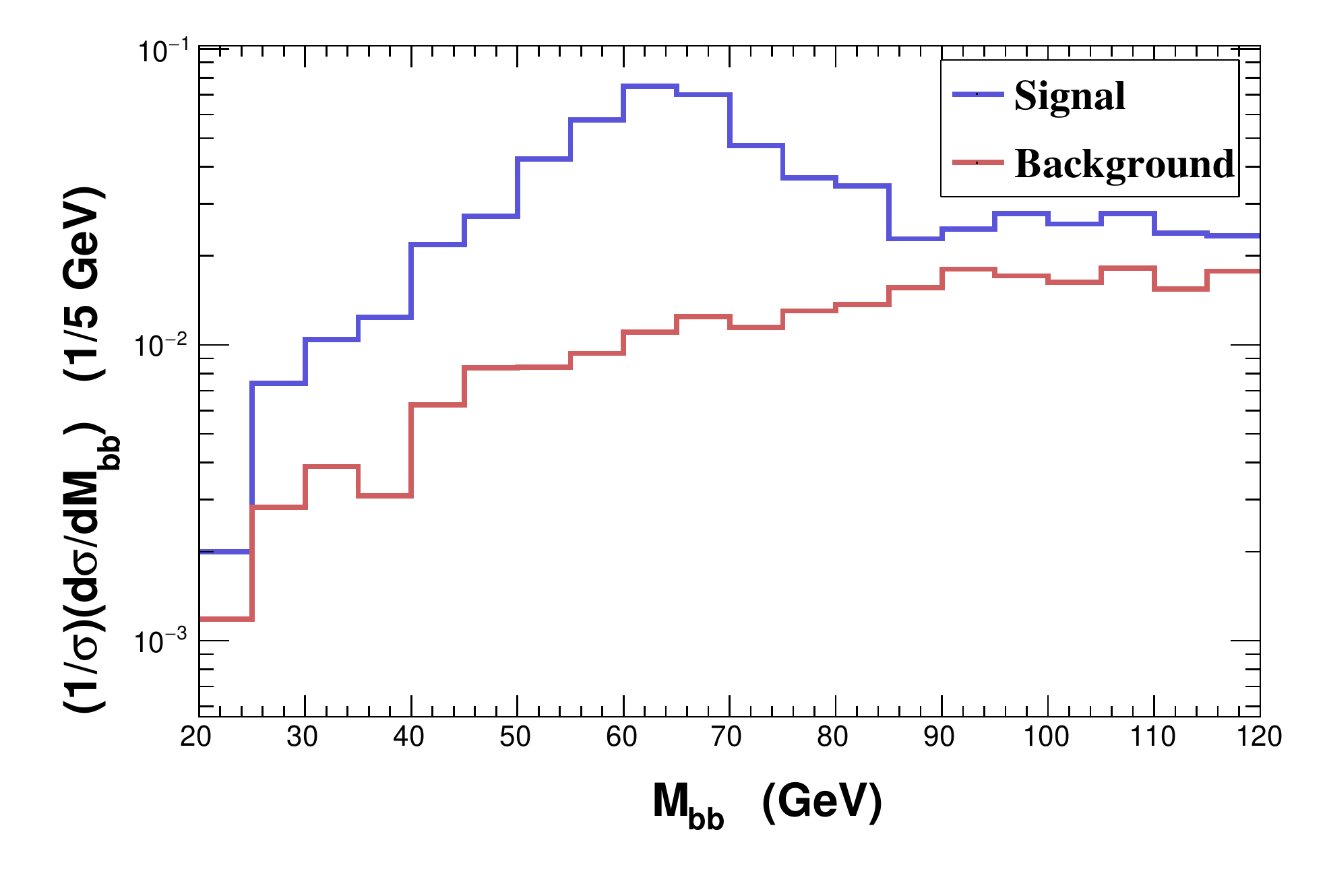}}
\caption{\small
  Invariant mass distributions of $ M_{bbl^{\pm}} $ (left panel) and
  $ M_{bb} $ (right panel) for the signal with 
$m_{H^{\pm}}=205 $ GeV, $m_{A^0} = 65 $ GeV, $\Delta m = 200 $ GeV,
  $\tan\beta =5 $ and $ \sin(\beta-\alpha)=0.97 $, and the
  total background at $ \sqrt{s}=14 $ TeV.
  Preselection cuts in Eqs.~(\ref{eq:trigger}) and
  (\ref{eq:jf-cut}) are imposed.
}
\label{fig:kin1}
\end{figure}

\item Finally, in order to further reduce the contributions from SM backgrounds,
the following selection cuts are imposed on both signal and background events. 
For $m_{H^{\pm}}$ cuts  at least two bottom quarks and a lepton
have to satisfy
\beqa
M_{bbl^{\pm}}\leq M_{H^{\pm}} +45 \;{\rm GeV}.
\label{eq:MCHcut}
\eeqa
For $ m_{A^0} $ cuts  at least a pair of bottom quarks
are required to be around the mass of $A^0$:
\beqa
\quad m_{A^0} - 15 \;{\rm GeV} \leq M_{bb} \leq m_{A^0} +25\; {\rm GeV}.
\label{eq:MAcut}
\eeqa
Again, the cross sections for both signal and background events after
this sequence of event selection cuts are shown in the last two rows
of Table~\ref{tab:SSCF-1}.

\end{enumerate}
After all selection cuts the signal-to-background ratio is almost
close to 1. With a luminosity of 3000 fb$^{-1}$ we expect about 4
signal and 5 background events.  The major background comes from
$t\bar t t\bar t$ production while the other backgrounds listed in
Table~\ref{tab:SSCF-1} are much suppressed.

\begin{table}[t]
\centering     
\caption{ \small \label{tab:SSCF-1-27}
  Cut flow table for the Type-I 2HDM signal
  $ p p\rightarrow H^{\pm} H^{\pm} j_F j_F $ with
  $m_{H^{\pm}}=205 $ GeV, $m_{A^0} = 65 $ GeV, $\Delta m = 200 $ GeV,
  $\tan\beta =5 $ and $ \sin(\beta-\alpha)=0.97 $,
  and various backgrounds
  at $ \sqrt{s}=27 $ TeV.  }
\begin{ruledtabular}
\begin{tabular}{l c c c c c}
Cross section (fb) & signal & $t\bar{t}t\bar{t}$ & $ t\bar{t}b\bar{b}l^+l^- $ & $3t1b$ & $2t2b2j$ \\ \hline
Preselection & $6.88\times 10^{-2}$ & $5.67\times 10^{-1}$ & $5.60\times 10^{-2}$ & $2.40\times 10^{-4}$ & $6.76\times 10^{-4}$ \\ \hline
$N(b,l^{\pm})\geq 3,2$, &&&&& \\
$P^{b,l^{\pm}}_T>20$GeV,$|\eta^{b,l}|<2.5$ & $5.15\times 10^{-3}$ & $5.67\times 10^{-2}$ & $4.43\times 10^{-3}$ & $2.44\times 10^{-5}$ & $6.42\times 10^{-5}$ \\ \hline
$N(j)\geq 2$, &&&&& \\
$P^{j}_T>30$GeV,$M_{jj}>500$GeV & $4.54\times 10^{-3}$ & $5.22\times 10^{-2}$ & $2.49\times 10^{-3}$ & $9.67\times 10^{-6}$ & $3.27\times 10^{-5}$ \\ \hline
$m_{H^{\pm}}$ Cuts &&&&& \\
$M_{bbl^{\pm}} < 200 $GeV & $4.10\times 10^{-3}$ & $2.28\times 10^{-2}$ & $1.08\times 10^{-3}$ & $4.29\times 10^{-6}$ & $1.45\times 10^{-5}$ \\
$m_{A}$ Cuts &&&&& \\
$ 50 < M_{bb} < 80 $GeV & $3.76\times 10^{-3}$ & $1.12\times 10^{-2}$ & $6.09\times 10^{-4}$ & $1.91\times 10^{-6}$ & $7.15\times 10^{-6}$ \\
\end{tabular}
\end{ruledtabular}
\end{table}

  Even though the signal-to-background ratio is close to one 
for the analysis at the HL-LHC, the total number of events is small and
the fluctuations of SM backgrounds may also be an issue.
Since we cannot draw any concrete conclusion for this situation, 
we further extend the signal-background analysis to the proposed 27
TeV $pp$ collider(HE-LHC). 
The SM background cross sections
grow faster than the signal one from $\sqrt{s}= $ 14 to 27 TeV. In
order to reduce the enhanced background cross sections, both
$m_{H^{\pm}}$ and $ m_{A^0} $ cuts are tightened relative to those in 
Eqs.~(\ref{eq:MCHcut}) and (\ref{eq:MAcut}).  For
$m_{H^{\pm}}$ cuts at least two bottom quarks and a lepton have to
satisfy \beqa M_{bbl^{\pm}}\leq M_{H^{\pm}} -5 \;{\rm GeV}.
\label{eq:MCHcut27}
\eeqa
For $ m_{A^0} $ cuts  at least a pair of bottom quarks is
required to be around the mass of $A^0$:
\beqa
\quad |M_{bb}-m_A|\leq 15\; {\rm GeV}. 
\label{eq:MAcut27}
\eeqa
Other preselection cuts,
given in Eqs.~(\ref{eq:trigger}) and (\ref{eq:jf-cut}),
are the same as before.
On the other hand, the shape of kinematical distributions for $ M_{bbl^{\pm}} $ and $ M_{bb} $
with $m_{H^\pm} =205$ GeV and $m_{A^0} = 65$ GeV
at $\sqrt{s}= 27 $ TeV for the signal and backgrounds
are similar to Fig.~\ref{fig:kin1}, so we do not repeat displaying
them here.
We choose the same signal benchmark point to
illustrate the cut flow under a sequence of
selection cuts at $\sqrt{s}= 27 $ TeV in Table~\ref{tab:SSCF-1-27}. 

\begin{figure}[t!]
  \centering
\includegraphics[width=0.48\textwidth]{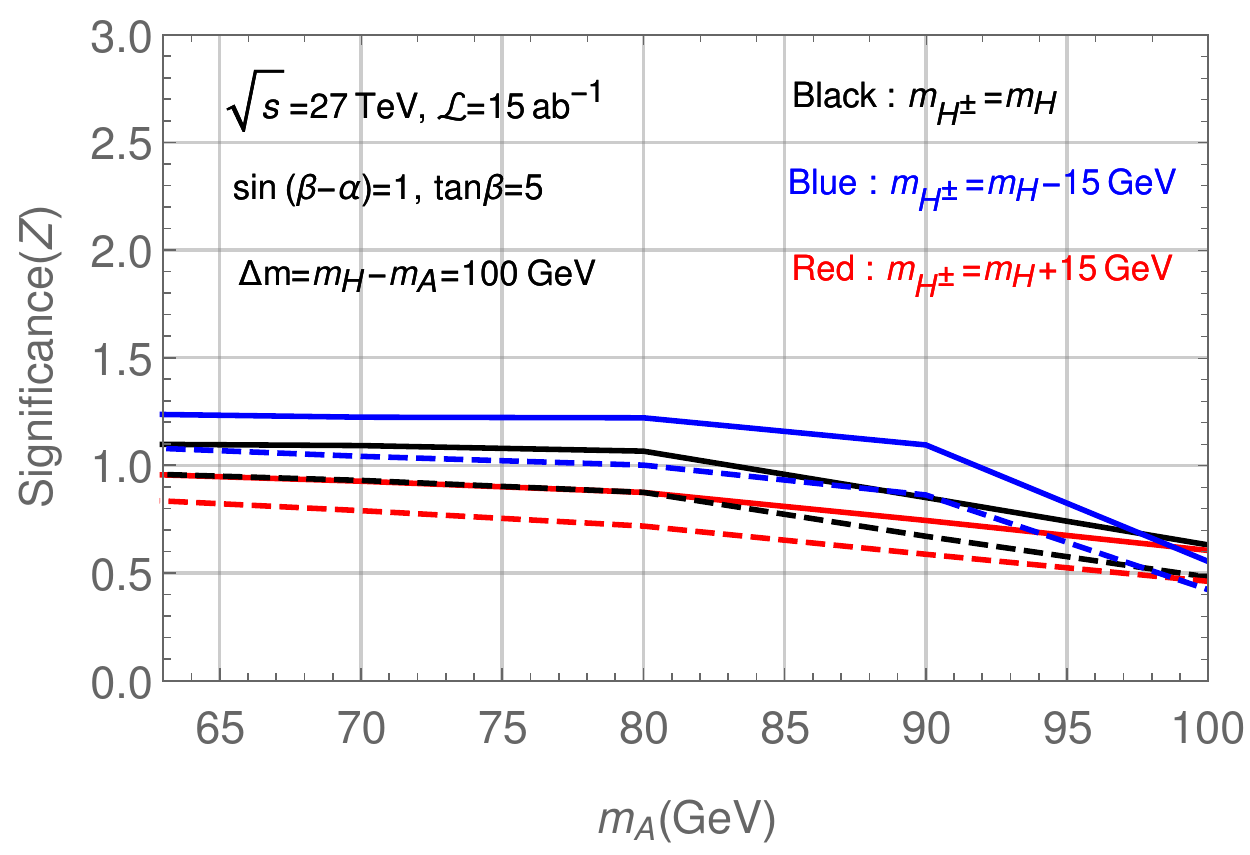}
\includegraphics[width=0.48\textwidth]{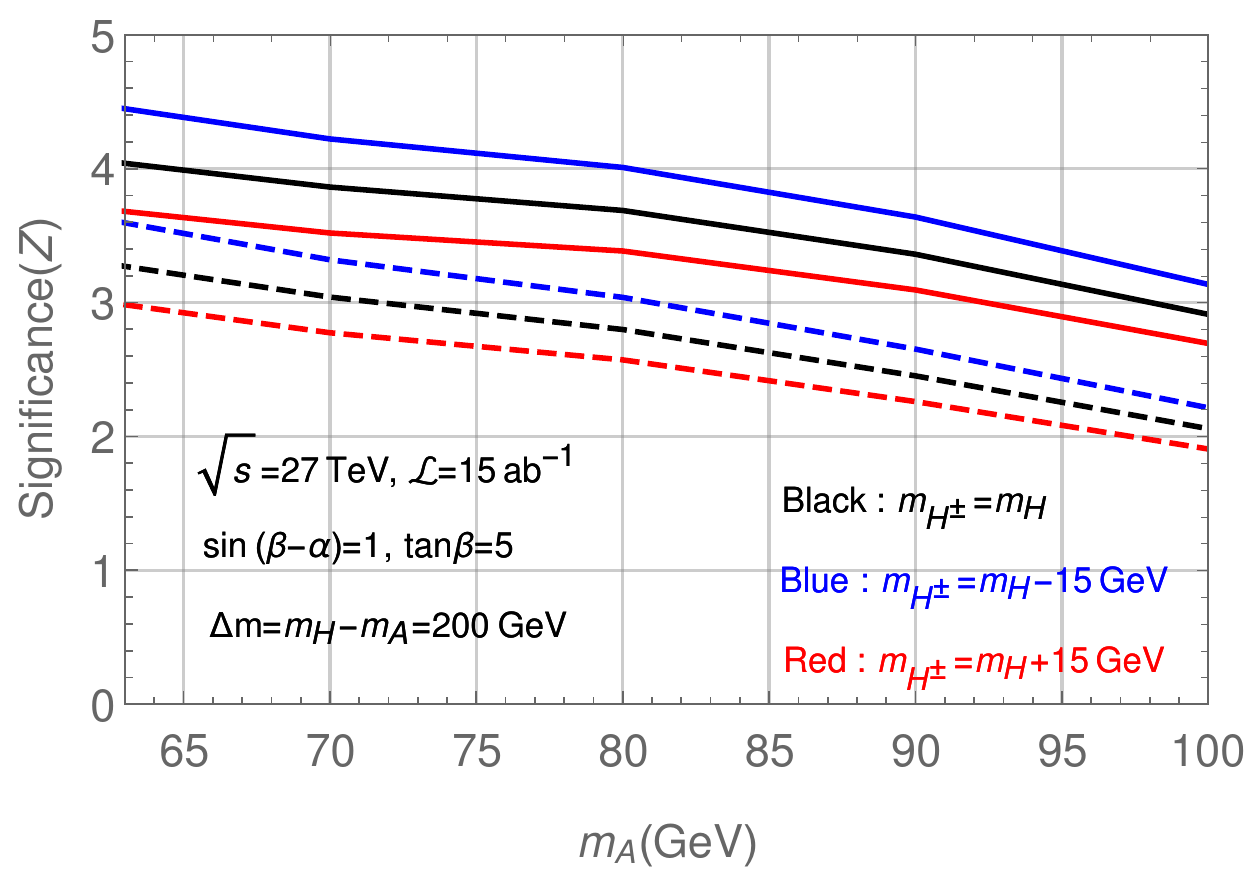}
\includegraphics[width=0.48\textwidth]{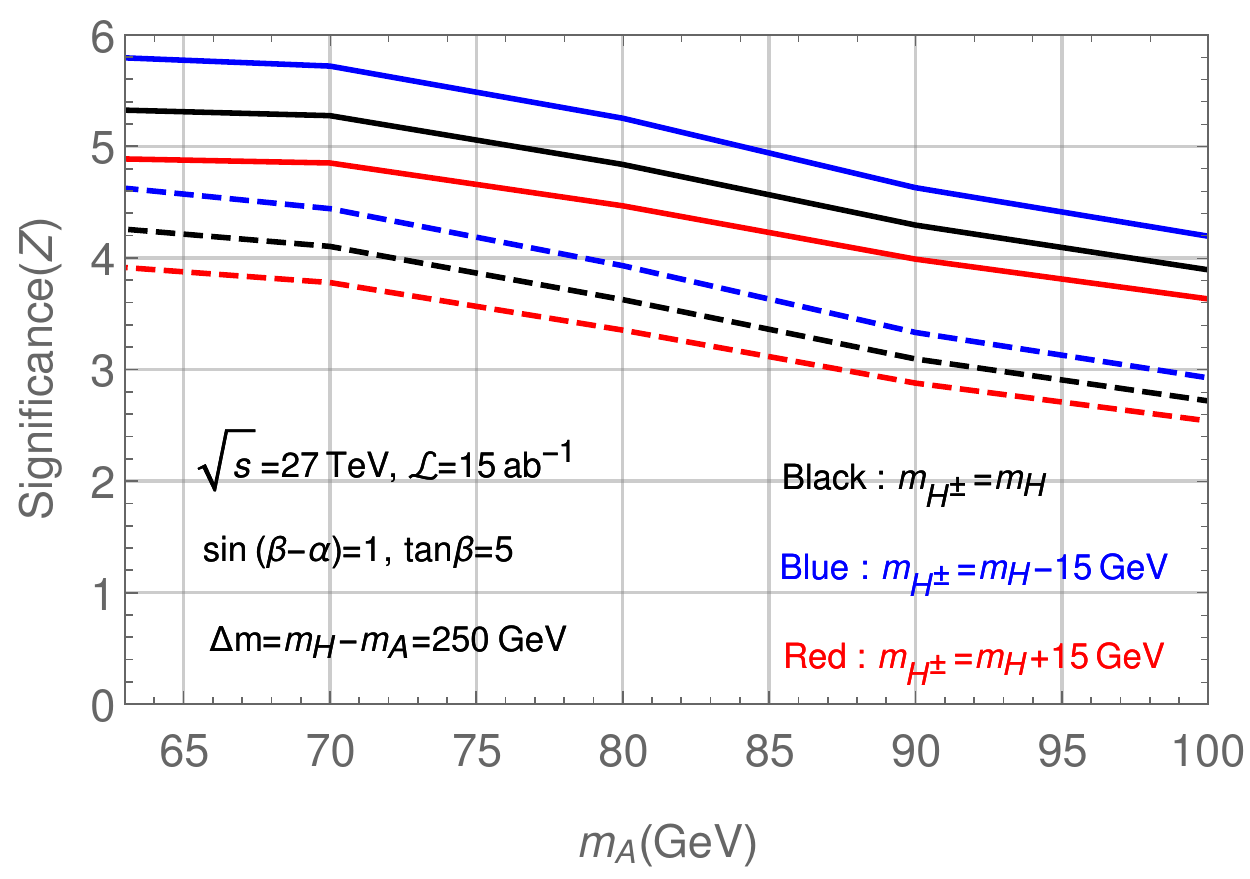}
\caption{\small
  The significance $ Z $ versus $ m_{A^0} $ from 63 to 100 GeV in
  Type-I 2HDM at $\sqrt{s}= 27 $ TeV with luminosity
  $ \mathcal L = 15 \, {\rm ab}^{-1} $.
  We have fixed $\sin(\beta -\alpha) = 1 $ and $\tan\beta = 5 $
  with $ \Delta m \equiv m_{H^0} - m_{A^0} = $ 100 GeV
  (upper-left panel), 200 GeV (upper-right panel), and
  250 GeV (lower panel).
  The dashed lines correspond to additional $5\%$ systematic
    uncertainties of the SM background events in Eq.~(\ref{eq:uncertainty}).
}
\label{fig:TP1SUM}
\end{figure}

Finally, we summarize our signal-background analysis for Type-I 2HDM
at $\sqrt{s}= 27 $ TeV with luminosity $ \mathcal L = 15 ab^{-1} $ in
Fig.~\ref{fig:TP1SUM}.  The preselection cuts in
Eqs.~(\ref{eq:trigger}), (\ref{eq:jf-cut}), (\ref{eq:MCHcut27}) and
(\ref{eq:MAcut27}) are imposed as before.  We vary $ m_{A^0} $ from 63 to
100 GeV with fixed $\sin(\beta -\alpha) = 1 $ and $\tan\beta = 5 $
for $ \Delta m = m_{H^0} - m_{A^0} = $ 100 GeV (upper-left panel), 200 GeV
(upper-right panel), and 250 GeV (lower panel) in
Fig.~\ref{fig:TP1SUM} as the illustrative examples.
The black lines are $ m_{H^{\pm}} = m_{H^0} $, the
blue lines are $ m_{H^{\pm}} = m_{H^0} - 15 $ GeV, and the red lines are
$m_{H^{\pm}} = m_{H^0} + 15 $ GeV.  We first define the significance by
\beqa
Z = \sqrt{2\cdot\left[(s+b)\cdot ln(1+s/b)-s\right]} \;,
\eeqa
where $s$ and
$b$ represent the numbers of signal and background events,
respectively.
According to the production cross sections of same-sign
charged Higgs in the right panel of Fig.~\ref{fig:Xsec1}, it is obvious
that the cases with moderate mass splittings $ \Delta m $ are difficult to
be detected even at HE-LHC with high luminosities. The maximum
significance is only about $Z=1.2$ for $ \Delta m $ = 100 GeV. We need
other charged Higgs production channels to detect this kind of
moderate mass splitting $ \Delta m $ cases. However, this same-sign charged
Higgs production channel is sensitive to the cases with large mass
splitting $ \Delta m $. The average significance is about $Z=3.5$ for
$m_{A^0} $ from 63 GeV to 100 GeV with $ \Delta m $ = 200 GeV, and its
maximum can reach to more than $Z=4.4$ at $ m_{A^0} = $ 63 GeV. Moreover,
the average significance can grow to about $Z=4.5$ for $ m_{A^0} $ from 63
GeV to 100 GeV with $ \Delta m $ = 250 GeV, and its maximum can
further reach to $Z=5.8$ for $ m_{A^0}\leq $ 70 GeV.

We further consider a $5\%$ systematic uncertainty in estimation
of the SM background. The significance of the signal is modified to 
\beqa
Z = \sqrt{2\cdot\left[(s+b)\cdot ln\left(\frac{(s+b)(b+\sigma^2_b)}{b^2 +(s+b)\sigma^2_b}\right)-\frac{b^2}{\sigma^2_b}\cdot ln\left(1+\frac{\sigma^2_b s}{b(b+\sigma^2_b)}\right)\right]} \;,
\label{eq:uncertainty}
\eeqa 
where $\sigma_b$ is the systematic uncertainty of the SM background $b$.
We show the effect of including systematic uncertainties
as dashed lines in Fig.~\ref{fig:TP1SUM} for comparisons.
  \footnote{
Notice that the $5\%$ systematic uncertainty that we have assumed in
estimation of the SM background is an optimistic choice.
Even it is not trivial, this level of systematics might be still
achievable at HE-LHC with luminosity $\mathcal L = 15 ab^{-1}$.
}
The reduction of the systematic uncertainty in future collider experiments is a long shot, but with
better understanding of the SM backgrounds and theoretical calculations, a level of less than $10\%$ 
systematic uncertainty is not beyond reach. If we take the number of signal and background events
with the cross sections shown in the last row of Table~\ref{tab:SSCF-1-27}, $s = 56, b = 168$ with 15 ab$^{-1}$ 
integrated luminosity. The significance $Z = 4.1$ with $0\%$ systematic uncertainty, but reduces to 
3.4, 2.4, 1.4 with $5\%$, $10\%$, $20\%$ systematic uncertainties. Therefore, one can see that a 
systematic uncertainty better than $10\%$ is needed to see a significant excess. In order to preserve 
a significant excess it is better to achieve as good as $5\%$ systematic uncertainty.

\subsection{Signal-background analysis for Type-X 2HDM}

In type X 2HDM, the major decay of the pseudoscalar $A^0$ is $A^0
\to \tau \tau$.  Therefore, we modify the above signal-background
analysis to
two pairs of tau leptons, instead of two pairs of bottom quarks, in
the final state.  The decay chain is shown in Eq.~(\ref{eq:VBF-2}).
Therefore, we are considering the following set of backgrounds at LO :
\beqa
pp\rightarrow t\overline{t}Zjj \rightarrow
(bW^+)(\overline{b}W^-)(\tau^+\tau^-)jj\rightarrow l^{\pm}2b3\tau 2j,
\label{eq:2t2jzBG}
\eeqa
\beqa
pp\rightarrow t\overline{t}W^{\pm}jj\rightarrow (bW^+)(\overline{b}W^-)(\tau^{\pm}\nu_{\tau})jj\rightarrow l^{\pm}2b2\tau 2j,
\label{eq:2t2jwBG}
\eeqa
\beqa
pp\rightarrow W^{\pm}W^{\mp}Zjj\rightarrow (l^{\pm}\nu_l)(\tau^{\mp}\nu_{\tau})(\tau^+\tau^-)jj
\rightarrow l^{\pm}3\tau 2j,
\label{eq:2w2jzBG}
\eeqa
\beqa
pp\rightarrow W^{\pm}ZZjj\rightarrow (l^{\pm}\nu_l)(\tau^+\tau^-)(\tau^+\tau^-)jj
\rightarrow l^{\pm}4\tau 2j \;.
\label{eq:2z2jwBG}
\eeqa
The extra same-sign charged leptons may come from some cascade
decays of the tau leptons, B mesons, or showering. Similarly, the
extra tau leptons can also come from B meson cascade decays,
showering, or jet misidentification.

\begin{table}[t]
\centering     
\caption{ \small \label{tab:SSCF-3}
  Cut flow table for the Type-X 2HDM signal
  $ p p\rightarrow H^{\pm} H^{\pm} j_F j_F $ with
  $m_{H^{\pm}}=205 $ GeV, $m_{A^0} = 65 $ GeV, $\Delta m = 200 $ GeV,
  $\tan\beta =5 $ and $ \sin(\beta-\alpha)=0.97 $, and various backgrounds
  at $ \sqrt{s}=14 $ TeV. }
\begin{ruledtabular}
\begin{tabular}{l c c c c c}
Cross section (fb) & signal & $t\overline{t}Zjj$ & $t\overline{t}W^{\pm}jj$ & $W^{\pm}W^{\mp}Zjj$ & $W^{\pm}ZZjj$ \\ \hline
Preselection & $2.98\times 10^{-2}$ & $ 3.60\times 10^{-1} $ & $ 2.44\times 10^{-1} $ & $ 3.28\times 10^{-2} $ & $ 1.87\times 10^{-3} $ \\ \hline
$N(\tau,l^{\pm})\geq 3,2$, &&&&& \\
$P^{\tau,l^{\pm}}_T>20$GeV,$|\eta^{\tau,l}|<2.5$ &$1.23\times 10^{-3}$ & $ 7.42\times 10^{-3} $ & $ 1.07\times 10^{-3} $ & $ 3.89\times 10^{-4} $ & $ 9.61\times 10^{-5} $ \\ \hline
$N(j)\geq 2$, &&&&& \\
$P^{j}_T>30$GeV,$M_{jj}>500$GeV & $9.81\times 10^{-4}$ & $ 4.63\times 10^{-3} $ & $ 6.19\times 10^{-4} $ & $ 1.97\times 10^{-4} $ & $ 5.08\times 10^{-5} $ \\ \hline
b-jet veto & $9.15\times 10^{-4}$ & $1.15\times 10^{-3}$ & $2.03\times 10^{-4}$ & $1.71\times 10^{-4}$ & $4.32\times 10^{-5}$ \\ \hline
$m_{H^{\pm}}$ Cut  &&&&& \\
$M_{\tau^+\tau^-l^{\pm}} < 250 $GeV & $8.24\times 10^{-4}$ & $ 7.52\times 10^{-4} $ & $ 9.18\times 10^{-5} $ & $ 1.15\times 10^{-4} $ & $ 2.98\times 10^{-5} $ \\
$m_{A^0}$ Cut &&&&& \\
$ 40 < M_{\tau^+\tau^-} < 100 $GeV & $7.95\times 10^{-4}$ & $ 6.28\times 10^{-4} $ & $ 5.81\times 10^{-5} $ & $ 1.04\times 10^{-4} $ & $ 2.73\times 10^{-5} $ \\
\end{tabular}
\end{ruledtabular}
\end{table}

Again, we choose $m_{H^\pm} = 205$ GeV and $m_{A^0} = 65$ GeV to
illustrate the cut flow under a sequence of selection cuts.
\begin{enumerate}
\item
  We apply the same VBF cut $ \eta_{j_1}\times\eta_{j_2} < 0 $ and
  $|\Delta \eta_{jj}| > 2.5$ for the minimum
rapidity difference between the forward jet pair at parton level for
all signal and SM background events. Their cross sections after this
pre-selection cut are shown in the first row of
Table~\ref{tab:SSCF-3}.

\item
  After parton showering and hadronization with Pythia8 and detector simulation
  by Delphes3, we apply the selections cuts for 
  a pair of same-sign dileptons and at least $3 \tau$:
\beqa
N(\tau,l^{\pm})\geq 3,2,\quad P^{l^{\pm}}_T > 20\;{\rm  GeV},
\quad |\eta^{l^{\pm}}| < 2.5,\quad
P^{\tau}_T > 20\; {\rm GeV},\quad |\eta^{\tau}|<2.5.
\label{eq:trigger-3}
\eeqa
Notice we take the hadronic decays of the tau leptons. The tau tagging
in Delphes3 is encoded with the origin of jets from hadronic decay
modes of the tau lepton with an efficiency 0.6 and the misidentification
rate from light-jet to tau-jet 0.01.
The charge of tau-jet can be determined and reconstructed from the charged
pions in the final state according to the algorithm inside Delphes3.
  The cross sections for both signal and backgrounds 
  are shown in the second row of Table~\ref{tab:SSCF-3}.

\item
  The forward jet pair is also required to be energetic with the following selection cuts
  \begin{equation}
    N(j) >2, \qquad p_T^j > 30\;{\rm GeV},\qquad |\eta^j| < 5,\qquad
    m_{jj} > 500 \; {\rm GeV} \;.
    \label{eq:jf-cut-3}
  \end{equation}
  The cross sections after this step for both signal and
  backgrounds are shown in the third row of Table~\ref{tab:SSCF-3}.

\item Since the major background comes from the
  $ t\overline{t} $ associated processes, we
  apply b-jet veto to suppress background events:
\beqa
N(b)=0\quad {\rm with} \quad
P^b_T > 20\; {\rm GeV},\quad |\eta^b|<2.5.
\label{eq:b-veto}
\eeqa
  The cross sections after this step for both signal and background events are shown in
  the fourth row of Table~\ref{tab:SSCF-3}.

\item
  The kinematical distributions of $ M_{l^{\pm}\tau^+\tau^-} $ and $
  M_{\tau^+\tau^-} $ with $m_{H^\pm} =205$ GeV and $m_{A^0} = 65$ GeV
  for the signal and backgrounds are shown in Fig.~\ref{fig:kin2}.
  Note that we have applied all the selection cuts except for
  $m_{H^{\pm}}$ and $ m_{A^0} $ cuts in these two kinematical
  distributions.  The signal and background distributions of $
  M_{l^{\pm}\tau^+\tau^-} $ are similar to $M_{bbl^\pm}$ in
  Fig.~\ref{fig:kin1}. However, the peak shape at 65 GeV in $
  M_{\tau^+\tau^-} $ distribution for the signal from the resonance of
  $A^0 $ is not so obvious compared with $ M_{bb} $ distribution in
  Fig.~\ref{fig:kin1}. The reason is that the $ \tau $-tagging is
  not as effective as b-tagging.
  On the other hand, since there are always neutrinos
  in $ \tau $ lepton decays, the $ \tau $ lepton cannot be fully
  reconstructed.  This also explains why the shift of fat peak shape from 65
  GeV to a slightly lower $ M_{\tau^+\tau^-} $.

\begin{figure}[t!]
  \centering
\includegraphics[width=0.48\textwidth]{{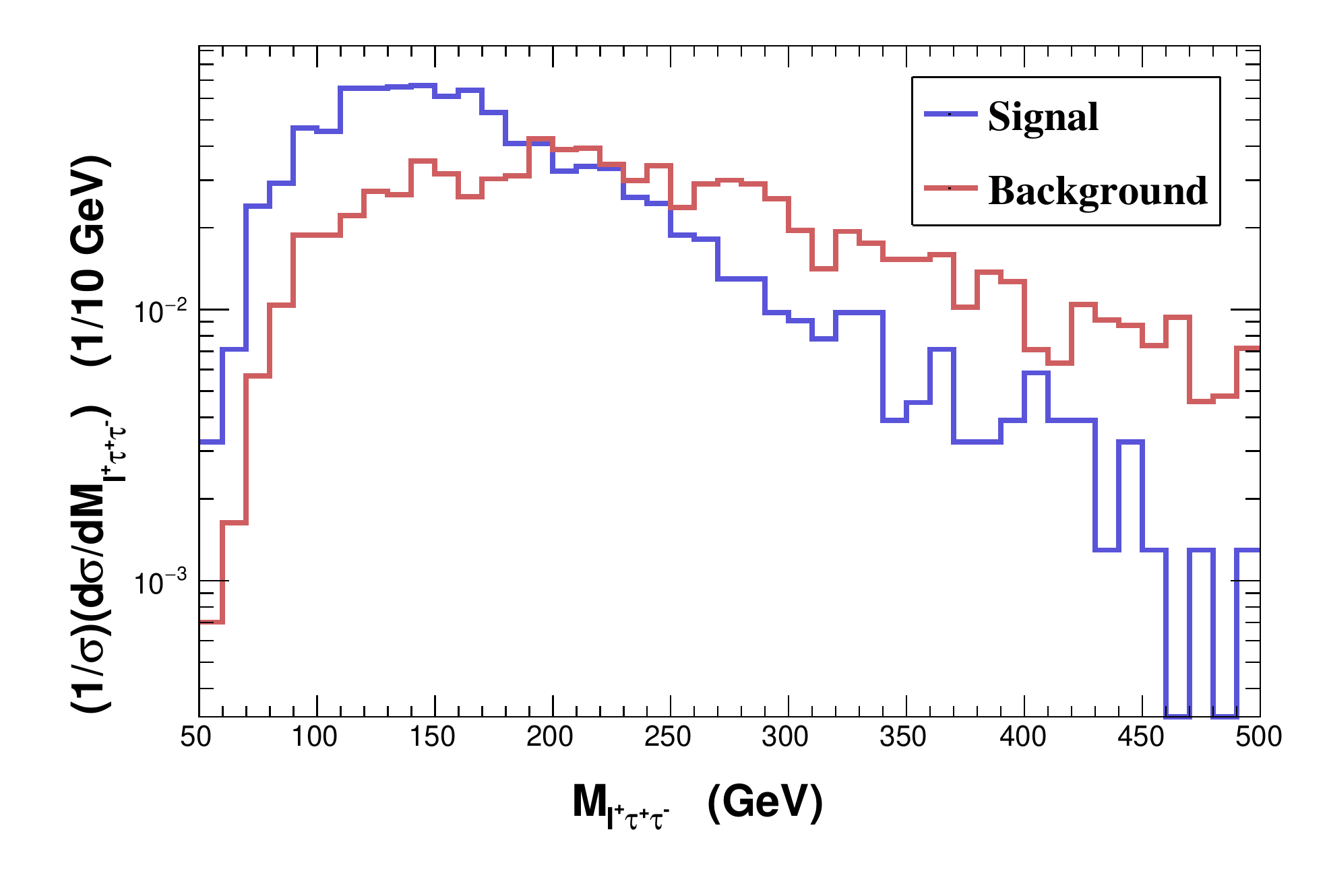}}
\includegraphics[width=0.48\textwidth]{{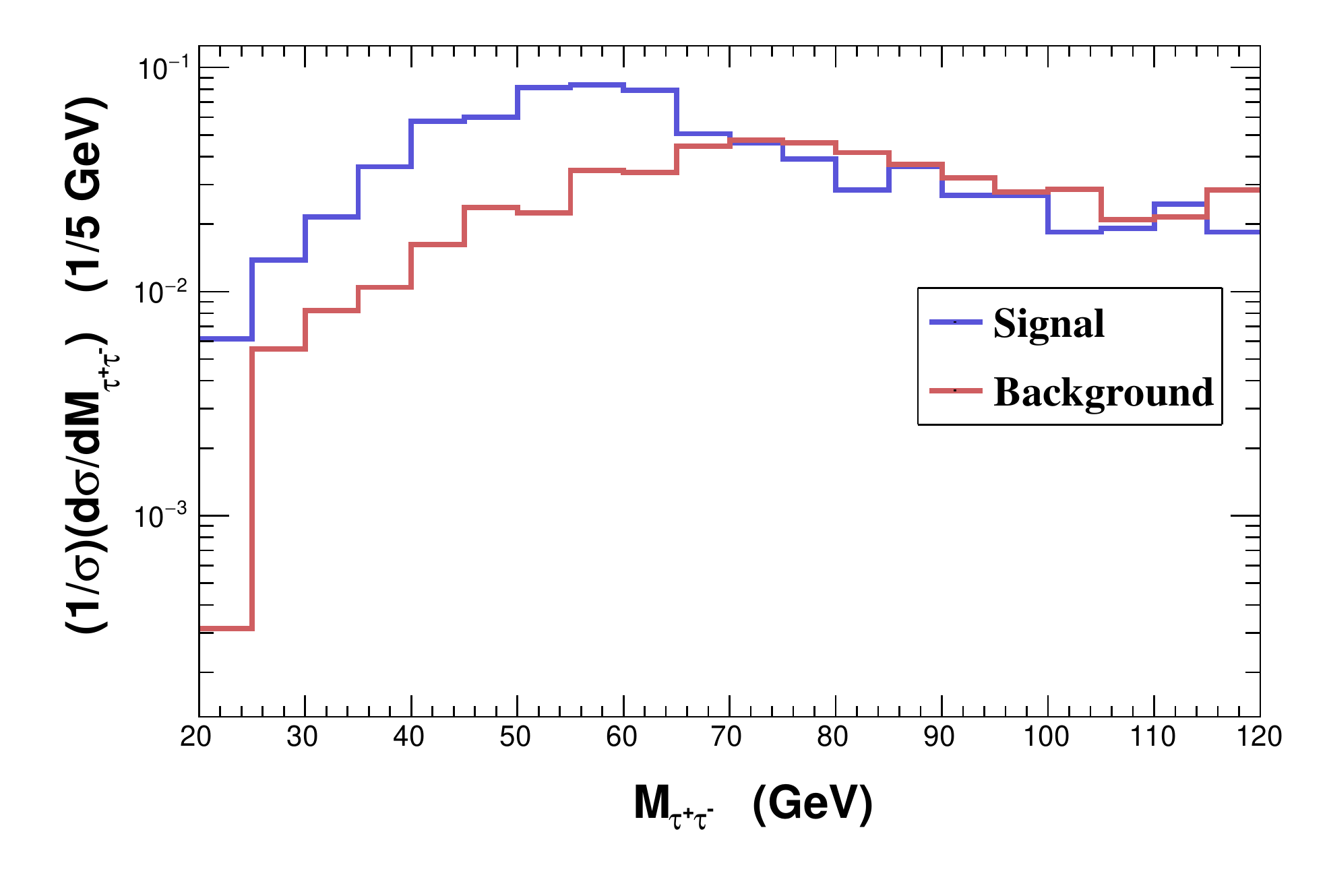}}
\caption{\small
  Invariant mass distributions of $ M_{l^{\pm}\tau^+\tau^-}$ (left panel)
  and $ M_{\tau^+\tau^-} $ (right panel) for the signal with 
  $m_{H^{\pm}}=205 $ GeV, $m_{A^0} = 65 $ GeV, $\Delta m = 200 $ GeV,
  $\tan\beta =5 $ and $ \sin(\beta-\alpha)=0.97 $,
  and the total background at $ \sqrt{s}=14 $ TeV.
  Preselection cuts in Eqs.~(\ref{eq:trigger-3}), (\ref{eq:jf-cut-3})
  and (\ref{eq:b-veto}) are imposed.
}
\label{fig:kin2}
\end{figure}

\item Finally, in order to further reduce the contributions from SM backgrounds,
the following selection cuts are imposed on both signal and background events. 
For $m_{H^{\pm}}$ cuts at least two opposite-sign tau leptons and a lepton
have to satisfy
\beqa
M_{l^{\pm}\tau^+\tau^-}\leq M_{H^{\pm}} +45 \;{\rm GeV}.
\label{eq:MCHcu-3}
\eeqa
For the $ m_{A^0} $ cut  at least a pair of opposite-sign tau leptons
is required to around the mass of $A^0$:
\beqa
\quad m_{A^0} - 25 \;{\rm GeV} \leq M_{\tau^+\tau^-} \leq m_{A^0} +35\; {\rm GeV}.
\label{eq:MAcut-3}
\eeqa
The cross sections for both signal and background events after
this sequence of event selection cuts are shown in the last two rows
of Table~\ref{tab:SSCF-3}.
 \end{enumerate}

\begin{table}[t]
\centering     
\caption{ \small \label{tab:SSCF-3-27}
  Cut flow table for the Type-X 2HDM signal
  $ p p\rightarrow H^{\pm} H^{\pm} j_F j_F $ with
  $m_{H^{\pm}}=205 $ GeV, $m_{A^0} = 65 $ GeV, $\Delta m = 200 $ GeV,
  $\tan\beta =5 $ and $ \sin(\beta-\alpha)=0.97 $, and
  various backgrounds at $ \sqrt{s}=27 $ TeV.
}
\begin{ruledtabular}
\begin{tabular}{l c c c c c}
Cross section (fb) & signal & $t\overline{t}Zjj$ & $t\overline{t}W^{\pm}jj$ & $W^{\pm}W^{\mp}Zjj$ & $W^{\pm}ZZjj$ \\ \hline
Preselection & $9.93\times 10^{-2}$ & $ 2.51 $ & $ 1.49 $ & $ 1.51\times 10^{-1} $ & $ 8.62\times 10^{-3} $ \\ \hline
$N(\tau,l^{\pm})\geq 3,2$, &&&&& \\
$P^{\tau,l^{\pm}}_T>20$GeV,$|\eta^{\tau,l}|<2.5$ &$4.27\times 10^{-3}$ & $ 4.96\times 10^{-2} $ & $ 6.14\times 10^{-3} $ & $ 1.71\times 10^{-3} $ & $ 4.04\times 10^{-4} $ \\ \hline
$N(j)\geq 2$, &&&&& \\
$P^{j}_T>30$GeV,$M_{jj}>500$GeV & $3.71\times 10^{-3}$ & $ 3.69\times 10^{-2} $ & $ 4.50\times 10^{-3} $ & $ 1.08\times 10^{-3} $ & $ 2.67\times 10^{-4} $ \\ \hline
b-jet veto & $3.40\times 10^{-3}$ & $ 9.23\times 10^{-3} $ & $ 1.41\times 10^{-3} $ & $ 9.23\times 10^{-4} $ & $ 2.19\times 10^{-4} $ \\ \hline
$m_{H^{\pm}}$ Cut  &&&&& \\
$M_{\tau^+\tau^-l^{\pm}} < 200 $GeV & $2.75\times 10^{-3}$ & $ 4.04\times 10^{-3} $ & $ 4.17\times 10^{-4} $ & $ 3.94\times 10^{-4} $ & $ 1.09\times 10^{-4} $ \\
$m_{A^0}$ Cut &&&&& \\
$ 40 < M_{\tau^+\tau^-} < 70 $GeV & $2.35\times 10^{-3}$ & $ 2.20\times 10^{-3} $ & $ 1.96\times 10^{-4} $ & $ 2.29\times 10^{-4} $ & $ 6.63\times 10^{-5} $ \\
\end{tabular}
\end{ruledtabular}
\end{table}

Again, even we can get a good signal-to-background ratio, 
the total number of events is still small.
We further extend the signal-background analysis to the proposed 27
TeV $pp$ collider (HE-LHC). Similar as before, we tighten both
$m_{H^{\pm}}$ and $ m_{A^0} $ cuts relative to those in 
Eqs.~(\ref{eq:MCHcu-3}) and (\ref{eq:MAcut-3}).  For $m_{H^{\pm}}$
cuts at least two tau leptons and a lepton have to satisfy
\beqa
M_{\tau^+\tau^- l^{\pm}}\leq M_{H^{\pm}} -5 \;{\rm GeV}.
\label{eq:MCHcut27-3}
\eeqa
For $ m_{A^0} $ cuts at least a pair of opposite-sign tau
leptons is required to around the mass of $A^0$:\footnote{Here we
  apply an asymmetric mass window cut for $ M_{\tau^+\tau^-} $ 
  based on the shift of peak shape in the right panel of 
  Fig.~\ref{fig:kin2} and in order to veto the pair of
  opposite-sign tau leptons from the Z-pole.}
\beqa \quad m_{A^0} -
25 \;{\rm GeV} \leq M_{\tau^+\tau^-} \leq m_{A^0} +5\; {\rm GeV}.
\label{eq:MAcut27-3}
\eeqa
Other preselection cuts in Eqs.~(\ref{eq:trigger-3}),
(\ref{eq:jf-cut-3}) and (\ref{eq:b-veto}) are imposed, as before.
We choose the same signal benchmark point to
illustrate the cut flow under a sequence of selection cuts at $\sqrt{s}= 27 $ TeV in Table~\ref{tab:SSCF-3-27}.

\begin{figure}[t!]
  \centering
\includegraphics[width=0.48\textwidth]{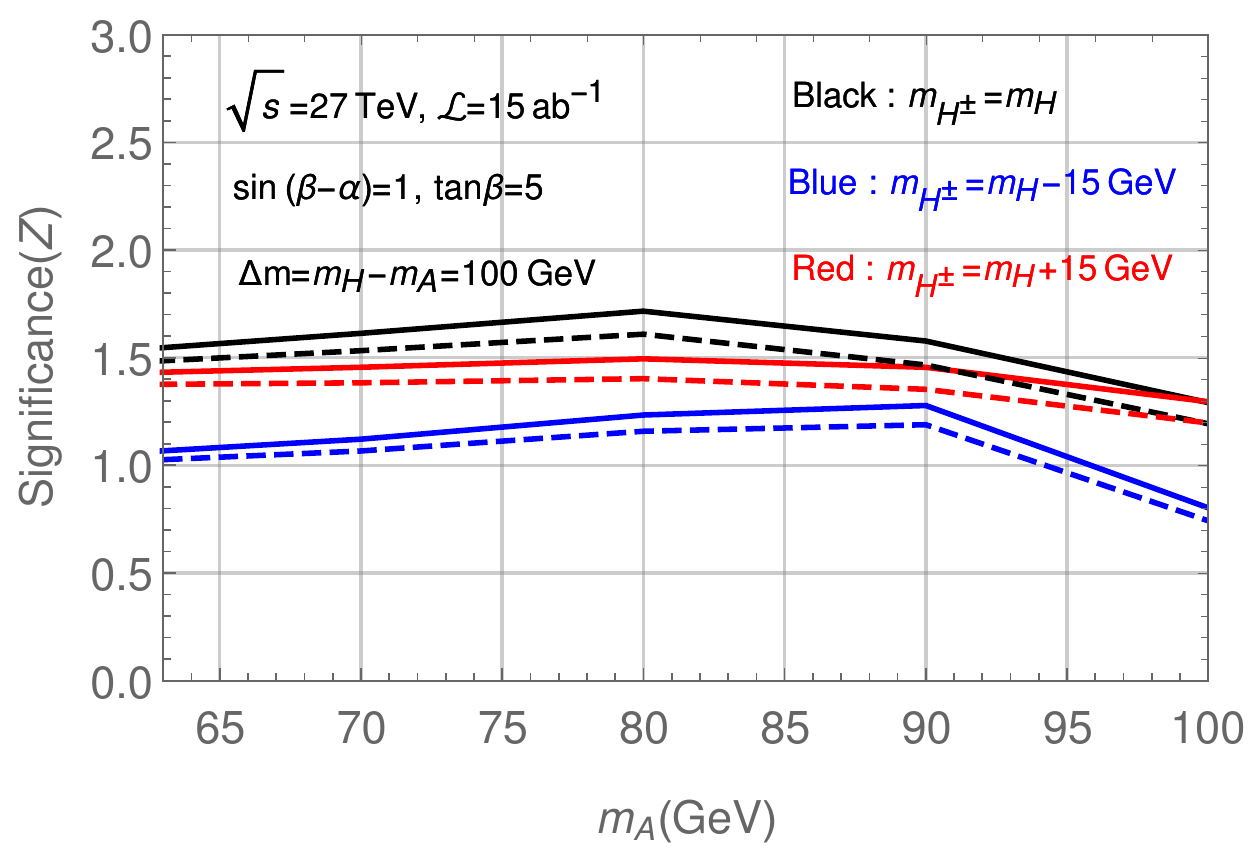}
\includegraphics[width=0.48\textwidth]{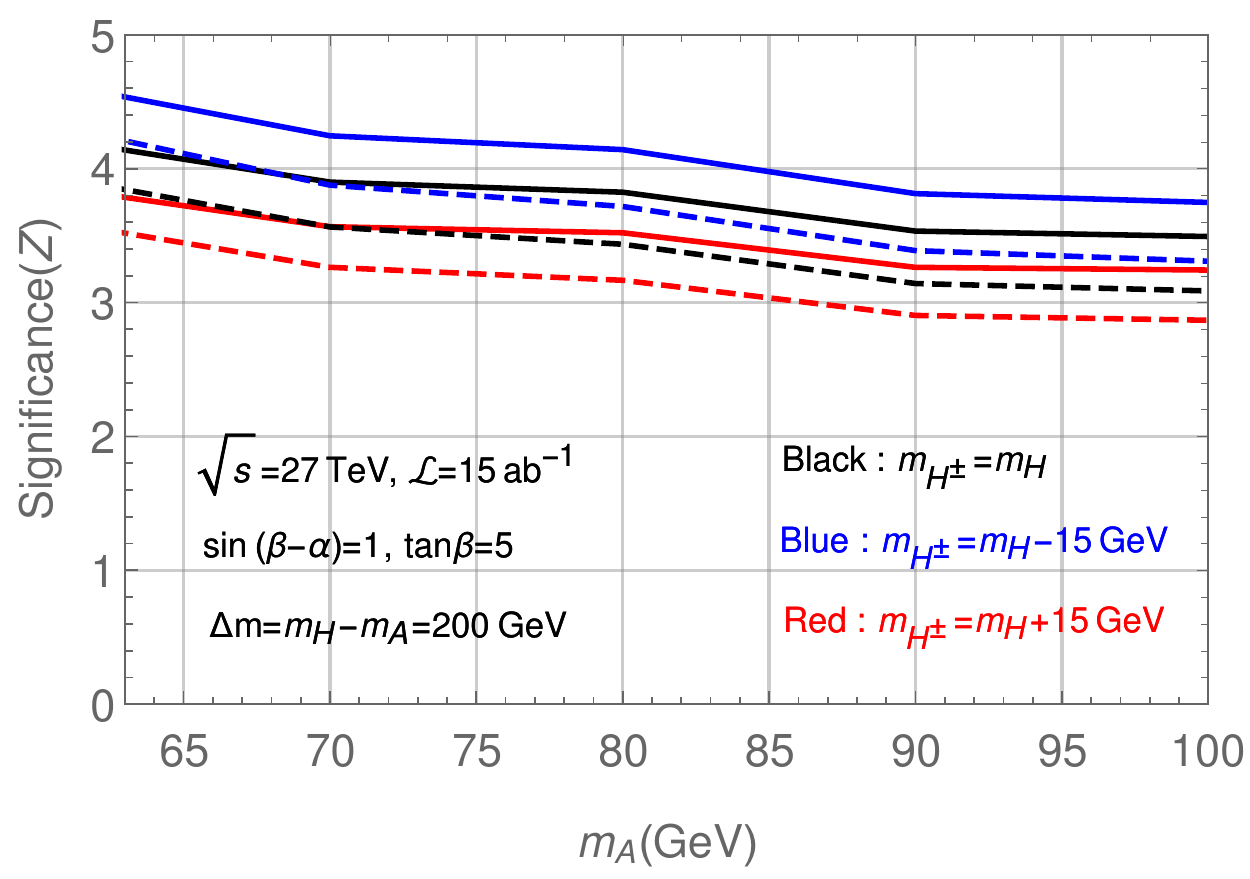}
\includegraphics[width=0.48\textwidth]{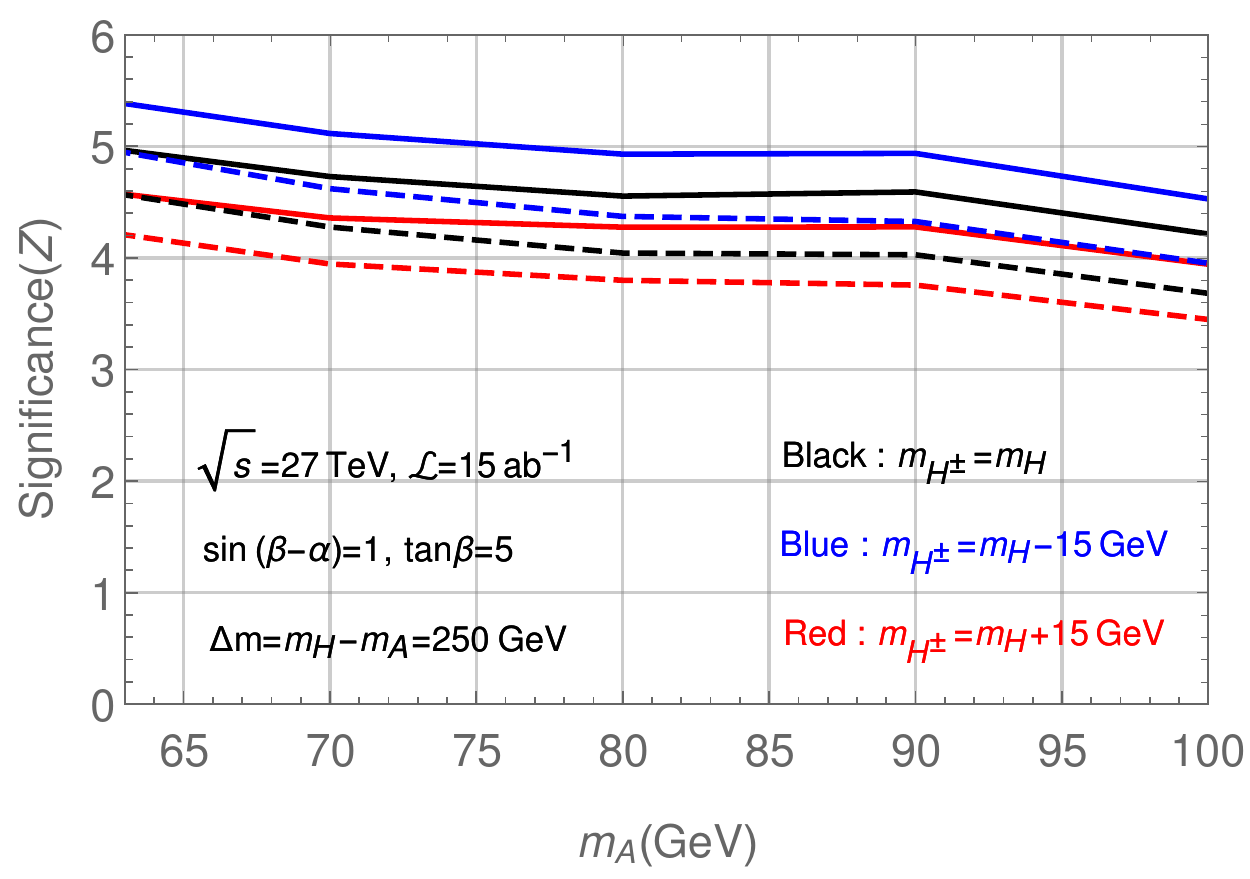}
\caption{\small
The same as Fig.~\ref{fig:TP1SUM}, but in Type-X 2HDM.
}
\label{fig:TP3SUM}
\end{figure}

Finally, we summarize the results for signal-background analysis
of Type-X 2HDM
at $\sqrt{s}= 27 $ TeV with luminosity $ \mathcal L = 15 ab^{-1} $ in
Fig.~\ref{fig:TP3SUM}.  The preselection cuts in
Eqs.~(\ref{eq:trigger-3}), (\ref{eq:jf-cut-3}), (\ref{eq:MCHcut27-3})
and (\ref{eq:MAcut27-3}) are imposed as before.  We vary $ m_{A^0} $ from
63 to 100 GeV with fixed $\sin(\beta -\alpha) = 1 $ and
$\tan\beta = 5 $ for $ \Delta m = m_{H^0} - m_{A^0} = $ 100 GeV
(upper-left panel), 200 GeV
(upper-right panel), and 250 GeV (lower panel) in
Fig.~\ref{fig:TP3SUM} as the illustrative examples. The black lines are $ m_{H^{\pm}} = m_{H^0} $, the
blue lines are $ m_{H^{\pm}} = m_{H^0} - 15 $ GeV, and the red lines are
$m_{H^{\pm}} = m_{H^0} + 15 $ GeV.  The maximum significance can reach to
about $Z=1.7$ at $ m_{A^0} = $ 80 GeV for $ \Delta m $ = 100 GeV. Notice
that the mass spectrum with $ \Delta m $ = 100 GeV and
$ m_{H^{\pm}} = m_{H^0} - 15 $ GeV in Type-X 2HDM will produce sizable
$B(H^{\pm}\rightarrow\tau\nu_{\tau}) $ and suppress
$B(H^{\pm}\rightarrow W^{\pm}A^0) $. That makes reduction of the
significance for the blue line in the upper-left panel in
Fig.~\ref{fig:TP3SUM}. On the other hand, the significance can reach to
more than $Z=3$ for $ m_{A^0} $ from 63 GeV to 100 GeV with
$ \Delta m $ =
200 GeV, and its maximum is about $Z=4.5$ at $ m_{A^0} = $ 63 GeV. Moreover, the significance can grow to more than $Z=4$
for $ m_{A^0} $ from 63 GeV to 100 GeV with $ \Delta m $ = 250 GeV, and
its maximum can further reach to $Z=5.4$ at $ m_{A^0} = $ 63 GeV.
Again, the $5\%$ systematic errors of the SM background events in Eq.~(\ref{eq:uncertainty})
are shown as dashed lines in Fig.~\ref{fig:TP3SUM} for comparisons.

\section{Conclusions}

Extending the minimal Higgs sector is one of the approaches to address
some weakness of the SM. Such extensions can give rise to rich
phenomenology.
The 2HDM is one of the most popular extended models in literature.
Exploring the whole mass 
spectrum in 2HDM is undoubtedly an important mission to help us 
understand the mystery of electroweak symmetry breaking. There are only 
a few examples that can cover the effects of all new scalar masses 
in a single process.
We have studied a novel process -- production of same-sign charged Higgs
production shown in Eq.~(\ref{eq:2W2A2j}),
which was first proposed in Ref.~\cite{Aiko:2019mww}.
It allows one to probe the whole mass spectrum in the 2HDM for
some specific mass relations.

We have investigated same-sign charged Higgs-boson production via
vector-boson-fusion at the HL-LHC and HE-LHC (27 TeV) in
Type I and X 2HDM's. The dependence 
of the production cross section
on the mass difference $\Delta m \equiv m_{H^0} - m_{A^0}$ between
the heavier scalar boson $H^0$ and the pseudoscalar boson $A^0$ is studied.
The scattering amplitude of the key subprocess $W^+ W^+ \to H^+ H^+$ is
proportional to $\Delta m$ as shown in Eq.~(\ref{amp}),
such that the production cross section
nearly vanishes in the limit $\Delta m \to 0$.  
As we mentioned
before, even if the mass splitting $\Delta m$ can be determined by
separately measuring $m_{H^0}$ and $m_{A^0}$ from other production
channels of $H^0$ and $A^0$, the measurement of same-sign charged
Higgs-boson production cross section can be used to understand the
mass spectrum of the heavier scalar and pseudoscalar bosons in the 2HDMs.

%

Given the constraints from electroweak precision, B physics, and
direct searches at colliders, we have explored the allowed parameter
space in $m_{H^{\pm}},\,\tan\beta,\,\Delta m$. Then we investigated
the sensitivity to the allowed parameter space at the HL-LHC and
HE-LHC, especially we have made use of the bosonic channel $W^\pm A^0$
of the charged Higgs boson, which is complementary to the study in
Ref.~\cite{Aiko:2019mww}.

In type I 2HDM, we used the decay channel 
$H^\pm H^\pm \to (W^\pm A^0) ( W^\pm A^0) \to (l^\pm \nu b\bar b)\,
(l^\pm \nu b\bar b)$ together with a pair of forward jets to perform
the signal-background analysis. At the end, we found about 4 signal
events versus 5 background events at HL-LHC with luminosity of
3000 fb$^{-1}$ for a typical benchmark point. At the HE-LHC,
significance level of $2.6-5.8$ can be achieved for
$\Delta m = 200 - 250$ GeV.

On the other hand, in type X 2HDM we used the decay channel
$H^\pm H^\pm \to (W^\pm A^0) ( W^\pm A^0) \to (l^\pm \nu \tau^+ \tau^-)\,
(l^\pm \nu \tau^+\tau^-)$ together with a pair of forward jets to
perform the signal-background analysis. At the HL-LHC, we can
achieve the signal-to-background ratio equal to 1, and the number of
signal events is about 2 for a luminosity of 3000 fb$^{-1}$. Nevertheless,
at the HE-LHC the significance can rise to the level of $3.2-5.4$ for $\Delta m = 200 - 250$ GeV. 

The main purpose of this study focuses on the search for a possible
large mass splitting between the neutral scalar and pseudoscalar
through same-sign charged-Higgs-boson production in 2HDMs via the
vector-boson fusion.  It is easy to see that this is not the discovery
mode of the charged Higgs because the event rate is much lower than
other direct processes, e.g., $g b \to t H^-$ or $gg \to t \bar b H^-$.  "
According to Ref.~\cite{CMS2, Sanyal:2019xcp} for the search of
$g b\rightarrow tH^-$ or $gg\rightarrow t\overline{b}H^-$ with 
$H^-\rightarrow\tau^-\overline{\nu}_{\tau}$, the constrained region is
$\tan\beta\lesssim$ 2\,(4) in type-I (type-X) 2HDM for the mass
range $160\leq M_{H^{\pm}}\leq 180$ GeV and there is no 
constraint for $M_{H^{\pm}} > 180$ GeV.

Notice that the process in Eq.~(\ref{eq:2W2A2j}) can be used to distinguish 
between the charged Higgs boson from a doublet and a triplet. Here, we take the Y=2    triplet model (type II seesaw) as an example.  In this model, the triplet VEV is highly constrained from electroweak precision measurement to be less than a few GeV \cite{Perez:2008ha,Melfo:2011nx,Arhrib:2011uy}. On the other hand, both $m_{H^\pm}$ and $m_{A^0}$ in this model are close to degenerate, therefore $H^\pm \to W^\pm A^0$ is very suppressed.  The observation of such decay would exclude type II seesaw model.

For the VEV of triplet around 
1 GeV and $ m_{H^{\pm}} < 400 $ GeV, the three dominant $H^{\pm}$ decay modes, $ H^{\pm}\rightarrow \{W^{\pm}h, W^{\pm}Z, tb\} $, are competitive \cite{Perez:2008ha}. If one can 
reconstruct the $Z/h$ invariant mass in the final state, it would be viewed as a clear 
signal beyond 2HDMs in the alignment limit.  
In the process in Eq.~(\ref{eq:2W2A2j}), besides   the t-channel $Z$ boson exchange 
and s-channel doubly charged Higgs contributions for the same-sign charged Higgs 
pair production would also show  the differences between the triplet model and 2HDMs.

However, in the case of tiny triplet VEV and high triplet mass scale, $ H^{\pm}\rightarrow \{ W^{\pm}h, W^{\pm}Z\}$ and also $H^{\pm\pm}W^\mp W^\mp$ couplings would be very suppressed. Therefore, the dominant decay  mode of $H^{\pm}$ turns out to be $ H^{\pm}\rightarrow l\nu_l $ and the doubly charged Higgs 
contribution in same-sign charged Higgs pair production would be small. 
Besides, because of the mass degeneracy between $A^0$ and $H^{\pm}$ 
in the Y=2 triplet model, we would not have 
$H^{\pm}\rightarrow W^{\pm}A^0$ decay mode. 
In the end, for the 
case of tiny triplet VEV, even the triplet model can mimic the same-sign charged 
higgs pair production with $ H^{\pm}\rightarrow \tau\nu_{\tau} $ in 2HDMs as shown in 
Ref.~\cite{Aiko:2019mww}.  The $H^{\pm} \rightarrow W^{\pm}A^0$ decay mode in this work  can help us to distinguish 2HDMs from the triplet model. 

One can also advocate $H^{\pm} \rightarrow W^{\pm}A^0$ decay channel 
to distinguish the 2HDM from the Minimal Supersymmetric Standard Model (MSSM)
which is a 2HDM of type II. However, because of the MSSM sum rules \cite{Gunion:1990kf}, we have $m_{H\pm}^2=m_{A^0}^2 + m_W^2$ which makes the decay channel
$H^{\pm} \rightarrow W^{\pm}A^0$ not open very often and turn out to be rather small. In fact, in the MSSM, $Br(H^{\pm} \rightarrow W^{\pm}A^0)$ 
is very suppressed (less than $10^{-2}$)
while $Br(H^{\pm} \rightarrow W^{\pm}h^0)$ is of the order of a few percent \cite{Djouadi:2005gj}. Therefore, the dominant decay of $H^\pm$ are $\tau\nu$ for low charged Higgs mass and $tb$ for $m_{H^\pm}>m_t+m_b$. In this case also,
the MSSM will mimic the same-sign charged higgs pair production with $ H^{\pm}\rightarrow \{\tau\nu_{\tau}, tb\} $ in 2HDMs as shown in  Ref.~\cite{Aiko:2019mww}. 

%
%

In summary, 
the process in Eq.~(\ref{eq:2W2A2j}) can be an additional check
of the mass relation between same-sign charged Higgs-boson production and 
$\Delta m$, especially,
if the 2HDM mass spectrum has the following relations:
\begin{itemize}
\item one light (pseudo)scalar, say $ A^0 $,
\item a large mass splitting between two neutral scalars, 
$\Delta m = (m_{H^0} - m_{A^0})$, and 
\item the charged Higgs mass is above the $W^\pm A^0$ threshold,
\end{itemize}
then this scenario in the 2HDMs can be either
pinned down or ruled out in the future.

\section*{Acknowledgment}  
The work of K.C. was supported by the National Science
Council of Taiwan under Grants Nos. MOST-105-2112-M-007-028-MY3 and
MOST-107-2112-M-007-029-MY3.
AA is supported in part by the Moroccan Ministry of 
Higher Education and Scientific Research under Contract N'PPR/2015/6 .
AA would like to thank NCTS for hospitality, where this work has been done.



\begin{thebibliography}{99}

\bibitem{Lee:1973iz} 
  T.~D.~Lee,
  Phys.\ Rev.\ D {\bf 8}, 1226 (1973).
  doi:10.1103/PhysRevD.8.1226



\bibitem{Branco:2011iw} 
  G.~C.~Branco, P.~M.~Ferreira, L.~Lavoura, M.~N.~Rebelo, M.~Sher and J.~P.~Silva,
  Phys.\ Rept.\  {\bf 516}, 1 (2012)
  doi:10.1016/j.physrep.2012.02.002
  [arXiv:1106.0034 [hep-ph]].



\bibitem{Glashow:1976nt} 
  S.~L.~Glashow and S.~Weinberg,
  Phys.\ Rev.\ D {\bf 15}, 1958 (1977).
  doi:10.1103/PhysRevD.15.1958



\bibitem{Barger:1993th} 
  V.~D.~Barger, R.~J.~N.~Phillips and D.~P.~Roy,
  Phys.\ Lett.\ B {\bf 324}, 236 (1994)
  doi:10.1016/0370-2693(94)90413-8
  [hep-ph/9311372].



\bibitem{Akeroyd:2016ymd} 
  A.~G.~Akeroyd {\it et al.},
  Eur.\ Phys.\ J.\ C {\bf 77}, no. 5, 276 (2017)
  doi:10.1140/epjc/s10052-017-4829-2
  [arXiv:1607.01320 [hep-ph]].



\bibitem{Abbiendi:2013hk} 
  G.~Abbiendi {\it et al.} [ALEPH and DELPHI and L3 and OPAL and LEP Collaborations],
  Eur.\ Phys.\ J.\ C {\bf 73}, 2463 (2013)
  doi:10.1140/epjc/s10052-013-2463-1
  [arXiv:1301.6065 [hep-ex]].



\bibitem{Abazov:2009wy} 
  V.~M.~Abazov {\it et al.} [D0 Collaboration],
  Phys.\ Rev.\ D {\bf 80}, 051107 (2009)
  doi:10.1103/PhysRevD.80.051107
  [arXiv:0906.5326 [hep-ex]].



\bibitem{ATLAS1}
  JHEP {\bf 1503}, 088 (2015)
  [arXiv:1412.6663 [hep-ex]].
\bibitem{ATLAS2}
  M.~Aaboud {\it et al.} [ATLAS Collaboration],
  JHEP {\bf 1809} (2018) 139
  doi:10.1007/JHEP09(2018)139
  [arXiv:1807.07915 [hep-ex]].
  \bibitem{ATLAS3}
  M.~Aaboud {\it et al.} [ATLAS Collaboration],
  Phys.\ Lett.\ B {\bf 759}, 555 (2016)
  [arXiv:1603.09203 [hep-ex]].
 
\bibitem{CMS1}
  V.~Khachatryan {\it et al.} [CMS Collaboration],
  JHEP {\bf 1511}, 018 (2015)
  [arXiv:1508.07774 [hep-ex]].
  \bibitem{CMS2}
  A.~M.~Sirunyan {\it et al.} [CMS Collaboration],
  arXiv:1903.04560 [hep-ex].
  \bibitem{CMS3}
  CMS Collaboration [CMS Collaboration],
  CMS-PAS-HIG-16-031.

\bibitem{Aad:2013hla} 
  G.~Aad {\it et al.} [ATLAS Collaboration],
  Eur.\ Phys.\ J.\ C {\bf 73}, no. 6, 2465 (2013)
  doi:10.1140/epjc/s10052-013-2465-z
  [arXiv:1302.3694 [hep-ex]].



\bibitem{ATLAS:2016qiq} 
  The ATLAS collaboration [ATLAS Collaboration],
  ATLAS-CONF-2016-089.



\bibitem{Aaboud:2018cwk} 
  M.~Aaboud {\it et al.} [ATLAS Collaboration],
  JHEP {\bf 1811}, 085 (2018)
  doi:10.1007/JHEP11(2018)085
  [arXiv:1808.03599 [hep-ex]].



\bibitem{CMS:1900zym} 
  CMS Collaboration [CMS Collaboration],
  CMS-PAS-HIG-18-004.



\bibitem{CMS:2019yat} 
  CMS Collaboration [CMS Collaboration],
  CMS-PAS-HIG-18-015.



\bibitem{Arhrib:2016wpw} 
  A.~Arhrib, R.~Benbrik and S.~Moretti,
  Eur.\ Phys.\ J.\ C {\bf 77}, no. 9, 621 (2017)
  doi:10.1140/epjc/s10052-017-5197-7
  [arXiv:1607.02402 [hep-ph]].



\bibitem{Kling:2015uba} 
  F.~Kling, A.~Pyarelal and S.~Su,
  JHEP {\bf 1511}, 051 (2015)
  doi:10.1007/JHEP11(2015)051
  [arXiv:1504.06624 [hep-ph]].



\bibitem{Coleppa:2014cca} 
  B.~Coleppa, F.~Kling and S.~Su,
  JHEP {\bf 1412}, 148 (2014)
  doi:10.1007/JHEP12(2014)148
  [arXiv:1408.4119 [hep-ph]].



\bibitem{Dermisek:2012cn} 
  R.~Dermisek, E.~Lunghi and A.~Raval,
  JHEP {\bf 1304}, 063 (2013)
  doi:10.1007/JHEP04(2013)063
  [arXiv:1212.5021 [hep-ph]].



\bibitem{Akeroyd:2007yj} 
  A.~G.~Akeroyd, A.~Arhrib and Q.~S.~Yan,
  Eur.\ Phys.\ J.\ C {\bf 55}, 653 (2008)
  doi:10.1140/epjc/s10052-008-0617-3
  [arXiv:0712.3933 [hep-ph]].



\bibitem{Abdallah:2003wd} 
  J.~Abdallah {\it et al.} [DELPHI Collaboration],
  Eur.\ Phys.\ J.\ C {\bf 34}, 399 (2004)
  doi:10.1140/epjc/s2004-01732-6
  [hep-ex/0404012].



\bibitem{Sirunyan:2019zdq} 
  A.~M.~Sirunyan {\it et al.} [CMS Collaboration],
  Phys.\ Rev.\ Lett.\  {\bf 123}, no. 13, 131802 (2019)
  doi:10.1103/PhysRevLett.123.131802
  [arXiv:1905.07453 [hep-ex]].



\bibitem{Aiko:2019mww} 
  M.~Aiko, S.~Kanemura and K.~Mawatari,
  Phys.\ Lett.\ B {\bf 797}, 134854 (2019)
  doi:10.1016/j.physletb.2019.134854
  [arXiv:1906.09101 [hep-ph]].



\bibitem{Aoki:2009ha} 
  M.~Aoki, S.~Kanemura, K.~Tsumura and K.~Yagyu,
  Phys.\ Rev.\ D {\bf 80}, 015017 (2009)
  doi:10.1103/PhysRevD.80.015017
  [arXiv:0902.4665 [hep-ph]].



\bibitem{Barger:1989fj} 
  V.~D.~Barger, J.~L.~Hewett and R.~J.~N.~Phillips,
  Phys.\ Rev.\ D {\bf 41}, 3421 (1990).
  doi:10.1103/PhysRevD.41.3421



\bibitem{Lee:1977eg} 
  B.~W.~Lee, C.~Quigg and H.~B.~Thacker,
  Phys.\ Rev.\ D {\bf 16}, 1519 (1977).
  doi:10.1103/PhysRevD.16.1519



\bibitem{Kanemura:1993hm} 
  S.~Kanemura, T.~Kubota and E.~Takasugi,
  Phys.\ Lett.\ B {\bf 313}, 155 (1993)
  doi:10.1016/0370-2693(93)91205-2
  [hep-ph/9303263].



\bibitem{Kanemura:2015ska} 
  S.~Kanemura and K.~Yagyu,
  Phys.\ Lett.\ B {\bf 751}, 289 (2015)
  doi:10.1016/j.physletb.2015.10.047
  [arXiv:1509.06060 [hep-ph]].



\bibitem{Tanabashi:2018oca} 
  M.~Tanabashi {\it et al.} [Particle Data Group],
  Phys.\ Rev.\ D {\bf 98}, no. 3, 030001 (2018).
  doi:10.1103/PhysRevD.98.030001



\bibitem{Haber:1992py} 
  H.~E.~Haber and A.~Pomarol,
  Phys.\ Lett.\ B {\bf 302}, 435 (1993)
  doi:10.1016/0370-2693(93)90423-F
  [hep-ph/9207267].



\bibitem{Misiak:2017bgg} 
  M.~Misiak and M.~Steinhauser,
  Eur.\ Phys.\ J.\ C {\bf 77}, no. 3, 201 (2017)
  doi:10.1140/epjc/s10052-017-4776-y
  [arXiv:1702.04571 [hep-ph]].



\bibitem{Misiak:2015xwa} 
  M.~Misiak {\it et al.},
  Phys.\ Rev.\ Lett.\  {\bf 114}, no. 22, 221801 (2015)
  doi:10.1103/PhysRevLett.114.221801
  [arXiv:1503.01789 [hep-ph]].



\bibitem{Enomoto:2015wbn} 
  T.~Enomoto and R.~Watanabe,
  JHEP {\bf 1605}, 002 (2016)
  doi:10.1007/JHEP05(2016)002
  [arXiv:1511.05066 [hep-ph]].



\bibitem{Haller:2018nnx} 
  J.~Haller, A.~Hoecker, R.~Kogler, K.~Mönig, T.~Peiffer and J.~Stelzer,
  Eur.\ Phys.\ J.\ C {\bf 78}, no. 8, 675 (2018)
  doi:10.1140/epjc/s10052-018-6131-3
  [arXiv:1803.01853 [hep-ph]].



\bibitem{Eriksson:2009ws} 
  D.~Eriksson, J.~Rathsman and O.~Stal,
  Comput.\ Phys.\ Commun.\  {\bf 181}, 189 (2010)
  doi:10.1016/j.cpc.2009.09.011
  [arXiv:0902.0851 [hep-ph]].



\bibitem{Aad:2015typ} 
  G.~Aad {\it et al.} [ATLAS Collaboration],
  JHEP {\bf 1603}, 127 (2016)
  doi:10.1007/JHEP03(2016)127
  [arXiv:1512.03704 [hep-ex]].



\bibitem{Sanyal:2019xcp} 
  P.~Sanyal,
  Eur.\ Phys.\ J.\ C {\bf 79}, no. 11, 913 (2019)
  doi:10.1140/epjc/s10052-019-7431-y
  [arXiv:1906.02520 [hep-ph]].



\bibitem{Sirunyan:2018wim} 
  A.~M.~Sirunyan {\it et al.} [CMS Collaboration],
  JHEP {\bf 1811}, 161 (2018)
  doi:10.1007/JHEP11(2018)161
  [arXiv:1808.01890 [hep-ex]].



\bibitem{CMS:2019hvr} 
  A.~M.~Sirunyan {\it et al.} [CMS Collaboration],
  JHEP {\bf 1905}, 210 (2019)
  doi:10.1007/JHEP05(2019)210
  [arXiv:1903.10228 [hep-ex]].



\bibitem{Sirunyan:2018aui} 
  A.~M.~Sirunyan {\it et al.} [CMS Collaboration],
  Phys.\ Lett.\ B {\bf 793}, 320 (2019)
  doi:10.1016/j.physletb.2019.03.064
  [arXiv:1811.08459 [hep-ex]].



\bibitem{ATLAS:2018xad} 
  The ATLAS collaboration [ATLAS Collaboration],
  ATLAS-CONF-2018-025.



\bibitem{Aaboud:2018esj} 
  M.~Aaboud {\it et al.} [ATLAS Collaboration],
  Phys.\ Lett.\ B {\bf 790}, 1 (2019)
  doi:10.1016/j.physletb.2018.10.073
  [arXiv:1807.00539 [hep-ex]].



\bibitem{Sirunyan:2018pzn} 
  A.~M.~Sirunyan {\it et al.} [CMS Collaboration],
  Phys.\ Lett.\ B {\bf 785}, 462 (2018)
  doi:10.1016/j.physletb.2018.08.057
  [arXiv:1805.10191 [hep-ex]].



\bibitem{Aaboud:2018iil} 
  M.~Aaboud {\it et al.} [ATLAS Collaboration],
  JHEP {\bf 1810}, 031 (2018)
  doi:10.1007/JHEP10(2018)031
  [arXiv:1806.07355 [hep-ex]].



\bibitem{Sirunyan:2018mot} 
  A.~M.~Sirunyan {\it et al.} [CMS Collaboration],
  Phys.\ Lett.\ B {\bf 795}, 398 (2019)
  doi:10.1016/j.physletb.2019.06.021
  [arXiv:1812.06359 [hep-ex]].



\bibitem{Sirunyan:2018mbx} 
  A.~M.~Sirunyan {\it et al.} [CMS Collaboration],
  JHEP {\bf 1811}, 018 (2018)
  doi:10.1007/JHEP11(2018)018
  [arXiv:1805.04865 [hep-ex]].



\bibitem{Aaboud:2018eoy} 
  M.~Aaboud {\it et al.} [ATLAS Collaboration],
  Phys.\ Lett.\ B {\bf 783}, 392 (2018)
  doi:10.1016/j.physletb.2018.07.006
  [arXiv:1804.01126 [hep-ex]].



\bibitem{Bechtle:2013wla} 
  P.~Bechtle, O.~Brein, S.~Heinemeyer, O.~Stål, T.~Stefaniak, G.~Weiglein and K.~E.~Williams,
  Eur.\ Phys.\ J.\ C {\bf 74}, no. 3, 2693 (2014)
  doi:10.1140/epjc/s10052-013-2693-2
  [arXiv:1311.0055 [hep-ph]].



\bibitem{Bechtle:2013xfa} 
  P.~Bechtle, S.~Heinemeyer, O.~Stål, T.~Stefaniak and G.~Weiglein,
  Eur.\ Phys.\ J.\ C {\bf 74}, no. 2, 2711 (2014)
  doi:10.1140/epjc/s10052-013-2711-4
  [arXiv:1305.1933 [hep-ph]].



\bibitem{Degrande:2014vpa} 
  C.~Degrande,
  Comput.\ Phys.\ Commun.\  {\bf 197}, 239 (2015)
  doi:10.1016/j.cpc.2015.08.015
  [arXiv:1406.3030 [hep-ph]].



\bibitem{Alwall:2014hca} 
  J.~Alwall {\it et al.},
  JHEP {\bf 1407}, 079 (2014)
  doi:10.1007/JHEP07(2014)079
  [arXiv:1405.0301 [hep-ph]].



\bibitem{Sjostrand:2007gs} 
  T.~Sjostrand, S.~Mrenna and P.~Z.~Skands,
  Comput.\ Phys.\ Commun.\  {\bf 178}, 852 (2008)
  doi:10.1016/j.cpc.2008.01.036
  [arXiv:0710.3820 [hep-ph]].



\bibitem{delphes3}
J.~de Favereau {\it et al.}  [DELPHES 3 Collaboration],
  JHEP {\bf 1402}, 057 (2014)
  [arXiv:1307.6346 [hep-ex]].

\bibitem{MA5}
 E. Conte, B. Fuks and G. Serret,
    Comput. Phys. Commun. 184 (2013) 222, [arXiv:1206.1599 [hep-ph]]
;E. Conte, B. Dumont, B. Fuks and C. Wymant,
     Eur. Phys. J. C 74 (2014) 10, 3103, [arXiv:1405.3982 [hep-ph]]
;B. Dumont, B. Fuks, S. Kraml et al.,
      Eur. Phys. J. C 75 (2015) 2, 56, [arXiv:1407.3278 [hep-ph]].

\bibitem{Perez:2008ha} 
  P.~Fileviez Perez, T.~Han, G.~y.~Huang, T.~Li and K.~Wang,
  Phys.\ Rev.\ D {\bf 78}, 015018 (2008)
  doi:10.1103/PhysRevD.78.015018
  [arXiv:0805.3536 [hep-ph]].



\bibitem{Melfo:2011nx} 
  A.~Melfo, M.~Nemevsek, F.~Nesti, G.~Senjanovic and Y.~Zhang,
  Phys.\ Rev.\ D {\bf 85}, 055018 (2012)
  doi:10.1103/PhysRevD.85.055018
  [arXiv:1108.4416 [hep-ph]].



\bibitem{Arhrib:2011uy} 
  A.~Arhrib, R.~Benbrik, M.~Chabab, G.~Moultaka, M.~C.~Peyranere, L.~Rahili and J.~Ramadan,
  Phys.\ Rev.\ D {\bf 84}, 095005 (2011)
  doi:10.1103/PhysRevD.84.095005
  [arXiv:1105.1925 [hep-ph]].



\bibitem{Gunion:1990kf} 
  J.~F.~Gunion, H.~E.~Haber and J.~Wudka,
  Phys.\ Rev.\ D {\bf 43}, 904 (1991).
  doi:10.1103/PhysRevD.43.904



\bibitem{Djouadi:2005gj} 
  A.~Djouadi,
  Phys.\ Rept.\  {\bf 459}, 1 (2008)
  doi:10.1016/j.physrep.2007.10.005
  [hep-ph/0503173].
\end{thebibliography}
\end{document}